\documentclass[12pt]{iopart}

\expandafter\let\csname equation*\endcsname\relax
\expandafter\let\csname endequation*\endcsname\relax

\usepackage{amsfonts}
\usepackage{amsmath}
\usepackage{amssymb}
\usepackage{bm}
\usepackage{dcolumn}
\usepackage{graphicx}
\usepackage{graphics}
\usepackage[T1]{fontenc}
\usepackage[utf8]{inputenc}
\usepackage{latexsym}
\usepackage{rotating}
\usepackage{cite}
\usepackage[perpage,marginal]{footmisc}
\usepackage[colorlinks=true]{hyperref}
\usepackage[all]{hypcap}
\usepackage{xspace} 
\usepackage[usenames]{color}
\usepackage{mathrsfs}
\usepackage{microtype}

\graphicspath{{fig/}}

\makeatletter
\renewcommand\footnoterule{%
  \kern-3\p@
  \hrule\@width2.5cm
  \kern2.6\p@}
\makeatother

\definecolor {darkgreen}{rgb}{0.2,0.7,0.2}

\newcommand\be{\begin{equation}}
\newcommand\ba{\begin{eqnarray}}
\newcommand\ee{\end{equation}}
\newcommand\ea{\end{eqnarray}}

\newcommand\bw{\begin{widetext}}
\newcommand\ew{\end{widetext}}

\newcommand{\nn}{\nonumber}

\newcommand{\Schw}{{\mbox{\tiny Schw}}}

\newcommand{\ppE}{{\mbox{\tiny ppE}}}

\newcommand{\ISCO}{{\mbox{\tiny ISCO}}}

\newcommand{\GR}{{\mbox{\tiny GR}}}
\newcommand{\MG}{{\mbox{\tiny MG}}}

\newcommand{\MAT}{{\mbox{\tiny mat}}}

\newcommand{\CS}{{\mbox{\tiny CS}}}

\newcommand{\pd}{\partial}
\newcommand{\cd}{\nabla}

\newcommand{\BH}{{\mbox{\tiny BH}}}
\newcommand{\NS}{{\mbox{\tiny NS}}}

\newcommand{\cmdot}{C_{\dot{M}}}
\newcommand{\so}{M_\odot}
\newcommand{\lla}{\left\langle}

\newcommand{\rra}{\right\rangle}

\newcommand{\rch}{{\cal R}}
\newcommand{\beq}{\begin{equation}}
\newcommand{\beqa}{\begin{eqnarray}}
\newcommand{\eeq}{\end{equation}}
\newcommand{\eeqa}{\end{eqnarray}}
\newcommand{\et}{{\it et al.}}

\newcommand{\mc}{\mathcal{M}}

\newcommand{\HH}{{\mbox{\tiny H}}}

\newcommand{\inter}{\mathrm{int}}

\newcommand{\mrm}{\mathrm}

\newcommand{\GB}{{\mbox{\tiny GB}}}
\newcommand{\DQG}{{\mbox{\tiny dQG}}}
\newcommand{\EH}{{\mbox{\tiny EH}}}

\newcommand{\EDGB}{{\mbox{\tiny EdGB}}}
\newcommand{\BD}{{\mbox{\tiny BD}}}

\newcommand{\RL}{{\mbox{\tiny R,L}}}
\newcommand{\R}{{\mbox{\tiny R}}}
\newcommand{\LL}{{\mbox{\tiny L}}}

\newcommand{\txt}[1]{{\textrm{\tiny{#1}}}}
\newcommand{\mpl}{m_\txt{pl}}
\newcommand{\avg}[1]{\left< #1 \right>}
\newcommand{\lstf}{\langle}
\newcommand{\rstf}{\rangle}
\newcommand{\plusonetwo}{+\left( 1\leftrightarrow 2 \right)}

\begin{document}

\newcommand{\ourtitle}{Black Hole Based Tests of General Relativity}
\title{\ourtitle}

\author{Kent Yagi}

\address{Department of Physics, Princeton University, Princeton, NJ 08544, USA.}
\address{Department of Physics, Montana State University, Bozeman, MT 59717, USA.}
\ead{kyagi@princeton.edu}

\author{Leo C.~Stein}

\address{Theoretical Astrophysics,
Walter Burke Institute for Theoretical Physics,
California Institute of Technology, Pasadena, CA
91125, USA}
\address{Cornell Center for Astrophysics and
Planetary Science (CCAPS), Cornell University, Ithaca, NY 14853, USA}

\vspace{10pt}

\begin{abstract}

General relativity has passed all solar system experiments and
neutron star based tests, such as binary pulsar observations, with
flying colors.
A more exotic arena for testing general relativity is in systems
that contain one or more black holes.
Black holes are the most compact objects in the universe, providing
probes of the strongest-possible gravitational fields.
We are motivated to study strong-field gravity since many theories
give large deviations from general relativity only at large field
strengths, while recovering the weak-field behavior.
In this article, we review how one can probe general relativity and
various alternative theories of gravity by using electromagnetic waves
from a black hole with an accretion disk, and gravitational waves from
black hole binaries.
We first review model-independent ways of testing gravity with
electromagnetic/gravitational waves from a black hole system.
We then focus on selected examples of theories that extend general
relativity in rather simple ways.
Some important characteristics of general relativity include 
(but are not limited to)
(i) only tensor gravitational degrees of freedom,
(ii) the graviton is massless,
(iii) no quadratic or higher curvatures in the action, and
(iv) the theory is 4 dimensional.
Altering a characteristic leads to a different extension of general relativity:
(i) scalar-tensor theories,
(ii) massive gravity theories,
(iii) quadratic gravity, and
(iv) theories with large extra dimensions.
Within each theory, we describe black hole solutions, their properties,
and current and projected constraints on each theory using black
hole-based tests of gravity.
We close this review by listing some of the open problems in
model-independent tests and within each specific theory.

\end{abstract}
\newpage{}
\tableofcontents
\markboth{\ourtitle}{\ourtitle}
\newpage{}
\section{Introduction}
\label{sec:introduction}

This past year, 2015, was the centennial anniversary of Einstein proposing
general relativity (GR).
So far, GR has passed all tests of gravity with flying colors.
Such tests include solar system experiments~\cite{TEGP,Will:2014kxa},
binary pulsar observations~\cite{lrr-2003-5,Wex:2014nva}
and table-top experiments~\cite{Murata:2014nra}.
These tests are restricted to either weak-field, nondynamical
or mildly dynamical situations. 
Precise tests of gravity in the strong-field/dynamical regime are 
necessary to probe certain alternative theories of gravity
that can show a large deviation only in such a regime.
Black holes (BHs) offer an excellent testbed to probe 
strong-field gravity due to their large internal gravity.
BH based tests of gravity have advantages over 
neutron star (NS) based tests since the latter in general
have large systematic errors due to uncertainties in nuclear physics
(though see e.g.~\cite{I-Love-Q-Science,I-Love-Q-PRD} that discusses how one can
in principle project out such uncertainties and perform strong field
tests of gravity with NSs).

\subsection{Observations}

One can either use gravitational wave (GW)~\cite{Gair:2012nm,Yunes:2013dva} or electromagnetic wave (EMW)~\cite{lrr-2008-9,Bambi:2015kza} 
observations to probe strong-field gravity with BHs.

\subsubsection{GWs}

Regarding the former, the gravitational waveform of an inspiraling compact
binary depends on the binary's binding energy and 
the energy flux carried out to infinity. In alternative theories of gravity,
BH solutions are in general different from the Kerr one,\footnote{%
Though many theories exist in which the Kerr BH is also a solution to
the modified field equations~\cite{Psaltis:2007cw}.}
which modifies the 
binding energy of a binary system relative to GR. Furthermore, the additional gravitational degrees of freedom
(e.g.~a scalar field) also affect the binding energy, and generate
additional radiation, which modifies the energy flux, and in turn,
the binary's evolution. One can either pick a specific theory and calculate the correction to
the waveform from GR~\cite{will1994,damour-GW-ST,scharre,willyunes,bertibuonanno,yagiLISA,
yagiDECIGO,Arun:2013bp,Yagi:2013du,Berti:2012bp,Yunes:2011aa,Healy:2011ef,Berti:2013gfa,
will1998,arunwill,stavridis,keppel,delpozzo,bertisesana,DeFelice:2013nba,Narikawa:2014fua,Alexander:2007kv,
Yunes:2010yf,Sopuerta:2009iy,pani-DCS-EMRI,Yagi:2011xp,Yagi:2012vf,Canizares:2012is,inoue,mc,
yagi:brane}, or consider a parameterized waveform that captures deviations from 
GR in the waveform in a generic way~\cite{Arun:2006yw,arun-model-indep,Yunes:2009ke,cornish-PPE,Mirshekari:2011yq,Li:2011cg,
Huwyler:2011iq,Vallisneri:2012qq,Arun:2012hf,
mishra,Chatziioannou:2012rf,Vallisneri:2013rc,Sampson:2013lpa,Sampson:2013jpa,Sampson:2013wia,
Stein:2013wza,Vitale:2013bma,Agathos:2013upa,Huwyler:2014vva,DelPozzo:2014cla,
Loutrel:2014vja}. 
One then carries out a matched filtering analysis, in which
one takes a correlation between the GW signal and the theoretical template, to see how well one can
constrain non-GR theories with future GW interferometers. 
One can also carry out a Bayesian model selection analysis to see whether a hypothetical observation
favors GR or a specific non-GR model~\cite{Agathos:2013upa}.

Another interesting way of testing GR through GWs is to probe properties of a BH spacetime directly.
In particular, Kerr BHs enjoy the no-hair property
where higher order multipole moments are completely determined by the first two, namely the BH mass
and spin angular momentum~\cite{israel,israel2,carter-uniqueness,hawking-uniqueness,
hawking-uniqueness0,hansen:46,robinson}.
After a merger, one can use the BH ringdown frequencies and damping
times of different modes to check the consistency
of the no-hair relation and constrain possible non-GR effects~\cite{Detweiler:1980gk,Echeverria:1989hg,Kojima:1991np,Yoshida:1994xi,
finn1992,Flanagan:1997sx,Dreyer:2003bv,Hughes:2004vw,Berti:2005ys,Berti:2006qt,Balakrishna:2006ru,Berti:2007zu,
Chirenti:2007mk,Pani:2009ss,Kamaretsos:2011um,Gossan:2011ha,
Macedo:2013jja,Meidam:2014jpa,Nakano:2015uja}.
One can also carry out such no-hair tests with GWs from binary
inspirals, such as extreme mass ratio inspirals
(EMRIs)~\cite{Ryan:1995wh,Ryan:1997hg,Kesden:2004qx,Barack:2006pq,glampedakis,
Gair:2007kr,Apostolatos:2009vu,LukesGerakopoulos:2010rc,Contopoulos:2011dz,
Pani:2010em,Gair:2011ym,Wade:2013hoa,Macedo:2013qea,Macedo:2013jja}. 
The BH ringdown also allows us to 
probe the "firewall" effect~\cite{Barausse:2014tra,Barausse:2014pra}. 

\begin{figure}[htb]
\begin{center}
\includegraphics[width=12.cm,clip=true]{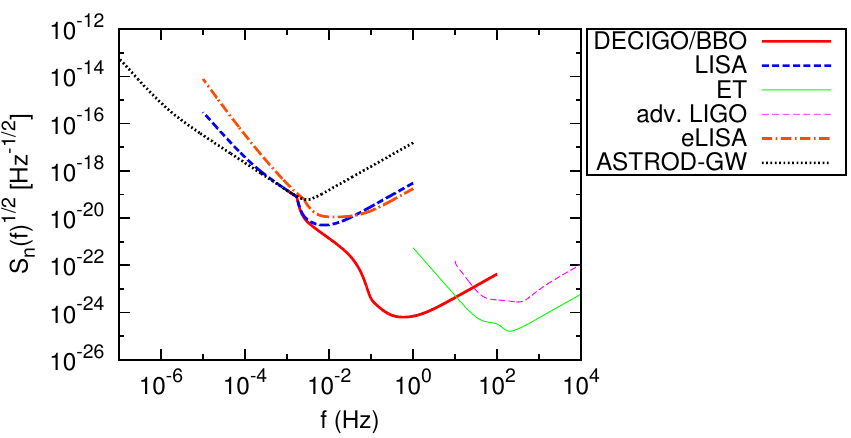}
\caption{\label{fig:noise}
  (Color online) Noise spectral density $S_n(f)$ of various GW interferometers. Adv.~LIGO
  and ET are the second- and third-generation ground-based detectors respectively.
  All the other detectors are the space-borne ones.  
  This figure is taken from~\cite{Yagi:2013du}.
  }
\end{center}
\end{figure}

Let us now describe various GW interferometers that are considered in this review. 
Figure~\ref{fig:noise}
presents the sensitivity curve of these detectors.
Currently, Adv.~LIGO is in operation, and other second-generation ground-based interferometers,
such as Adv.~Virgo or KAGRA, will come online shortly. These detectors have armlengths of 3--4km
and their best sensitivities
at around 100--1000Hz. The third-generation ground-based
interferometer, Einstein Telescope (ET), is an 
underground and cryogenic detector with armlengths of 10km, and may be realized 
in the second half of the 2020--2030 decade. ET roughly has a sensitivity that is one order of
magnitude better than second-generation ones.
Regarding space missions, eLISA was proposed for the theme of the ``Gravitational Wave Universe''
for the Cosmic Vision L3 mission of the European Space Agency. The base configuration has 
one mother and two daughter spacecrafts separated by $10^6$km, and eLISA is aimed to be 
launched in $2034$. It has its best sensitivity at around 1 mHz. DECIGO is similar to eLISA, but has
four triangular constellations in total, with two of them forming a star-of-David. DECIGO has its best sensitivity at around 0.1--1Hz, armlengths of $10^3$km, and is aimed to be launched in the
2030s.

Astrophysical environment around BH binaries, such as accretion disks, another SMBH, 
magnetic fields, non-vanishing charges, etc., is 
unlikely to spoil strong-field tests of GR. Although thin disks around EMRIs for 
eLISA~\cite{narayan,Barausse:2007dy,yunes-massive-perturber,yunes-disk,kocsis-disk,
Barausse:2014tra,Barausse:2014pra} 
and circumbinary disks around stellar-mass and intermediate-mass BH binaries for 
DECIGO~\cite{hayasaki:yagi} may have detectable effects, event rates are estimated to be 
rather small. 
Other non-astrophysical effects, such as cosmological constant, are also 
small~\cite{Barausse:2014tra,Barausse:2014pra}. Mismodelling of the waveform due to unknown
self-force effects~\cite{Barack:2009ux,Poisson:2011nh}
 might be problematic for testing GR with EMRIs and intermediate-mass ratio inspirals.

\subsubsection{EMWs}

Regarding EMW observations, one can use X-ray continuum spectra~\cite{Harko:2009rp,
Harko:2009gc,Harko:2010ua,
Bambi:2011jq,Bambi:2013sha,Vincent:2013uea,
Johannsen:2014wba,
Kong:2014wha,Bambi:2014sfa} and
Fe K$\alpha$ line emissions~\cite{Bambi:2013sha,Vincent:2013uea,Johannsen:2014wba,Johannsen:2012ng,Bambi:2012at,
Bambi:2013hza,Bambi:2013jda,Jiang:2014loa,Jiang:2015dla} 
that have been used to measure BH spins in GR. Since these spectra
depend strongly on the inner edge of the accretion disks, which are usually taken to be 
the innermost stable circular orbit (ISCO), such spectra can also be used to probe
deviations from the Kerr BH. One can also use BH shadow observations at (sub-)millimeter wavelengths
using very long baseline interferometry (VLBI)~\cite{Vincent:2013uea,Broderick:2005at,
Johannsen:2010ru,Amarilla:2010zq,Amarilla:2011fx,
Bambi:2011yz,Amarilla:2013sj,Li:2013jra,
Loeb:2013lfa,Atamurotov:2013dpa,Wei:2013kza,Atamurotov:2013sca,Johannsen:2015qca,
Tsukamoto:2014tja,Bambi:2014mla,Psaltis:2014mca,Wei:2015dua,Moffat:2015kva}. 
The size and shape of the shadow depends not only on the BH 
spin and inclination, but also on possible deviations from the Kerr solution. Although systematic errors are
quite large, one can in principle use quasi-periodic oscillation (QPO) frequencies to probe the 
BH spacetime~\cite{Vincent:2013uea,Johannsen:2010bi,Bambi:2012pa,Bambi:2013fea,Maselli:2014fca}.
We note that although EMW tests can, in principle, probe deviations from the Kerr spacetime,
this does not always mean that such tests can probe modified theories of gravity as
there are many theories that allow the Kerr BH as a solution to the modified field equations (see the footnote above).
On the other hand, GW tests can probe non-GR effects under the Kerr background since the perturbation
is in general modified from the GR one in alternative theories of gravity, including those that admit the Kerr solution~\cite{Barausse:2008xv}.

Regarding detector facilities, the continuum spectra, iron lines and QPOs have already been
observed by X-ray missions, including RXTE, Chandra, NuSTAR, ASCA, 
Suzaku, XMM-Newton and Beppo-SAX. Future X-ray missions include Astro-H, LOFT and Athena.
Astro-H is planned to be launched in 2016, while other two are aimed in the 2020s. 
Regarding (sub-)millimeter telescopes, Event Horizon Telescope 
is suitable for detecting BH shadows. It combines existing and planned (sub-)millimeter facilities into a high-sensitivity and high-angular-resolution telescope over the next decade.

Systematics on EMW observations for testing GR are more problematic than those in 
GW observations. One major origin is the correct modeling of the accretion disk. Other systematics
are due to uncertainties in e.g.~emissivities, hardening factors due to deviations from blackbody radiation and inclinations. QPOs also suffer from knowing the correct model \textit{a priori}.

\subsection{Alternative Theories of Gravity}

In this article, we first describe model-independent ways of testing GR using
GW and EMW observations as mentioned above. We then focus on testing four specific 
alternative theories of gravity as selected examples that 
extend one or more of the important characteristics of GR\footnote{%
BHs can be useful for probing theories other than those mentioned in
this review, e.g.~Einstein-\AE{}ther theory~\cite{Jacobson:2000xp,
  Jacobson:2008aj, Eling:2006ec, Barausse:2011pu, Barausse:2013nwa}
and string axiverse~\cite{Arvanitaki:2009fg, Arvanitaki:2010sy,
  Yoshino:2012kn, Yoshino:2013ofa, Yoshino:2014wwa,
  Arvanitaki:2014wva}.
}, namely
(i) gravity has tensor degrees of freedom, (ii) the graviton is massless,
(iii) no higher order curvature terms in the action, and (iv) a 4 dimensional theory.

The first example that we consider is scalar-tensor theories~\cite{fujii}. 
These theories are one of the most well-studied alternative theories of gravity and
 one of the simplest extensions of GR, namely introducing 
scalar gravitational degrees of freedom via scalar fields.
Such theories are motivated from the low energy effective theory of
string theory~\cite{polchinski1,polchinski2},
the late-time accelerated expansion of the Universe~\cite{perrotta},
and inflation~\cite{steinhardt}.
As in GR, the no-hair theorems also apply to BHs in the Jordan-Brans-Dicke type of scalar-tensor theories.
Namely, stationary, asymptotically flat BHs with a single real scalar
field in vacuum are the same as the Kerr one and do not possess
scalar hairs~\cite{hawking-no-hair,1971ApJ...166L..35T,
1970CMaPh..19..276C,Heusler:1995qj,Sotiriou:2011dz,Graham:2014ina}.
A similar theorem holds for more generic scalar-tensor theories and f(R) theories~\cite{Sotiriou:2011dz},
and in shift-symmetric Horndeski theories~\cite{Hui:2012qt} 
modulo some exceptions~\cite{1992PhLB..285..199C,Mignemi:1992nt,
Yunes:2011we,Sotiriou:2013qea,Sotiriou:2014pfa}.
Proposed constraints on scalar-tensor theories with GWs from BH/NS binaries
have been derived in~\cite{will1994,scharre,willyunes,bertibuonanno,yagiLISA,yagiDECIGO,
Arun:2013bp,Yagi:2013du,Berti:2012bp,Cardoso:2011xi,Yunes:2011aa}, while
late inspiral and merger simulations of compact binaries in these theories were performed 
in~\cite{Healy:2011ef,Barausse:2012da,
Palenzuela:2013hsa,Shibata:2013pra,Berti:2013gfa,Taniguchi:2014fqa,Sampson:2014qqa}. 
Strong constraints on scalar-tensor theories have already been placed from 
binary pulsar observations~\cite{Wex:2014nva,bhat,Freire:2012mg,2.01NS,alsing,Berti:2015itd} and
solar system experiments~\cite{cassini,peri,alsing}.
Future BH/pulsar observations may allow us to further constrain such theories~\cite{Liu:2014uka}.

The second example is massive gravity theories, which are another
well-motivated extension of GR, adding a finite mass to the graviton.
At the linear level, such an extension was originally proposed by Fierz and Pauli~\cite{fierz}, though it suffers from 
pathologies such as the van Dam-Veltman-Zakharov (vDVZ) discontinuity~\cite{vdv,z}
and the Boulware-Deser ghost modes~\cite{boulware}. Recently, massive
graviton theories have been generalized in a nonlinear way so that the
theories are ghost free to all orders~\cite{derham1,derham2,hassan-generic}.
The mass of the graviton has been constrained by observations of orbits of planets in 
the solar system~\cite{talmadge}.
Weaker bounds on Fierz-Pauli type theory were derived from binary pulsar observations~\cite{finnsutton}.
Stronger but less robust bounds were also derived from tidal interactions between 
galaxies~\cite{goldhaber}, weak lensing~\cite{choudhury} and galactic velocity dispersion~\cite{sjors}.
The dispersion relation of the massive graviton modifies the phase of the gravitational waveform of 
compact binaries~\cite{will1998}. Projected constraints on the mass of
the graviton with future GW interferometers
have been derived in~\cite{will1998,willyunes,bertibuonanno,arunwill,stavridis,
yagiLISA,yagiDECIGO,keppel,delpozzo,cornish-PPE,bertisesana,
Huwyler:2011iq,Arun:2013bp,Yagi:2013du,Narikawa:2014fua}. 
If one measures both GW signals and their EM counterparts, one can also
constrain the graviton mass from the difference in the arrival time of 
gravitons and photons~\cite{kocsis-mg,Hazboun:2013pea,Nishizawa:2014zna}.

Our third example is quadratic gravity, which introduces generic
quadratic curvature terms
to the Einstein-Hilbert action that are coupled to dynamical scalar fields~\cite{Yunes:2011we}. 
Such a theory includes well-known theories motivated from string theory,
such as Einstein-dilaton-Gauss-Bonnet (EdGB) theory~\cite{Metsaev:1987zx,Maeda:2009uy}
and dynamical Chern-Simons (dCS) gravity~\cite{Jackiw:2003pm,Smith:2007jm,Alexander:2009tp}.
BHs play a crucial role in constraining quadratic gravity.
Regarding EdGB gravity, BHs possess a scalar hair~\cite{1992PhLB..285..199C,Mignemi:1992nt,
Yunes:2011we,Yagi:2011xp,Sotiriou:2014pfa} while ordinary stars
such as NSs do not~\cite{Yagi:2011xp} in the so-called decoupled limit 
where the scalar field linearly couples to the Gauss-Bonnet density in the action.
This means that binary pulsar observations with two NSs cannot place stringent constraints on this theory
due to the absence of the scalar dipole radiation, while BH binaries allow us
to place strong constraints on the theory with future radio and GW observations~\cite{Yagi:2011xp,kent-LMXB,Yagi:2015oca}.
Regarding dCS gravity, spherically symmetric solutions are the same as in GR 
due to parity considerations.  Therefore, one needs to consider spinning compact objects in order to place
meaningful constraints on the theory. Since BHs generally spin faster than NSs, BHs in general place
stronger constraints on dCS gravity than NSs. For example,
future GW interferometers, such as Adv.~LIGO, may place constraints that are
six orders of magnitude stronger than the current solar system and table-top experiments~\cite{Yagi:2012vf}.

Our fourth example is a large extra dimension model, in particular the braneworld model
proposed by Randall and Sundrum~\cite{randall2} (RS-II model), in which we live
on a 4-dimensional brane with a positive tension and the bulk is
anti-de Sitter (AdS) spacetime. The size of the extra dimension is characterized 
by the AdS curvature length $\ell$. Gravity is localized on the brane and
the gravitational potential acquires a correction that is proportional to 
$\ell^2$~\cite{randall2,garriga}, 
which has been constrained from table-top experiments~\cite{adelberger}.
Emparan \textit{et~al.}~\cite{emparan-conj} and Tanaka~\cite{tanaka-conj}
applied the anti-de Sitter/conformal field theory (AdS/CFT)
correspondence to brane-localized BHs, where the correspondence
states that the gravity in the $AdS_5 \times S^5$ spacetime can be interpreted as 
the four-dimensional $\mathcal{N}=4$ $U(N)$ super Yang-Mills theory on the
 AdS boundary~\cite{maldacena,aharony}. 
They proposed that such brane-localized BHs effectively lose their mass classically 
due to the universal acceleration in the bulk away from the brane~\cite{gregory} 
and the Gregory-Laflamme instability~\cite{gregory}. 
On the CFT side, such an effect can be understood as 
an enhanced Hawking radiation due to the large number of degrees of freedom 
in CFT~\cite{aharony}.
Such an enhanced Hawking radiation has been constrained from astrophysical
BH observations, in particular, the orbital decay rate of low-mass 
X-ray binaries (LMXBs)\footnote{%
A LMXB consists of a NS or a BH as a primary and a companion being
e.g.~a post-main sequence star. Gas from the companion fills its Roche lobe 
and accretes onto the primary. X-rays are emitted from the accretion disk around
the primary, and in the case of an NS primary, also from the NS
surface.} with BHs~\cite{johannsen1,johannsen2}.
Proposed constraints with GW observations on the size of extra dimension are 
derived in~\cite{mc,yagi:brane}.
Notice that BHs are crucial to probe such an effect since it is absent for 
ordinary stars like NSs.

Since static brane-localized BH solutions have now been constructed numerically
in~\cite{Figueras:2011gd,Abdolrahimi:2012qi}, the correctness of the conjecture 
proposed in~\cite{emparan-conj,tanaka-conj} is unclear.
Nevertheless, we cover this theory in this review for the following reasons:
(i) the conjecture itself is interesting and may stimulate other work
in the context of the AdS/CFT correspondence;
(ii) the gravitational waveform for BH binaries with the mass loss
effect can have other applications, such as to varying $G$
theories~\cite{yunespretorius} and to the BH binary system with dark
energy accretion~\cite{Babichev:2004yx,Babichev:2005py}; and
(iii) the RS-II model serves as an interesting example to see how GWs
in higher-dimensional theories are modified from those in GR.

\subsection{Organizations}

The organization of this article is as follows. 
In Sec.~\ref{sec:generic-ways-test}, we review how one can carry out 
model-independent tests of GR with BHs. We first explain bumpy BH spacetimes
that parameterize deviations away from the Kerr spacetime in a generic manner. We then
summarize how to perform model-independent tests using GWs and EMWs
from a system that contains at least one BH.
We then turn to the four specific types of modifications discussed above.
In Secs.~\ref{sec:scal-tens-theor}--\ref{sec:large-extra-dimens}, 
we review (respectively) scalar-tensor theories, massive gravity theories,
quadratic gravity, and extra dimension theories.
In each section, we describe BH solutions within each theory and their properties.
Then, we explain current and projected constraints with GWs and EMWs
in each theory.
We conclude in Sec.~\ref{sec:open-questions} by giving some of the
open questions related to testing GR with BHs generically, in the
specific theories we have discussed, and in theories beyond the scope
of this review.
We mostly use geometrical units with $c=1=G$.

\section{Theory-independent tests of GR with BHs}
\label{sec:generic-ways-test}

Scientists give many names to their approaches to experimentally
testing a physical theory $X$, but in essence there are just two
methods.  First, there are the theory-specific tests, and second,
there are the theory-independent tests.  There are advantages and
disadvantages to each approach.  Theory-specific tests (sometimes
called ``external'' tests) naturally require specific alternative
physical theories $Y_{i}$ which compete with $X$ at giving a more
faithful mathematical description of nature.  Meanwhile,
theory-independent tests (sometimes called ``internal'' or ``null''
tests) focus on whether or not observations agree with the predictions
of theory $X$, without regard to specific alternatives.

The advantage of a theory-specific test is that it is straightforward
to quantify whether theory $X$ or $Y_{i}$ is a more faithful
description of nature.  This quantification comes from Bayesian model
comparison~\cite{Agathos:2013upa}.  Model comparison is easy to understand: it
quantifies the degree of belief you should place in theory $X$ over
theory $Y_{i}$, given some prior beliefs, and
observational/experimental data whose signatures may be computed in
each theory.  The disadvantage is that you need to have a specific
calculation in each theory under consideration, of e.g.~the spectrum
of an accretion disk, or the precession of an orbit, or a
gravitational waveform.  Each theory $Y_{i}$ is different and thus
requires unique calculations which can not be reused for theory
$Y_{j\ne i}$.

This same argument is an advantage for theory-independent tests.  In
theory-independent tests, one tries to parametrize just enough of the
physics to create parametric tests of some experiment or
observation. Examples are to parametrize the spacetime metric (the
parameterized post-Newtonian or PPN formalism~\cite{Nordtvedt:1968qs,
  TEGP, Will:2014kxa}), parametrize orbital post-Keplerian (PK)
parameters~\cite{DD1986, Damour:1991rd, Taylor:1989sw, lrr-2003-5},
parametrize perturbations about the FLRW space (the parameterized
post-Friedmann (PPF) framework~\cite{Hu:2007pj, Baker:2012zs,
  Li:2014eha}), or parametrize a gravitational waveform (the
parameterized post-Einstein or PPE framework~\cite{Yunes:2009ke,
  Yunes:2013dva}).  However, it is difficult to create a
parametrization that is sufficiently general that it captures the
signatures of all alternative theories.  For example, the original PPE
framework failed to capture the GW signature in
dCS gravity, where the leading deformation from GR
is sourced by spin effects.  Similarly, many bumpy BH
spacetimes (which we discuss in Sec.~\ref{sec:bumpy-spacetime}) fail
to capture BH solutions in deformations of GR.

In this Section we turn our attention to theory-independent tests of
GR with BHs.
In Sec.~\ref{sec:bumpy-spacetime} we will discuss parametrizing the
spacetime around a BH.
In Sec.~\ref{sec:gw-tests} we will focus on parameterized gravitational
wave tests, including the parameterized post-Einstein framework.
In Sec.~\ref{sec:gener-scal-corr} we will discuss a
quasi-theory-independent framework for parametrizing scalar
theories which deform away from GR.
In Sec.~\ref{sec:electr-wave-tests} we will discuss electromagnetic
observations of BHs which can be used for theory-independent and
theory-specific tests.

\subsection{Bumpy BHs}
\label{sec:bumpy-spacetime}

One of the most astounding predictions of classical GR
is that of BHs---regions of spacetime where gravity is so
strong that light, matter, and any information which enters may never
exit.  This is perhaps the most non-Newtonian phenomenon in general
relativity, and exists because GR is nonlinear (while Newtonian
gravity is linear).\footnote{%
  Michell and Laplace developed a black-hole-like concept that
  predates Maxwell theory, treating light as
  corpuscular~\cite{2009JAHH...12...90M}.
}
Nonlinearity is essential because a
gravitational field may become so strong that gravity itself
gravitates---thus a BH consists of nothing besides gravity.
In fact, though real astrophysical BHs are formed from the
collapse of a massive star, once a BH forms, it very rapidly
settles down~\cite{Price:1971fb} to the unique vacuum solution in GR:
the Kerr metric.
The BH has no ``memory'' of any light, matter, or information
that falls in, save for the total mass $M$ and spin angular momentum
$S$.  BHs in GR are thus very simple, forming a two-parameter
family of spacetimes, with all of the (Geroch-Hansen) multipole
moments given by the simple
formula~\cite{Geroch:1970cc,Geroch:1970cd,hansen:46}
\begin{equation}
  \label{eq:geroch-hansen}
  M_{\ell} + i S_{\ell} = M(ia)^{\ell}\,,
\end{equation}
where $a\equiv S/M$, $-M\le a \le M$ is the Kerr spin parameter.

An obvious theory-independent test of GR is to ask: Are real
astrophysical BHs described by the Kerr metric?  This is
sometimes called the \emph{Kerr hypothesis}.

One may pose several analogous theory-independent questions in
different contexts of testing GR.  As mentioned above, the
parameterized post-Newtonian, post-Keplerian, and parameterized
post-Einstein frameworks are such examples.  Let us look at the
weak-field limit of GR, where tests of GR are addressed by the
parameterized post-Newtonian (PPN) formalism~\cite{TEGP,Will:2014kxa}.
There, the metric may be parameterized in standard harmonic coordinates
via~\cite{TEGP}
\begin{align}
  g_{00} &= -1 + 2U-2\beta U^{2} + \ldots \\
  g_{jk} &= (1+2\gamma U)\delta_{jk} + \ldots
\end{align}
where $U$ is the potential sourced by mass density, and $\beta,\gamma$
are two of the ten PPN parameters.  PPN is specially tailored to the
weak-field regime where the metric is close to the flat Minkowski
metric.  PPN is perturbative, and in order for it to be applicable,
the potential $U\sim GM/rc^{2}$ must be small compared to 1.  This
condition is obviously violated in extreme spacetimes such as those
harboring NSs and BHs, which have dimensionless
compactnesses $M_{\star}/R_{\star}\sim 0.1-1$.

Therefore a different approach is needed to formulate a null test of
the Kerr hypothesis.  The approach we discuss here is to parametrize
BH metrics or deviations away from the Kerr metric.  To
parametrize the weak-field metric up to 1PN order requires only 10
numbers, the PPN parameters
$(\gamma, \beta, \xi, \alpha_{1,2,3}, \zeta_{1,2,3,4})$.  In contrast,
an infinite number of parameters are required to specify deviations
from the Kerr
spacetime.  In fact, it is still unknown if the parameter space of
interest is countably infinite or if it is uncountable.  Parameterized
deviations from the Kerr metric are commonly referred to as
\emph{bumpy BHs}.

The starting point for most bumpy BH investigations is the
Kerr metric, which in Boyer-Lindquist coordinates $(t,r,\theta,\phi)$
reads~\cite{Wald:1984cw}
\begin{equation}
  \label{eq:kerr-metric}
  \begin{aligned}
    ds^{2} = & -\left(1-\frac{2Mr}{\Sigma}\right) dt^{2}
    - \frac{4Mar \sin^{2}\theta}{\Sigma} dt d\phi
    + \frac{\Sigma}{\Delta} dr^{2} + \Sigma d\theta^{2} \\
    &{}+\left( r^{2} + a^{2} + \frac{2Ma^{2}r\sin^{2}\theta}{\Sigma} \right)
      d\phi^{2}\,,
  \end{aligned}
\end{equation}
with the usual definitions $\Sigma\equiv r^{2}+a^{2}\cos^{2}\theta$
and $\Delta\equiv r^{2}-2Mr+a^{2} = (r-r_{+})(r-r_{-})$, which has the
two roots of the equation $\Delta=0$ at $r_{\pm}\equiv
M\pm\sqrt{M^{2}-a^{2}}$.  These roots are the locations of null
surfaces of constant $r$ which coincide with outer ($+$) and inner
($-$) horizons.

The Kerr metric is a special case of a stationary, axisymmetric,
asymptotically flat (SAAF) spacetime, which is the expectation for an
isolated BH geometry at late times, after perturbations have
settled down.  The general form for a SAAF spacetime is given in
cylindrical coordinates $(t,\phi,\rho,z)$
as~\cite{Wald:1984cw},
\begin{equation}
  \label{eq:pre-weyl-papa}
    ds^{2} = - V(dt-wd\phi)^{2}+V^{-1}\rho^{2}d\phi^{2} +\Omega^{2}
    (d\rho^{2} + \Lambda dz^{2}) \,,
\end{equation}
where $(V,w,\Omega,\Lambda)$ are functions of $(\rho,z)$ only.
This form requires certain integrability conditions to be satisfied
(see~\cite{Wald:1984cw} for details); these are automatically
satisfied when working in vacuum in GR.  They may also be satisfied in
other theories of gravity, but a deeper analysis would be required.
When using Ricci-flatness, Eq.~\eqref{eq:pre-weyl-papa} can be further
simplified to the Weyl-Lewis-Papapetrou form,
\begin{equation}
  \label{eq:weyl-papa}
    ds^{2} = - V(dt-wd\phi)^{2}+V^{-1}[\rho^{2}d\phi^{2} +
    e^{2\gamma}(d\rho^{2} + dz^{2})] \,,
\end{equation}
where $\gamma=\frac{1}{2}\ln(V\Omega^{2})$.

The solutions for functions $(V,w,\gamma)$ needed to express the Kerr
metric can be found in Eq.~(51) of ref.~\cite{sarahleo}.  Of course
this solution enjoys the multipole relation given
in Eq.~\eqref{eq:geroch-hansen}, and the Kerr metric is said to have ``no
hairs'' (though two hairs would be more appropriate).  Other solutions
for $(V,w,\gamma)$ will have different multipole moments (the
Geroch-Hansen multipole moments~\cite{Geroch:1970cd,hansen:46} are
defined for stationary spacetimes via derivatives of analytic metric
functions taken at the point at spatial infinity).  This generalization was
carried out by Manko and Novikov~\cite{1992CQGra...9.2477M} and can
describe an exterior vacuum spacetime with arbitrary mass multipole
moments.  This same spirit was taken up more recently by
Backdahl~\cite{Backdahl:2005uz, Backdahl:2005be, Backdahl:2006ed},
paying special attention to when a prescribed set of multipole moments
leads to the existence of a solution.  Essentially, one can not choose
arbitrary multipole moments for $(M_{\ell},S_{\ell})$; in order for a
solution to exist, the choice of multipole moments must satisfy a
convergence criterion.

It is not enough to simply describe the metric of a bumpy BH spacetime.  To
test the Kerr hypothesis, we also need gauge-invariant observables to
compare between models and data.  The first such calculations were
performed by Ryan~\cite{Ryan:1995wh}.  Ryan considered the motion of a
small compact object in the field of a large bumpy BH, leaving
$(V,w,\gamma)$ free, focusing on nearly-circular and nearly-equatorial
trajectories.  Ryan treated the motion as being geodesic except for
back-reaction.

Let us comment here on the applicability of the geodesic motion
assumption, because many papers on bumpy BHs include geodesic
calculations.  The assumption of geodesic motion restricts one to
bodies in theories of gravity which respect the strong equivalence
principle (SEP).  Note that the SEP is \emph{violated} in a large
number of theories, e.g.~dCS and EdGB.  In Jordan-Brans-Dicke,
BHs in the absence of a scalar field will respect the SEP, while NSs
violate the SEP.  If the degree of violation is sufficiently small
then the geodesic approximation may still be relevant.

Within this framework, Ryan computed the motion, calculated precession
frequencies and GWs in the adiabatic approximation,
and gave a prescription for finding the unknown multipole moments from
the observables.  Later, Ryan performed a Fisher matrix forecast of
how well LISA would be able to extract multipole
moments~\cite{Ryan:1997hg}.

One criticism of the preceding work is that it is tied to the
Geroch-Hansen multipoles, which are evaluated at the point at spatial
infinity.  This is a Newtonian-inspired formalism, but it might not be
relevant in the strong-field region, close to the horizon.
Indeed, some of the bumpy black hole spacetimes we discuss have naked
singularities, lack horizons, or have other
pathologies~\cite{Johannsen:2013rqa}.  A different
approach was taken in~\cite{collins,sarah,Vigeland:2010xe}.  Collins
and Hughes~\cite{collins} advocated to instead apply perturbation theory
about the Schwarzschild solution.  That is, we take the metric
functions to be
\begin{align}
  \label{eq:collins-hughes}
  (V,w,\gamma) = (V_{\Schw},w_{\Schw},\gamma_{\Schw}) + \epsilon
  (V^{(1)},w^{(1)},\gamma^{(1)}) + \mathcal{O}(\epsilon^{2})
\end{align}
and expand the Einstein field equations to linear order in
$\epsilon$.  In principle, this approach should be equivalent to the
linearization of the approach used by Manko, Novikov, Ryan, and
Backdahl, at least in the far-field.

For two examples of bumpy solutions, Collins and Hughes took as
sources (i) point masses along the North and South poles of the
coordinate system (inspired by~\cite{Suen:1988kq}), and (ii) adding an
axisymmetric equatorial ring of mass to the spacetime.  Both of these
perturbations affect the Geroch-Hansen quadrupole (and higher moments)
of the spacetime.  For each of these perturbations, Collins and Hughes
then computed (geodesic) orbital frequencies, which could in principle be
measured through pulsar timing or by a LISA-like mission.

This same approach was extended by Vigeland and Hughes (VH)~\cite{sarah}
and Vigeland~\cite{Vigeland:2010xe}.  Rather than using the artificial
point or ring sources of~\cite{collins}, VH found
everywhere-Ricci-flat solutions to order $\mathcal{O}(\epsilon)$ for a
number of low multipole orders.  For purely mass multipole moments,
the VH approach should be equivalent to a Manko-Novikov
solution~\cite{1992CQGra...9.2477M} near Schwarzschild, with the
higher mass moments perturbatively small.  In the Schwarzschild
background, VH's axially-symmetric solutions are all of the
qualitative form
\begin{align}
  \label{eq:vigeland-hughes}
  \gamma^{(1)}_{\ell} \sim B_{\ell} M^{\ell+1}
  \frac{P_{\ell}(\cos\theta_{\text{Weyl}})}{(\rho^{2}+z^{2})^{(\ell+1)/2}}
\end{align}
for some angular mode number $\ell$, where
$\cos\theta_{\text{Weyl}} = z/\sqrt{\rho^{2}+z^{2}}$, and where
$B_{\ell}$ is a dimensionless coefficient which describes how large
the $\ell$-mode bump is.  For these solutions they then went on to
compute the (geodesic) trajectories of orbits, including finding the excess
pericenter precession induced by the bumpy modes.

Further, VH went on to use the Newman-Janis
trick~\cite{Newman:1965tw} to perform a complex rotation on the
Schwarzschild solution in an attempt to construct bumpy rotating
solutions.\footnote{%
  Applying the NJ transformation to the non-rotating VH spacetime
  leads to a spacetime which is not Ricci-flat.
  The applicability of the NJ trick has been called into question by
  Hansen and Yunes~\cite{Hansen:2013owa}, especially in constructing
  metrics which are supposed to satisfy field equations other than
  those of GR.}
They also computed (geodesic) orbits and observables, such as the three
precession frequencies of an orbit, for the rotating bumpy metrics
with $\ell=2,3,4$.

Vigeland further extended this line of work
in~\cite{Vigeland:2010xe}.  Firstly, she generalized the types of
bumps which could be produced from only the mass-type moments
in~\cite{sarah} to both the mass- and spin-type moments.  Secondly,
she constructed the mapping between these linearized bumps and the
Geroch-Hansen multipole moments.  In particular, she found that a
linearized perturbation of order $\ell$ only affected Geroch-Hansen
multipole moments of order $\ell$ and above.

Another simple and commonly-used parameterization is the
\emph{quasi-Kerr} metric of Glampedakis and Babak~\cite{glampedakis}.
Unlike the previously-mentioned parameterizations, the quasi-Kerr
metric only has \emph{one} additional parameter instead of an infinite
number.  Furthermore, this free parameter is designed to affect the
quadrupole (though we are unaware of general $\ell$ Geroch-Hansen
moments being computed for the quasi-Kerr metric, so it likely affects
higher moments).  The principle behind constructing the quasi-Kerr
metric was to extract the quadrupole and spin-squared pieces of the
Hartle-Thorne slow rotation metric~\cite{Hartle:1968ht}, which
describes the exterior spacetime of any slowly-rotating body.  The
difference between the Hartle-Thorne metric for general quadrupole
moment (not tied to the spin, like in Kerr) and that for the Kerr
solution then gives a metric deformation to ``paste onto'' the Kerr
metric.  Glampedakis and Babak also studied (geodesic) motion in the
quasi-Kerr metric, especially for equatorial orbits, computing
``kludge'' gravitational waveforms.\footnote{%
  A ``kludge'' approach is one which mixes different approximations,
  not necessarily arising from a single consistent approximation
  scheme.  For example: using particles on Kerr geodesics as sources
  entering into PN waveform formulas; or using GR waveform formulas
  even in the context of other theories.  Kludge waveforms are
  easier to compute and should still capture the qualitative
  nature of full waveforms.
}

Many of the observational properties of Glampedakis and Babak's
proposed quasi-Kerr metric have been computed by Johannsen and Psaltis
in a series of papers~\cite{Johannsen:2010xs, Johannsen:2010ru,
  Johannsen:2010bi, Johannsen:2012ng}.  This series of papers focused
on (i): general properties, including the location of the horizon and
ISCO, redshifts, and null trajectories; (ii): numerically ray-tracing
images of the quasi-Kerr spacetime (assuming that photons still follow
null geodesics); (iii): computing geodesic and epicyclic (precession)
frequencies in the strong-field, which could be putative QPO
frequencies; and (iv): simulating the Fe-K$\alpha$ line spectrum from
the strong-field region of the quasi-Kerr spacetime, which includes
relativistic broadening and boosting (again assuming that photons and
matter follow respectively null and timelike geodesics).

Johannsen and Psaltis also introduced another proposal called the
\emph{modified Kerr} metric~\cite{Johannsen:2011dh}.  Their approach
was to start with a spherically symmetric and static metric,
\begin{equation}
  \label{eq:JP1}
  ds^{2} = -f(r)[1+h(r)]dt^{2} + f(r)^{-1}[1+h(r)]dr^{2} +
  r^{2}(d\theta^{2} +\sin^{2}\theta d\phi^{2})\,,
\end{equation}
where $f(r) = 1- 2M/r$ is the usual Schwarzschild metric function.  In
the limit of $h(r)\to 0$ this metric agrees with Schwarzschild.  For
$h(r)$ they take the ansatz
\begin{equation}
  \label{eq:JP-h}
  h(r) = \sum_{k=0}^{\infty} \epsilon_{k} \left(\frac{M}{r}\right)^{k}\,,
\end{equation}
which is inspired by a far-field expansion.  Johannsen and Psaltis
then arrive at the modified Kerr metric by applying the Newman-Janis
trick~\cite{Newman:1965tw}, as was earlier done by VH~\cite{sarah}.
As mentioned in the footnote on the previous
page, applying the NJ trick is very questionable here, especially
since Eq.~\eqref{eq:JP1} is not a solution to the Einstein
equations~\cite{Hansen:2013owa}.  Johannsen and Psaltis then go on to
find the event horizon, compute circular, equatorial geodesics, and
find the ISCO and circular, equatorial photon orbit (again assuming
photons follow null geodesics).

Cardoso, Pani, and Rico performed a thorough analysis and extended the
Johannsen and Psaltis metric in~\cite{Cardoso:2014rha}.  Their
generalization simply introduced $h^{t}$ and $h^{r}$ in place of $h$
in the $tt$ and $rr$ components in Eq.~\eqref{eq:JP1}, doubling the
infinite number of coefficients needed to parameterize the spacetime.
They also showed that all coefficients become equally important
in the strong-field, near the central object.  This is a strong
criticism of the Glampedakis and Babak metric, which only has one
parameter.  This same feature highlights a severe degeneracy in
strong-field observables in terms of the $\epsilon_{k}$ parameters.
Furthermore, Cardoso, Pani, and Rico also showed that known non-GR
solutions, such as those in dCS and EdGB, do not fit within either the
original or the extended modified Kerr parameterization.  For these
reasons, the authors urged extreme caution with this
parameterization.

More recently, Rezzolla and Zhidenko took a similarly-inspired yet
unique approach to parameterizing deviations from
Schwarzschild~\cite{Rezzolla:2014mua}.  Their starting point was to
write the spherically-symmetric, static metric as
\begin{equation}
  \label{eq:RZ1}
  ds^{2} = -N^{2}(r) dt^{2} + \frac{B^{2}(r)}{N^{2}(r)} dr^{2} + 
  r^{2}(d\theta^{2} +\sin^{2}\theta d\phi^{2})\,.
\end{equation}
At this point it is essentially the same as the starting point of
Johannsen and Psaltis.  However, to maintain regularity at the horizon
and to parameterize the near-horizon behavior, Rezzolla and Zhidenko
use the coordinate $x=1-r_{0}/r$ (where $N(r_{0})=0$ is the coordinate
of the horizon), and write
\begin{align}
  \label{eq:RZ-N2-eq}
  N^{2} &= x A(x), & A(x) &> 0, & 0\le x&\le 1\,.
\end{align}
They then expand $A(x)$ and $B(x)$ as
\begin{align}
  \label{eq:RZ-A-B}
  A(x) &= 1 - \epsilon(1-x) + (a_{0}-\epsilon)(1-x)^{2} +
         \tilde{A}(x)(1-x)^{3} \,, \\
  B(x) &= 1 + b_{0}(1-x) + \tilde{B}(x)(1-x)^{2} \,.
\end{align}
The functions $\tilde{A},\tilde{B}$ are then expanded as infinite
continued fractions,
\begin{align}
  \label{eq:RZ-A-B-tilde}
  \tilde{A}(x) &= \cfrac{a_{1}}{1
  + \cfrac{a_2 x}{1
  + \cfrac{a_3x}{1 + \ldots}}}\,, &
  \tilde{B}(x) &= \cfrac{b_{1}}{1
  + \cfrac{b_2 x}{1
  + \cfrac{b_3x}{1 + \ldots}}}\,.
\end{align}
In Eqs.~\eqref{eq:RZ-A-B}--\eqref{eq:RZ-A-B-tilde},
$\epsilon$, $a_n$ and $b_n$ (with $n$ a non-negative integer)
are real constants.
The continued fraction expansion may have desirable convergence
properties.  Rezzolla and Zhidenko showed how to map between their
metric and the non-rotating one of Johannsen and Psaltis.  They
computed the ISCO and circular photon orbit (again assuming that test
bodies and photons move on timelike and null geodesics, respectively),
and found the quasinormal mode frequencies for a test scalar field
living on this background spacetime.  However, these frequencies may
have little relation to the quasinormal modes of the spacetime itself,
which depends on the dynamics in the theory of gravity which gives
rise to these solutions.  Different theories could give rise to the
same solution but have distinct quasinormal frequencies.  Finally,
Rezzolla and Zhidenko considered a Hartle-Thorne-like slow-rotation
expansion to linear order in spin to additionally parameterize
spinning objects.

Some of the aforementioned approaches to bumpy BHs satisfied
the Einstein field equations; some used the NJ trick; and some were
either ad-hoc or general parameterizations of SAAF spacetimes.  A
completely different approach was taken in Vigeland, Yunes, and
Stein~\cite{sarahleo} (VYS).  The authors were motivated to find
spacetimes which retained the Liouville integrability of test-particle
motion that the Kerr spacetime enjoys.  VYS did start with a totally
general perturbative parameterization by linearizing the
Weyl-Lewis-Papapetrou metric about the Kerr solution.  From this
point, VYS then imposed the condition that the perturbed spacetime
retained a Carter-like second rank Killing tensor to leading order in
perturbation theory, which restricts the function space of
deformations to Kerr.  This space is still large enough that it was
possible to describe the slowly-rotating BH solution in
dCS~\cite{Yunes:2009hc} (it was later found that at order
$\mathcal{O}(a^{2})$, the dCS BH does not have a third
integral of motion, and thus does not fall under the VYS
parameterization~\cite{Yagi:2012ya}).  The metric of VYS was later
simplified in~\cite{Johannsen:2013rca} and reads
\begin{subequations}
  \label{eq:Johannsen-YVS}
  \begin{align}
    g_{tt} ={}& -\frac{\tilde{\Sigma}[\Delta-a^2A_2(r)^2\sin^2\theta]}{[(r^2+a^2)A_1(r)-a^2A_2(r)\sin^2\theta]^2},  \\
g_{t\phi} ={}& -\frac{a[(r^2+a^2)A_1(r)A_2(r)-\Delta]\tilde{\Sigma}\sin^2\theta}{[(r^2+a^2)A_1(r)-a^2A_2(r)\sin^2\theta]^2},  \\
g_{rr} ={}& \frac{\tilde{\Sigma}}{\Delta A_5(r)}, \\
g_{\theta \theta} ={}& \tilde{\Sigma}, \\
g_{\phi \phi} ={}& \frac{\tilde{\Sigma} \sin^2 \theta \left[(r^2 + a^2)^2 A_1(r)^2 - a^2 \Delta \sin^2 \theta \right]}{[(r^2+a^2)A_1(r)-a^2A_2(r)\sin^2\theta]^2},
\label{eq:metric}
  \end{align}
\end{subequations}
where Johannsen defined $\tilde{\Sigma}\equiv \Sigma + f(r)$.  This
metric has four functional degrees of freedom, $A_{1,2,5}(r)$ and
$f(r)$.  Still, this metric is of limited applicability because there
is no reason that the BH solution in some theory of gravity should
have Liouville integrable geodesics (for instance the
$\mathcal{O}(a^{2})$ BH in dCS).

Despite this plethora of bumpy BH frameworks, none of them are
completely satisfactory.  Most of them have pathologies, some of which
were investigated in~\cite{Johannsen:2013rqa}.  Johannsen showed that
some of these metrics include naked singularities and closed timelike
curves.  In addition, no one framework seems to be able to capture all
BH solutions from non-GR theories.
More research is required along this line of work to reach a
theoretically satisfactory description of bumpy BHs and the phenomena
which take place in these spacetimes.

\subsection{GW Tests}
\label{sec:gw-tests}

In this subsection, we review how one can perform a model independent tests of GR with GW observations.
We refer to recent reviews~\cite{Gair:2012nm,Yunes:2013dva} for more details on this topic.

\subsubsection{GWs in GR}

Let us first briefly explain how one can calculate GWs from a compact binary inspiral with masses $(m_1,m_2)$
and a separation $r_{12}$ to leading post-Newtonian (PN) order, where one expands in $m/r_{12}$ 
with $m=m_1+m_2$ representing the total mass. Such a PN approximation corresponds to assuming that
the velocity of binary constituents is much smaller than the speed of light.
The total energy of the system is given by
$E = - \eta m^2/(2r_{12})$ while the energy flux emitted from this binary is calculated from the 
quadrupole formula as $\dot{E} = - (32/5) \eta^2 m^5/r_{12}^5$, where $\eta \equiv m_1 m_2/m^2$
is the symmetric mass ratio. Since GW frequency $f$ is related to the orbital angular frequency 
$\Omega = \sqrt{m/r_{12}^3}$ by $f = \Omega/\pi$, one can derive the evolution of the GW frequency as
\be
\dot{f} = \frac{df}{dr_{12}} \, \frac{dr_{12}}{dE} \, \frac{dE}{dt} =  \frac{96}{5}\pi ^{8/3} \mathcal{M} ^{5/3}f^{11/3}\,,
\ee
where $\mc \equiv m \eta^{3/5}$ is the chirp mass.
One solves this equation to yield
\be
f=\left( \frac{5}{256} \right)^{3/8} \frac{1}{\pi \mathcal{M}^{5/8}} \frac{1}{(t_0-t)^{3/8}}\,, 
\qquad \phi(t)=\int 2\pi f dt = -2 \left( \frac{1}{5} \mathcal{M}^{-1} (t_0-t) \right)^{5/8} + \phi_0\,, \label{phi}
\ee
where $\phi(t)$ is the GW phase in the time domain and $t_0$ and $\phi_0$
correspond to the time and phase at coalescence respectively.

Next, we derive the gravitational waveform in the Fourier domain by applying the 
stationary phase approximation (see e.g.~\cite{maggiore2007gravitational}). Let us begin by
Fourier transform a function $B(t) \equiv A(t) \cos \phi(t)$ as
\ba
\tilde{B}(f) &\equiv & \int^{\infty}_{-\infty}  A(t) \cos \phi(t) e^{2\pi ift} dt \nn \\
& = & \frac{1}{2} \int^{\infty}_{-\infty} A(t) \left( e^{i \phi(t)} + e^{-i \phi(t)} \right) e^{2\pi ift}  dt\,.
\ea
Since $f>0$ and $d\phi/dt > 0$, the term proportional to $e^{i(2\pi ft - \phi)}$ has 
a stationary point while the term proportional to $e^{i(2\pi ft + \phi)}$ oscillates 
rapidly and becomes negligible upon integration. 
Requiring $d \ln A/dt \ll d \phi / dt$, the stationary point $t_*(f)$ is determined
by solving $2 \pi f = d\phi (t_*)/dt$, while one can approximate $A(t)$ with $A(t_*)$
and take it out of the integral. Expanding the exponent in the integrand around $t=t_*$, 
one finds
\be
\label{eq:Bf}
\tilde{B}(f) \approx \frac{1}{\sqrt{2}} A(t_*) \left( \frac{d^2\phi (t_*)}{dt^2} \right)^{-1/2} e^{i[2\pi f t_*-\phi(t_*)]} 
\int_{-\infty}^{\infty} e^{-i X^2} dX\,,
\ee
with $X \equiv \sqrt{(d^2\phi (t_*)/dt^2)/2} \; (t-t_*)$. Using 
the Fresnel integral $\int_{-\infty}^{\infty} e^{-i X^2} dX = \sqrt{\pi} e^{-i\pi/4}$,
one finally arrives at
\be
\label{eq:Bf}
\tilde{B}(f) \approx \frac{1}{2} A(t_*) \left( \frac{df(t_*)}{dt} \right)^{-1/2} e^{i[2\pi ft_*-\phi(f)-\pi/4]}\,,
\ee
where we remind that $t_*=t_*(f)$.

Let us now apply this result to the gravitational waveform.
From Eq.~\eqref{eq:Bf}, one finds that the phase in the Fourier domain is
$2\pi ft_*(f)-\phi[t_*(f)]-\pi/4$.
On the other hand, from Eq.~\eqref{phi}, one finds 
\be
\label{eq:t-phi-f}
t_*(f) = t_0-5\mc(8\pi \mc f)^{-8/3} \,, 
\qquad \phi(f) = \phi_0-2(8\pi \mathcal{M} f)^{-5/3}\,.
\ee
Combining Eqs.~\eqref{eq:Bf} and~\eqref{eq:t-phi-f}, the phase of the gravitational waveform 
in GR is given by
\be
\Psi_\GR (f) = 2\pi f t_0 - \phi_0 - \frac{\pi}{4} + \frac{3}{128}(\pi \mathcal{M}f)^{-5/3} \left\{ 1+ \mathcal{O}\left[ (\pi \mc f)^{2/3} \right] \right\}\,.
\ee
One can extend the above calculations by including higher PN corrections. 
For example, for a non-spinning binary, the waveform phase is known up to 3.5PN order~\cite{arun35} 
and are given in the form
\be
\label{eq:phase-3.5}
\Psi_\GR (f) = 2\pi f t_0 - \phi_0 - \frac{\pi}{4} + \sum_{i=0}^7 \left[ \psi_i^{(0)} + \psi_i^{(1)} \ln f \right] f^{(i-5)/3}\,,
\ee
where the coefficients $\psi_i^{(0)}$ and $\psi_i^{(1)}$ are functions of $m$ and $\eta$.

\subsubsection{Data Analysis}

We here review how one can carry out a GW data analysis to probe strong-field gravity with GWs.
One method is to use a Fisher analysis~\cite{finn,cutlerflanagan} 
and determine how well future GW interferometers can measure
parameters that characterize deviations away from GR. Such an analysis corresponds to performing 
a matched filtering analysis and take correlations between GW signals and templates.
Fisher analysis is only valid when the signal-to-noise ratio (SNR) is large.

Let us start by assuming that the detector noise is stationary and Gaussian.
%
Then, the noise follows a Gaussian probability distribution given by 
\begin{equation}
p(n_0) \propto \exp \left[-\frac{1}{2}(n_0|n_0)\right]\,, \qquad  
(A|B) \equiv 4 \mathrm{Re}\int ^{\infty}_{0}df \, \frac{\tilde{A}^{*}(f)\tilde{B}(f)}{S_n(f)}\,, \label{scalar-prod}
\end{equation}
where $S_n(f)$ is the noise power spectral density (see Fig.~\ref{fig:noise}).
SNR is defined via this definition of the inner product as $\rho = \sqrt{(h|h)}$, where
$h$ is the GW signal.
We denote the detected signal $s(t)=h(t;\bm{\theta}_t)+n_0(t)$,  
 where $\bm{\theta}_t$ is the true parameters of a binary. Then, one can rewrite Eq.~\eqref{scalar-prod} as
\begin{equation}
p(\bm{\theta}_t|s)\propto \exp\left[ (h_t|s)-\frac{1}{2}(h_t|h_t) \right]\,, \label{prob}
\end{equation}
with $h_t\equiv h(\bm{\theta}_t) $.
Determined binary parameters $\hat{\bm{\theta}}$ are those that 
maximize $p(\bm{\theta}_t|s)$, which satisfies $(\partial_ih_t|s)-(\partial_ih_t|h_t)=0$ with
$\partial_i \equiv \partial/\partial\theta^i_t$. Introducing $\Delta\theta^i$
by $\theta_t^i=\hat{\theta}^i+\Delta\theta^i$ and expanding
Eq.~(\ref{prob}) around $\Delta\theta^i=0$ and keeping up to quadratic order, one finds
\begin{equation}
p(\bm{\theta}|s)\propto \exp\left[ -\frac{1}{2}\Gamma_{ij}\Delta\theta^i\Delta\theta^j \right]\,,
\end{equation}
where $\Gamma_{ij}=(\partial_i\partial_jh|h-s)+(\partial_ih|\partial_jh)$ is the Fisher matrix.
Since $h-s=-n$, one can neglect the first term of $\Gamma_{ij}$ above in the large SNR limit and finds 
\begin{equation}
\Gamma_{ij}=(\partial_ih|\partial_jh)\,.
\label{gammaij}
\end{equation} 
The measurement error of $\theta^i$ then becomes
\begin{equation}
\sqrt{\left\langle (\Delta\theta^i)^2 \right\rangle} =\sqrt{(\Gamma^{-1})_{ii}}\,.
\end{equation}

A more sophisticated way of carrying out a parameter estimation study is to perform a Bayesian
Markov Chain Monte Carlo (MCMC) analysis~\cite{Cornish:2007ifz,Littenberg:2009bm}.
With this approach, one can directly calculate the posterior distribution $p(\bm{\theta}|d, \bar{\mathcal{M}})$ 
of a parameter for a given data $d$ within a model $\bar{\mathcal{M}}$ through the Bayes' theorem as
\be
p(\bm{\theta}|d, \bar{\mathcal{M}}) = \frac{p(d |\bm{\theta}, \bar{\mathcal{M}}) p(\bm{\theta}, \bar{\mathcal{M}})}{p(d, \bar{\mathcal{M}})}\,.
\ee
Here, $p(d |\bm{\theta}, \bar{\mathcal{M}})$ and $p(\bm{\theta}, \bar{\mathcal{M}})$ are the likelihood 
and prior distributions respectively, while $p(d, \bar{\mathcal{M}})$ is the evidence given by
\be
p(d, \bar{\mathcal{M}}) = \int d\bm{\theta} \, p(d |\bm{\theta}, \bar{\mathcal{M}}) \, p(\bm{\theta}, \bar{\mathcal{M}})\,.
\ee
One can also carry out a model selection study between models $\bar{\mathcal{M}}_1$ and $\bar{\mathcal{M}}_2$
by calculating the Bayes factor defined by taking the ratio of the evidence between the two models.
Such a Bayes factor can be calculated with e.g.~thermodynamic integration~\cite{Littenberg:2009bm,Veitch:2014wba},
nested sampling~\cite{Veitch:2014wba} and reverse jump, MCMC~\cite{Cornish:2007ifz}.
References~\cite{cornish-PPE,Vallisneri:2012qq} proposed a less computationally expensive way of calculating the Bayes factor 
between GR and non-GR models, whose validity was confirmed in~\cite{DelPozzo:2014cla}.

\subsubsection{Parameterized Tests of GR}
\label{sec:parameterized-GW}

We will next review how one can construct parameterized gravitational waveforms from compact binaries to 
perform strong-field tests of gravity. The first approach extends the PN waveform in Eq.~\eqref{eq:phase-3.5} to non-GR
theories, which allows us to carry out tests of gravity similar to parameterized post-Keplerian (PPK) tests 
with binary pulsar observations~\cite{lrr-2003-5,Wex:2014nva,Berti:2015itd}. PPK parameters depend on two masses
of the binary constituents. This means that any two independent measurement of two 
PPK parameters allows us to determine two masses, while an additional PPK parameter measurement 
enables us to perform a consistency test of GR. 

Arun \textit{et al.}~\cite{Arun:2006yw} proposed that since $\psi_i^{(0)}$ and $\psi_i^{(1)}$ in Eq.~\eqref{eq:phase-3.5} for gravitational waveforms from
a compact binary inspiral also 
depend on two masses of the binary constituents, 
independent measurement of such PN parameters allows us to perform a consistency test of GR. The authors treated all of the 
coefficients with different $i$s independently and found that large degeneracies exist among these parameters with a LISA observation.
The authors extended this analysis in~\cite{arun-model-indep,mishra} by treating $\psi_0^{(0)}$ and $\psi_2^{(0)}$ as fundamental parameters 
to determine two masses, and consider one of the remaining parameters as an additional parameter to perform consistency tests of GR. 
The authors used the restricted PN waveforms in~\cite{arun-model-indep} and the full PN waveform in~\cite{mishra}, where the former 
only keeps the leading PN order term in the amplitude, while the latter keeps up to 3PN order. 
In particular, Ref.~\cite{mishra} studied how well one can constrain deviations from GR by detecting GWs from BH binaries with
 future ground-based GW interferometers, such as Adv. LIGO and ET, 
with a Fisher analysis. The authors found that one can carry out such a type of tests with Adv. LIGO using $\psi_3^{(0)}$ 
as an additional parameter, while one can also use other parameters with ET. Such a Fisher analysis was extended 
in~\cite{Li:2011cg,Agathos:2013upa} to a Bayesian model selection study for binary NSs using a GW data analysis pipeline to test GR, called
Test Infrastructure for GEneral Relativity (TIGER).

One of the limitations in the above approach is that one can only consider corrections to non-vanishing terms in the waveform phase in GR.
This means that one cannot capture e.g.~the effect of scalar radiation entering at $-1$PN relative to GR and non-GR corrections to the waveform amplitude.
In order to construct a more generic parameterized waveform, Yunes and Pretorius~\cite{Yunes:2009ke} proposed a parameterized post-Einsteinian (PPE) waveform. 
They first introduce a generic parameterized correction to the binding energy (or Kepler's law) and energy flux. Then, the authors propagate such corrections to
the waveform of a compact binary inspiral in the Fourier domain and found the simplest PPE waveform given by
\be
\label{eq:simplest-PPE}
\tilde{h} (f) = A_\GR(f) (1 + \alpha_\ppE u^{a_\ppE}) \exp \left[i \Psi_\GR(f) + \beta_\ppE u^{b_\ppE} \right]\,,
\ee
where $u \equiv \pi \mc f$, $A_\GR$ and $\Psi_\GR$ are the waveform amplitude and phase in GR respectively, while 
$\alpha_\ppE$, $a_\ppE$, $\beta_\ppE$ and $b_\ppE$ are the PPE parameters that encodes the dominant deviations from GR
in the amplitude and phase. 
The waveform reduces to that in GR when $\alpha_\ppE = 0 = \beta_\ppE$ and the correction terms in the amplitude and phase
correspond to $(3/2) a_\ppE$PN and $[(3/2) b_\ppE + 5/2]$PN corrections relative to GR respectively.
Such a parameterization can capture gravitational waveforms in e.g.~scalar-tensor theories, massive gravity theories,
quadratic gravity (including EdGB and dCS gravity), variable $G$ theories, large extra dimension models, 
Lorentz violating theories (including Einstein-\AE{}ther and khronometric gravity). The authors also constructed PPE waveforms 
in the merger and ringdown phases.

Many follow-up papers were published that extends the original PPE framework. Cornish \textit{et al.}~\cite{cornish-PPE} 
performed a Bayesian inference and model selection study to reveal how well future GW interferometers, such as Adv. LIGO and LISA,
can measure PPE parameters $\alpha_\ppE$ and $\beta_\ppE$ for each $a_\ppE$ and $b_\ppE$ respectively.
The authors compared their results with~\cite{Yunes:2010qb}, where the latter studies the current constraints on $\beta_\ppE$ for each $b_\ppE$
from orbital decay rate of the double binary pulsar J0737-3039~\cite{burgay,lyne,kramer-double-pulsar,kramer-wex}, 
and found that Adv. LIGO will perform better than such binary pulsar observations
for $b_\ppE \geq -1.5$.
Chatziioannou \textit{et al.}~\cite{Chatziioannou:2012rf} extended Eq.~\eqref{eq:simplest-PPE} 
to include non-tensorial polarization modes. 
Sampson \textit{et al.}~\cite{Sampson:2013lpa} included subdominant PPE correction terms to the 
waveform in Eq.~\eqref{eq:simplest-PPE} and carried out a Bayesian inference study. They found that the simplest PPE waveform in
Eq.~\eqref{eq:simplest-PPE} is sufficient for measuring non-GR effects.
Reference~\cite{Sampson:2013wia} relates the PPN framework for solar system experiments and the
PPK framework for binary pulsar observations
to the PPE one.
Sampson \textit{et al.}~\cite{Sampson:2013jpa} studied the ``trouble with templates'' and found following: 
(i) LIGO/VIRGO network may miss signals using 
GR templates if the true signal is the non-GR one that is still consistent with existing constraints. (ii) The simplest PPE model of 
Eq.~\eqref{eq:simplest-PPE} can still capture non-GR signals that suddenly shows deviations from GR at a certain frequency, such as 
dynamical scalarization in quasi BD theory~\cite{Barausse:2012da,
Palenzuela:2013hsa,Shibata:2013pra,Taniguchi:2014fqa,Sampson:2014qqa} 
and massive scalar tensor theories~\cite{alsing,Berti:2012bp}. 
(iii) If one uses Eq.~\eqref{eq:simplest-PPE} that only contains the inspiral phase to filter the actual signals
that also contain merger and ringdown phases, the deviations from GR may be misidentified.
Loutrel \textit{et al.}~\cite{Loutrel:2014vja} constructed a non-GR, parameterized burst signals of 
GWs from binaries with highly eccentric orbits.
Huwler \textit{et al.}~\cite{Huwyler:2014vva} constructed a PPE waveform in the time domain.   
Vallisneri and Yunes~\cite{Vallisneri:2013rc} and Vitale and Del Pozzo~\cite{Vitale:2013bma} carried out a Bayesian 
analysis and studied \emph{stealth bias}~\cite{cornish-PPE} in detail, which refers to deviations from GR that are too small
to be detected but causes systematic errors that are larger than statistical errors.
They found that such systematic errors (in particular, in the mass measurements) can be significant 
even within the current bounds.
However, detailed analyses on systematic errors on PPE parameters themselves
due to the waveform mismodeling (similar to those carried out in 
GR in e.g.~\cite{Barausse:2014tra,Barausse:2014pra}) are still missing.

\subsubsection{BH No-hair Tests}


We now review testing BH no-hair relations among multipole moments with (i) inspiral and (ii) ringdown GWs.
Regarding the former, most literature considers EMRIs as a smaller compact object orbits around the central larger BH
many times (since the radiation reaction is suppressed by the mass ratio compared to comparable mass binaries), and
hence it can probe the central BH spacetime directly. As mentioned in Sec.~\ref{sec:bumpy-spacetime},
Ryan~\cite{Ryan:1995wh,Ryan:1997hg} derived a PN 
gravitational waveform from EMRIs with arbitrary multipole moments of the central object and carried out a Fisher 
analysis. He found that if LISA detects GW signals from an EMRI with masses $(10,10^5) M_\odot$ and SNR of 100, the 
dimensionless quadrupole moment can be measured to $\sim 10^{-3}$. Barack and Cutler~\cite{Barack:2006pq} improved 
Ryan's analysis by considering generic orbits and including the satellite's motion. They constructed the waveform using an 
analytic-kludge approach~\cite{Barack:2003fp}, where the authors model, at every instant, the orbit as Newtonian, emitting the leading 
order GWs, and solve the PN equations to secularly evolve the orbit's parameters.
They found constraints that are better than Ryan's.

One can also study future prospects of testing the BH no-hair property with GWs by studying test particle motion around
a bumpy metric (see Sec.~\ref{sec:bumpy-spacetime} for a detailed explanation of bumpy metrics). 
Glampedakis and Babak~\cite{glampedakis} studied
geodesic motion of a test particle around a quasi-Kerr object, which
was constructed based on the Hartle-Thorne
metric~\cite{hartle1967,Hartle:1968ht}.
They then constructed gravitational waveforms and found that the mismatch from the Kerr waveform can be significant for 
a modest deviation from Kerr, suggesting that using the Kerr waveform templates for extracting GWs from EMRIs around 
a non-Kerr metric may result in a significant loss of SNRs.
Gair \textit{et al.}~\cite{Gair:2007kr} studied orbits around a metric proposed by Manko and Novikov (MN)~\cite{1992CQGra...9.2477M}, 
whose multipole moments differ from the Kerr ones for $\ell \geq 2$. They found that certain orbits lead to an ergodic 
motion due to the loss of the Carter-like constant.
Orbital properties around the MN spacetime were more thoroughly studied by Apostolatos, Lukes-Gerakopoulos 
and Contopoulos~\cite{Apostolatos:2009vu,LukesGerakopoulos:2010rc,Contopoulos:2011dz}. 
According to the Poincar\'e-Birkhoff theorem, the resonant tori in the phase space of a perturbed integrable system disintegrate to form a chain
of islands, inside which the ratio of fundamental frequencies stays constant (also frequencies themselves evolve).
The appearance of such islands is a distinct feature of non-Kerr spacetime, and hence, a plateau in the evolution of the ratio of the fundamental 
frequencies, when the orbit crosses the island, is a smoking gun for a generic deviations from Kerr.
Gair and Yunes~\cite{Gair:2011ym} constructed EMRI waveforms from the VYS bumpy metric~\cite{sarahleo} 
using an analytic-kludge approach.

Yet, another approach is to study GWs from EMRIs of exotic compact objects.
Kesden \textit{et al.}~\cite{Kesden:2004qx} studied the orbital motion of a compact object around a supermassive boson stars 
(see Ref.~\cite{Liebling:2012fv} for a review on boson stars) and found that a stable orbit exists even inside the surface of a boson star.
They evolved the trajectory of a compact object from the exterior to the interior of the central boson star and found that GWs emitted from such a 
system is distinguishable from those from EMRIs with a central BH.
Macedo \textit{et al.}~\cite{Macedo:2013qea,Macedo:2013jja} extended the above analysis by calculating the gravitational and 
scalar radiation in a consistent and fully relativistic way for a few different boson star models.
They found that due to the absence of the event horizon, resonant oscillation modes are resonantly excited by orbiting compact objects and GW signals 
at the last stage of the inspiral becomes qualitatively different from that of an EMRI with a central BH.
GWs from merging boson stars were studied in~\cite{Palenzuela:2006wp,Palenzuela:2007dm}.
On the other hand,EMRI waveforms of gravastars~\cite{Mazur:2001fv} were calculated in~\cite{Pani:2010em}, where gravastars have the de Sitter and 
Schwarzschild metric in the interior and exterior respectively, with a thin-shell with a tensions that stitches these two metrics.
Similar to the boson star case, EMRI waveforms for gravastars have peaks that correspond to resonant modes excited during the inspiral phase 
by the resonant scattering of GWs due to the gravastar's surface. 
For a gravastar with a mass $10^6M_\odot$, such peak frequencies typically appear within the optimal sensitivity band of LISA.


Another type of BH no-hair tests is to use the ringdown signal of a BH e.g.~after a merger of two BHs.
Ringdown signals or quasi normal modes (QNMs) are exponentially damped sinusoids (see e.g.~\cite{Berti:2009kk} for a review on QNMs). 
Since the frequency and the damping time are given by the mass and spin of the ringing BH, a measurement of a single set of 
complex QNM frequencies allows one to extract the mass and spin, provided that one knows which particular normal mode such a measurement 
corresponds to. 
Then, an additional measurement allows one to perform a consistency test of the BH no-hair property~\cite{Dreyer:2003bv}.

Dreyer \textit{et al.}~\cite{Dreyer:2003bv} studied the prospects of testing the BH no-hair property with LISA by estimating the probability that one fails 
to identify an actual BH or one misidentifies non-BH QNMs to be the BH ones.
Flanagan and Hughes~\cite{Flanagan:1997sx} studied the detectability of these QNMs, while Berti \textit{et al.}~\cite{Berti:2005ys,Berti:2007zu} 
and Kamaretsos \textit{et al.}~\cite{Kamaretsos:2011um} carried out a Fisher analysis and investigated how accurately one can measure these QNM 
frequencies with future GW observations.
Such Fisher analyses were extended to a Bayesian inference study in~\cite{Gossan:2011ha} by carrying out a parameter estimation study and a model 
selection analysis between GR and non-GR models. 
The authors used the $(\ell,m) = (2,2)$ mode to extract the mass and spin of a BH and used the frequency of the $(\ell,m)=(3,3)$ mode as 
additional information to check the consistency in the mass-spin plane of a BH, just like the PPK test in binary pulsar
observations~\cite{lrr-2003-5,Wex:2014nva,Berti:2015itd} and GW tests with parameterized PN waveforms in~\cite{Arun:2006yw,arun-model-indep,mishra}.
They found that ET can measure a 10\% deviation in the frequency of the $(\ell,m) = (2,2)$ mode from GR for a 500$M_\odot$ BH if the luminosity 
distance is smaller than $\sim 6$Gpc or the redshift smaller than $z \sim 1$.
Meidam \textit{et al.}~\cite{Meidam:2014jpa} also carried out a Bayesian model selection study using the TIGER pipeline 
but for comparatively smaller SNR sources.
They found that a 10\% deviation in the frequency of the $(\ell,m) = (2,2)$ mode from GR for a 500$M_\odot$ BH can be detected with ET if one 
combines results from $\mathcal{O}(10)$ sources out to $\sim 50$Gpc ($z \leq 5$).
QNMs of boson stars were studied in~\cite{Kojima:1991np,Yoshida:1994xi,Berti:2006qt,Balakrishna:2006ru,Macedo:2013qea,Macedo:2013jja}, 
while those of gravastars were calculated in~\cite{Chirenti:2007mk,Pani:2009ss,Pani:2010em}.

\subsubsection{Systematics}

In this subsection, we explain possible systematic errors on probing GR with
GW observations of BHs. One origin of such systematics is some gas around BHs
that forms accretion disks. Narayan~\cite{narayan} studied such an effect 
on an EMRI embedded in an advection-dominated accretion flow disk.
He estimated the amount of the hydrodynamic drag on the small compact object and 
found that the effect is negligible.
Barausse and Rezzolla~\cite{Barausse:2007dy} studied the contribution of a non-self-gravitating torus around
a SMBH. They found that the effect of the hydrodynamic drag is typically subdominant compared to
the radiation reaction, though certain situations exist, where such an effect may be detectable
with LISA.
Yunes \et~\cite{yunes-disk} and Kocsis \et~\cite{kocsis-disk} investigated various effects of 
a geometrically thin, standard model accretion disk on an EMRI in great detail, including
the mass accretion, disk's self-energy, hydrodynamic drag, torques from spiral arms, 
and resonant interactions that are similar to planetary migrations~\cite{ward,armitage-disk}.
Among these, they found that the migration gives the dominant contribution to GWs, 
which may give large impacts on parameter estimation.  

Barausse \et~\cite{Barausse:2014tra,Barausse:2014pra} also studied 
astrophysical systematic errors on testing GR with GWs in detail. They found that such systematics
should not be important for eLISA, except for thin disks around EMRIs. However, they argue that 
EMRIs that can be detected by eLISA will likely to have thick disks instead of thin ones. This is 
because eLISA can only detect EMRI signals within the redshift $z \sim 0.7$, and galactic nuclei
within the local universe are typically quiescent~\cite{2004ApJ...613..109H,2010ApJ...719.1315K}.
The authors mention that only a few percent of 
all the EMRIs detected with eLISA may have thin disks. Given that current estimated event rate of 
EMRIs with eLISA is 5--50 per year, it is likely that matter effects do not significantly affect 
the tests of GR with EMRIs. They also derived intrinsic lower bounds on non-GR effects to
GWs, below which such effects are buried under astrophysical ones. Moreover, 
they estimated systematic 
errors on ringdown signals. For example, they found that a small matter distribution with 
mass $\delta M$ around an EMRI with central BH's mass $M$ can affect the ringdown 
frequency by roughly $0.05\% [\delta M / (10^{-3}M)]$, which places an intrinsic lower 
bound on the non-GR effect to be probed with GW ringdowns.

Regarding comparable-mass binaries, Ref.~\cite{hayasaki:yagi} studied the effect of  
circumbinary disks around such binaries. The orbital angular momentum of a binary is transferred 
to a circumbinary disk through the tidal/resonant interaction induced by the time-dependent gravitational potential of the binary~\cite{papaloizou,goldreich,artymowicz}, and
through the mass accretion of the gas from the 
inner edge of the circumbinary disk onto the central binary~\cite{hayasaki07}. 
Irrespective of details of the circumbinary disk model, the authors found that 
the ratio between the angular momentum transferred from the binary to the disk to the orbital
angular momentum of the binary as $C_1{\dot{M}/ \mu}$, where $\dot M$ is the mass accretion rate,
$\mu$ is the reduced mass
and $C_1$ is a factor of $\mathcal{O}(1)$. For example, the three-dimensional 
magnetohydrodynamic simulation~\cite{shi} shows $C_1 \sim 0.65$. The authors then derived the 
leading effect of such a circumbinary disk to GWs from the central binary, and found that such an
effect enters at $-4$PN order relative to the leading radiation reaction effect. Carrying out a
Fisher analysis, they found that the measurement accuracy of the accretion rate $\Delta \dot M$ with 
a 5yr observation of DECIGO at a source distance of 3Gpc leads to
\be
\label{eq:disk}
\frac{\Delta\dot{M}}{\dot{M}_{\mathrm{Edd}}} 
\approx 1.0 \times10^{-2} \ C_1^{-1}
\left( \frac{m}{10M_{\odot}} \right)^{10/23}\,,
\ee
for the total mass $m \lesssim 10^3 M_\odot$, 
where $\dot{M}_{\mathrm{Edd}}$ is the Eddington accretion rate. However, such an event
turns out to be rare. When a binary travels through a dense molecular cloud, the gas within the Bondi-Hoyle-Lyttleton radius~\cite{bondihoyle} will form a circumbinary disk. Assuming
such a situation, the authors derived the expected number of events with which the accretion rate
can be measured with DECIGO as
\be
N_\mrm{disk} \sim 
6.0 \times 10^{-2} \ {C_1}^{-9/5} \left( \frac{m}{10M_\odot} \right)^{117/115} \left( \frac{v_\infty}{20 \mrm{km} \ \mrm{s}^{-1}} \right)^{-27/5} \left( \frac{n_\mrm{gal}}{0.01 \mrm{Mpc}^{-3}} \right)\,,
\ee
where $v_\infty$ and $n_\mrm{gal}$ correspond to the bulk velocity and the number density
of galaxies respectively. 

Regarding systematics from effects other than accretion disks, 
Yunes \et~\cite{yunes-massive-perturber} 
studied the contribution of another SMBH on an EMRI. They found that the acceleration acting on 
a binary in a time-independent external gravitational field gives again a $-4$PN correction to the waveform
phase, and such an effect might be detected with LISA if the perturber is a few tenths of a parsec 
away from an EMRI. Such a $-4$PN correction degenerates with the correction in RS-II braneworld
model (see Sec.~\ref{sec:chirping-signals}) and that in varying $G$ theories~\cite{yunespretorius}.
However, notice that astrophysical environmental effects are unique to
each source, while coupling constants in non-GR theories are typically universal.
Hence, one may be able to distinguish the former from non-GR corrections to GWs
by detecting signals from multiple sources. A $-4$PN correction also arises from the time evolution
of the redshift (redshift drift)~\cite{setoDECIGO, takahashinakamura,
  Nishizawa:2011eq, yagi:inhom},
though such an effect will not be problematic as long as one has
a good cosmological model.
Barausse \et~\cite{Barausse:2014tra,Barausse:2014pra} also studied systematics 
due to non-vanishing charges of BHs, cosmological constant, magnetic fields and dark matter, 
but the effects are negligible.
EMRI GWs also suffer from systematics due to mismodeling of the waveform. For example,
the self-force contribution is not fully understood~\cite{Barack:2009ux,Poisson:2011nh}, 
though such an effect is suppressed 
by the mass ratio of a binary.

\subsection{Generic Scalar Corrections to GR}
\label{sec:gener-scal-corr}

A large portion of the literature on corrections to GR is focused on
theories whose field content in the gravity sector consists of the
metric and one additional dynamical, long-ranged scalar field.  There
is a good pragmatic reason for studying theories of this form: they
are the simplest possible corrections to GR.  Besides convenience,
there is also good theoretical motivation to study these theories.  In
order for an interaction to be ``gravitational'' in nature, it should be
long-ranged and couple weakly to matter.  Coupled with the desire to
have no preferred frame, these criteria motivate the study of
theories with the metric and a scalar.

Unfortunately, there are still an infinite space of theories which are
deformations of GR and include a long-ranged scalar field which
couples weakly to matter.  Rather than studying them one-by-one, here
we discuss how to parameterize over theory space.

Stein and Yagi~\cite{Stein:2013wza} provided one approach to
generically classify these theories and their phenomenology.  Unlike
the other approaches described in this Section, Stein and Yagi
(henceforth SY) did not only parameterize observables.  Rather, they
used a hybrid approach where they parameterized the theory, the
properties of compact objects, the multipoles of binaries and the
scalar radiation they emit; and from these to determine the
observables.  This approach more closely connects observables to the
theoretical corrections that give rise to them.

\subsubsection{Parameterization of Theories}
\label{sec:param-theor}
In Ref.~\cite{Stein:2013wza}, SY considered
theories in the Jordan frame, such that matter only couples to the
metric in this frame.  Thus the scalar field only interacts by
coupling non-minimally to curvature, with an action given by (in the
conventions of SY),
\begin{align}
\label{eq:action}
S &= S_{\txt{EH}} + S_{\txt{kin}} + S_{\txt{int}} + S_{\txt{mat}}\,,\\
\label{eq:S-EH-kin}
S_{\txt{EH}} &= \int \frac{1}{2} \mpl^{2} R \sqrt{-g} d^{4}x\,,
&
S_{\txt{kin}} &= \int -\frac{1}{2}(\pd_{a}\vartheta)(\pd^{a}\vartheta) \sqrt{-g} d^{4}x\,,\\
\label{eq:S-INT}
S_{\txt{int}} &= \int \mathcal{L}_{\txt{int}}[\vartheta,g,\epsilon,\cd,R]\sqrt{-g} d^{4}x \,,
&
S_{\txt{mat}} &= S_{\txt{mat}}[\Psi,g]\,,
\end{align}
where $\mpl^{2}=(8\pi G)^{-1}$, $\cd$ is the Levi-Civita connection of
the metric $g$ and $\epsilon$ its volume form, and $S_{\txt{mat}}$ is
the matter action, where $\vartheta$ does not appear.  Here there is
no potential for $\vartheta$, because a nontrivial potential would
lead to a short-ranged scalar field.

Of course there is still an infinite functional freedom in
$\mathcal{L}_{\txt{int}}$.  To make progress, SY use the effective
field theory (EFT) approach and expand $\mathcal{L}_{\txt{int}}$ in a
power series in $\vartheta$ and truncate at linear order.  This
expansion is not capable of expressing theories which rely on strong
self-interaction, such as nonlinear Galileons where the
self-interaction creates a screening mechanism.

Terms at
zeroth order do not source $\vartheta$. At first order in $\vartheta$
it is possible to integrate by parts and remove any derivative from
$\vartheta$.  Therefore the most general interaction Lagrangian which
is homogeneous of degree one in $\vartheta$ may be put into the form
of a sum of terms like
\begin{equation}
\mathcal{L}_{\txt{int}} \sim \vartheta \ T[g,\epsilon^{0,1},\cd^{d} ,R^{r}]\,,
\end{equation}
where $T[\cdot]$ is some scalar invariant built from the arguments.
Each term is constructed from $d$ derivatives, $r$ curvature tensors,
and $|\epsilon|=0$ or $|\epsilon|=1$ copies of the $\epsilon$ tensor
(no more copies may appear due to an identity).  This requires
introducing a new length scale, $\ell$, for dimensional correctness;
SY parameterize this as
\begin{equation}
\label{eq:L-INT}
\mathcal{L}_{\txt{int}} \sim (\mpl\ell)\, \ell^{\wp} \ \vartheta \
T[g,\epsilon^{0,1},\cd^{d} ,R^{r}]\,,
\end{equation}
where for dimensional correctness we have $\wp = d+2r-3$.  The new
length scale $\ell$ also controls the strength of this interaction.
This part of theory space is parameterized by the integers
$(|\epsilon|,d,r)$ and the length $\ell$.  This parameterization
includes other theories we discuss in this review, including some
scalar-tensor theories (Sec.~\ref{sec:scal-tens-theor}) and quadratic
gravity theories (Sec.~\ref{sec:modif-quadr-grav}); see
Table~\ref{tab:params} for the parameters of these theories.

\begin{table}[tb]
\begin{center}
\begin{tabular}{rccccccccc}
\hline
Theory&$|\epsilon|$&d&r&$\wp$$^{a}$&$\ell_\txt{BH}$&$\ell_\txt{NS}$&$\ell_{\txt{rad}}^\txt{HH}$&$\ell_{\txt{rad}}^\txt{HS}$&$\ell_{\txt{rad}}^\txt{SS}$\\
\hline
 ``Scalar-tensor'' &0& 0& 1 & $-1$ & ---$^{b}$ & 0 &---$^{b}$&1&1\\
 EDGB &0& 0 & 2 & 1 & 0 & 2$^{c}$&1&1&3$^{c}$\\
 dCS &1& 0& 2 & 1 & 1 & 1 &2&2&2\\
\hline
\end{tabular}
\end{center}
\caption{
\label{tab:params}
Parameters of three example theories with a long-ranged,
weakly-coupling gravitational scalar.  The parameters of the
Lagrangian are $(|\epsilon|,d,r)$, and the multipole parameters are
$(\ell_{\txt{BH}},\ell_{\txt{NS}},\ell_{\txt{rad}})$.
The combinations HH/HS/SS are the three binary combinations of BH (H)
and NS (S).
In this Table we have $\ell_{\txt{rad}}=1+\min(\ell_{1},\ell_{2})$.
Notes: ($a$)~Not independent, $\wp=2r+d-3$.  ($b$)~BHs have no
hair in classical scalar-tensor theories.  ($c$)~This is expected but
has not yet been calculated.
Table from Ref.~\cite{Stein:2013wza}
}
\end{table}

\subsubsection{Parameterization of Compact Object Properties}
\label{sec:param-comp-object}
Determining the leading effect of the scalar interaction on the
orbital dynamics of a binary system requires knowledge of the scalar
field profile sourced by each of the members of the binary system.
This profile is sometimes called the ``scalar hair'' of an object.
The structure of this hair must be determined through a strong-field
matching calculation in each theory of interest.  However, again in
the interest of creating a parameterized framework, SY simply
parameterize this hair by the lowest non-vanishing multipole number
$\ell_{\txt{body}}$ (an integer) which comes from this matching
calculation.  It is the lowest non-vanishing moments of the two bodies
which control the scalar-mediated interaction between the two bodies.
In several theories (e.g.~Brans-Dicke and EdGB), BHs
and NSs have different leading scalar multipole numbers.  Therefore
the body moments need to be parameterized through both
$\ell_{\txt{NS}}$ and $\ell_{\txt{BH}}$.  See Table~\ref{tab:params} for the
parameters in a number of theories.

For a given set of theory parameters $(|\epsilon|,d,r)$ and scalar
multipole number $\ell_{\txt{body}}$, it is possible to use curvature
and compactness scalings to determine the scaling law of the scalar
multipole moment $\mu^{Q}$ with
$Q=q_{1}q_{2}\cdots q_{\ell_{\txt{body}}}$ a multi-index of valence
$|Q|=q=\ell_{\txt{body}}$.  This scaling is determined up to some
dimensionless integrals of order unity.  These symmetric tracefree
(STF) tensors $\mu^{Q}$ are essentially Wilson coefficients for the
world-line effective action.

For parity-even theories ($|\epsilon|=0$), SY determine the scaling of
$\mu^{Q}$ as
\begin{equation}
\label{eq:even-mu-Q-general}
\mu^{Q} \sim
(\mpl\ell)
\left(\frac{\ell}{R_{*}} \right)^{\wp}
C_{*}^{r} R_{*}^{q}\,,
\end{equation}
where $R_{*}$ is the radius of the body, and $C_{*}=GM_{*}/R_{*}$ is
the compactness of the body, where $M_{*}$ is the mass of the body.
The scaling estimate of Eq.~\eqref{eq:even-mu-Q-general} agrees with
the strong-field matching calculation in EdGB~\cite{Yunes:2011we}.
For parity-odd
theories ($|\epsilon|=1$), SY determine the scaling as
\begin{align}
\label{eq:odd-mu-Q-general-Shat}
\mu^{Q} &\sim
(\mpl\ell)
v_\txt{eq}\hat{S}
\left(\frac{\ell}{R_{*}}\right)^{\wp} C_{*}^{r} R_{*}^{q}
= (\mpl\ell) S 
\left(\frac{\ell}{R_{*}}\right)^{\wp} C_{*}^{r-1} R_{*}^{q-2}\,,
\end{align}
where $S^{i}$
is the spin vector of the body
(with dimensions of length squared), $\hat{S}^{i}$ the unit normal
vector in the direction of $S^{i}$, and $v_{\txt{eq}}$ the velocity at
the equator of the rotating body.  The scaling estimate of
Eq.~\eqref{eq:odd-mu-Q-general-Shat} agrees with the strong-field
matching calculation in dCS~\cite{Yunes:2009hc,Konno:2009kg,Yagi:2013mbt}.

The scalings presented in Eq.~\eqref{eq:even-mu-Q-general} and
\eqref{eq:odd-mu-Q-general-Shat} may not capture effects such as
spontaneous, dynamical, or induced scalarization in scalar-tensor
theories.  However, the fact still remains that such scalar multipole
moments do exist, but in these cases they must be promoted to be
environment-dependent.  Therefore below we will present formulas with
the $\mu^{Q}$'s left general, and also substitute in the scalings
of~\eqref{eq:even-mu-Q-general} and \eqref{eq:odd-mu-Q-general-Shat}.

\subsubsection{Binary, Scalar Interaction, and Pulsar Timing Bounds}
\label{sec:binary-scal-inter}
Let us now consider a binary system of compact objects, either two
BHs (HH), two NSs (SS), or one of each (HS).  The
presence of a dynamical scalar and scalar ``hairs'' induces a new
interaction between the two bodies besides the interaction through the
metric (the Newtonian and post-Newtonian interactions).

In some theory where the scalar hairs of a BH and NS may differ
($\ell_{\txt{BH}} \ne \ell_{\txt{NS}}$), each of these three cases may
experience a different scalar pole-pole interaction.  The
electromagnetic analogy is that the electric monopole-monopole,
monopole-dipole, dipole-quadrupole, etc.~interactions all have
different radial and angular dependences.  Therefore to continue to be
generic, let the bodies with masses $m_{1,2}$ have scalar multipole
moment tensors $\mu_{1}^{S}$ and $\mu_{2}^{T}$, with $|S|=s$ the
scalar multipole number of body 1 and respectively $|T|=t$ for body 2.

Through the EFT technique of integrating out the
scalar field, SY show that the presence of these scalar hairs induces
an effective world-line interaction given by\footnote{%
This corrects a sign error in~\cite{Stein:2013wza} and a number of
other articles which made the same mistake in the integrating-out
procedure.  This sign error propagates through to the force,
pericenter precession, etc.
}
\begin{align}
L_{\times}[\mathbf{x}_{1},\mathbf{x}_{2}]&= (-)^{s}4\pi\mu_{1}^{S}\mu_{2}^{T}
\left( \pd_{ST}\frac{1}{r_{2}} \right) [\mathbf{x}_{1}] \\
&= (-)^{t}4\pi\mu_{1}^{S}\mu_{2}^{T}
\left( \pd_{ST}\frac{1}{r_{1}} \right) [\mathbf{x}_{2}]\,,\\
L_{\times}[\mathbf{x}_{1},\mathbf{x}_{2}] &=
(-)^{t}4\pi (2s+2t-1)!! \frac{\mu_{1}^{S}\mu_{2}^{T} n_{12}^{\lstf ST\rstf}}{r_{12}^{1+s+t}}\,.
\end{align}
Here $r_{A} \equiv |\mathbf{x} - \mathbf{x}_{A}|$ is the distance from
some field point $\mathbf{x}$ to body $A$,
$\mathbf{x}_{12} \equiv \mathbf{x}_{1} - \mathbf{x}_{2}$ is the vector
which points from body 2 to body 1, $r_{12} \equiv |\mathbf{x}_{12}|$
is the distance between the two bodies,
$\mathbf{n}_{12} \equiv \mathbf{x}_{12} / r_{12}$ is the unit normal
vector which points from body~2 to body~1, and $\lstf \cdots \rstf$ is
the STF operation on the enclosed indices.  Here we see that the sum
$s+t$ controls the radial and angular dependence of the interaction.
For $s=0=t$ we have the same structure as the Kepler interaction,
which simply renormalizes $G\to\mathcal{G}_{AB}$ pairwise for each
combination of bodies $A,B$.  This is similar to the Nordtvedt effect,
and the conservative effect can not be detected in an isolated binary.
However, all other choices of $s,t$ lead to post-Newtonian
corrections.

This effective interaction Lagrangian leads to a conservative shift in
the binding energy, which is simply $\delta E_{\txt{bind}}
=-L_{\times}[\mathbf{x}_{1},\mathbf{x}_{2}]$.  It also leads to an
additional scalar-mediated force between the two bodies, which SY find
to be
\begin{equation}
\label{eq:pole-pole-force}
F_{1}^{i}=(-)^{t+1}4\pi (2s+2t+1)!! \frac{\mu_{1}^{S}\mu_{2}^{T}
  n_{12}^{\lstf iST\rstf}}{r_{12}^{2+s+t}} \,,
\end{equation}
and similarly for the force on body 2 due to 1, which is easily seen
to be $F_{2}^{i}=-F_{1}^{i}$.

From this additional force it is possible to find the conservative
correction to the motion of the binary.  The most useful observable is
the correction to the pericenter precession.  Stein and Yagi compute
this via Gauss perturbation of the orbital elements.  The net result
of the calculation is that the excess pericenter precession is given
by
\begin{multline}
\label{eq:avg-omega-dot}
\avg{\delta\dot{\omega}} =
\frac{1}{T}\frac{p^{2}}{Gm}\mathcal{A}\mu_{1}^{S}\mu_{2}^{T}
\left[
-\frac{1}{e} \frac{s+t+1}{2s+2t+1} I_{1}^{ST}
+\frac{1}{e} \hat{L}^{i} I_{2}^{iST}
-\cot\iota \frac{s+t+1}{2s+2t+3} \epsilon_{ijk}\hat{L}^{j} I_{3}^{ikST}
\right]\,,
\end{multline}
where $p$ is the semi-latus rectum of the background (Keplerian)
orbit, $e$ its eccentricity, $T$ its period, $\iota$ the inclination
(relative to the line of sight), $\hat{L}^{i}$ the unit vector in the
direction of the angular momentum, $I_{1,2,3}$ are three tensor-valued
dimensionless integrals of order unity which depend on the
eccentricity, and defining the parameter
\begin{equation}
\mathcal{A} \equiv \frac{1}{\mu}(-)^{t+1}4\pi
(2s+2t+1)!! p^{-(2+s+t)}\,,
\end{equation}
where $\mu \equiv m_{1}m_{2}/m$ is the reduced mass, and
$m=m_{1}+m_{2}$ is the total mass of the binary.  Comparing this with
the leading GR pericenter precession, we see that this is a relative
$+(s+t-1)$~pN order correction.  Since $s$ and $t$ should have the
same parity, we see that this is always an odd relative pN order
correction.

By using the scaling laws from Eqs.~\eqref{eq:even-mu-Q-general} 
and~\eqref{eq:odd-mu-Q-general-Shat} it is possible to derive the
scaling of the pericenter precession in terms of the size and
compactness of the bodies in the binary.  For parity even
($|\epsilon|=0$) theories, SY find this to be
\begin{align}
\label{eq:avg-dot-omega-scaling}
\frac{\avg{\delta\dot\omega}}{\avg{\dot\omega}_{\txt{GR}}}
\sim{}&
(-)^{t+1}(2s+2t+1)!!\frac{\ell^{2}}{(Gm)(G\mu)}
\left(
\frac{\ell^{2}}{R_{1}R_{2}}
\right)^\wp
(C_{1}C_{2})^{r} \nn\\
&{}\times
\left(
\frac{R_{1}}{Gm}
\right)^{s}
\left(
\frac{R_{2}}{Gm}
\right)^{t}
f(e)
v^{2(s+t-1)}\,,
\end{align}
with $v=|v_{12}|$ the orbital velocity, and $f(e)$ is an
$\mathcal{O}(1)$ function that depends on eccentricity.  For parity
odd ($|\epsilon|=1$) theories, factors of $v_{\txt{eq}}$ for each body
must also be included.  This result reproduces the known result from
dCS~\cite{Yagi:2013mbt}.

\begin{figure}[tbp]
  \centering
  \includegraphics[height=10cm]{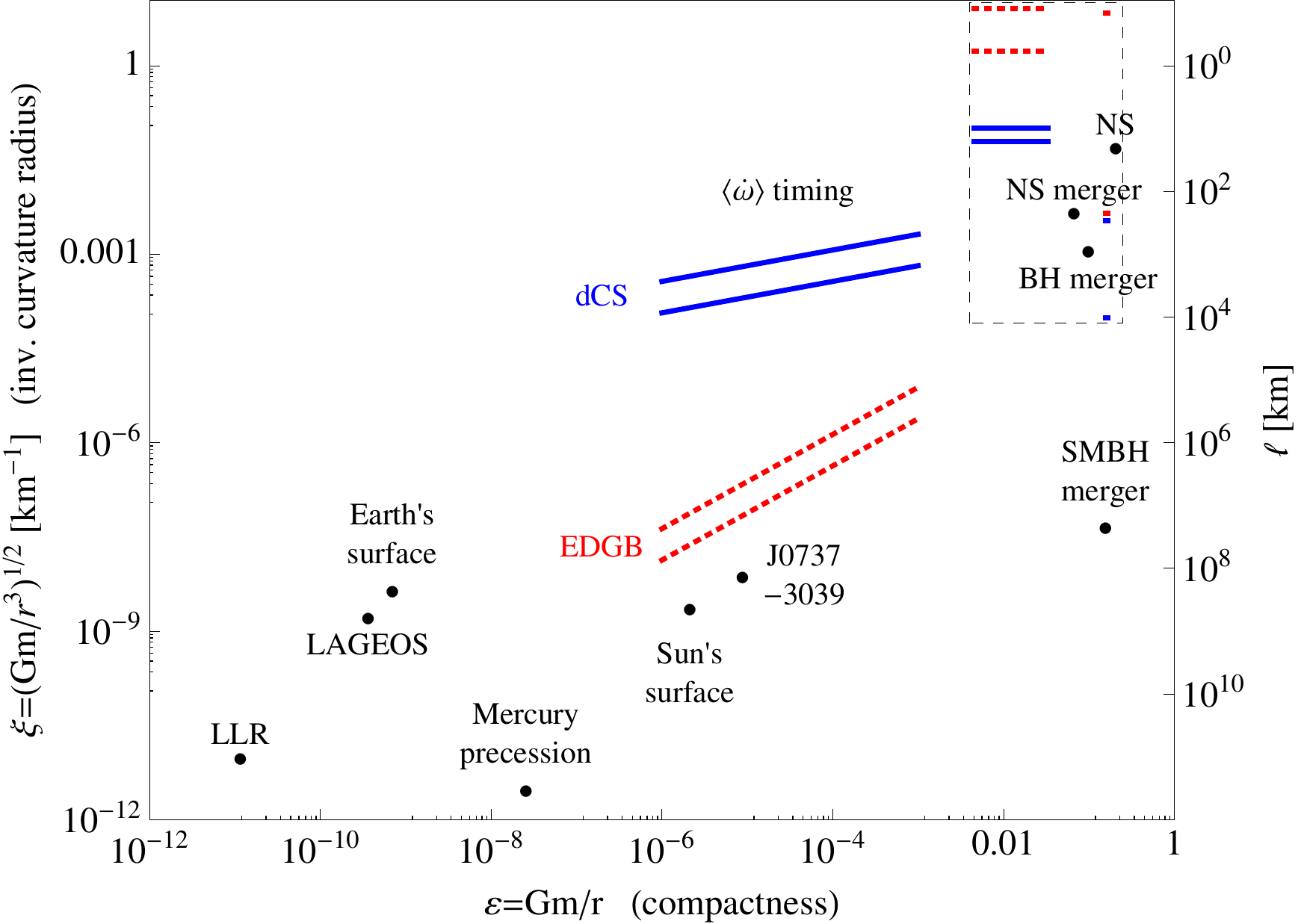}
  \caption{
    Estimated bounds on $\ell$, as shown on the right vertical axis.
    Tilted lines present estimates from pericenter precession
    in Eq.~\eqref{eq:ell-bound-est-pericenter}, while horizontal lines 
    correspond to those from GWs.
    The horizontal axis represents the compactness of a binary given 
    by  $(Gm/a)$.
    Solid (blue) curves are bounds for dCS, while dotted
    (red) curves are for EdGB.
    The lower (upper) curve for each theory is the estimate for
    $|\sigma/\avg{\dot{\omega}}|\sim 1$  
    ($|\sigma/\avg{\dot{\omega}}|\sim 10^{-2}$).
    The dashed region is expanded in Fig.~\ref{fig:est-bounds-beta}.
    Figure from~\cite{Stein:2013wza}.
    \label{fig:est-bounds-all}}
\end{figure}

From this calculation it is possible to estimate bounds which can be
placed on the dimensional coupling parameter, $\ell$, in some theory.
If a binary pulsar system is observed with sufficient precision to
measure $\avg{\dot\omega}$ and two other post-Keplerian parameters,
this constitutes a test of GR~\cite{lrr-2003-5,Wex:2014nva}.  Let us presume that the variance on
the measurement of $\avg{\dot\omega}$ is $\sigma$.  If the
measurements are consistent with GR, then we must have that the excess
precession [from Eq.~\eqref{eq:avg-dot-omega-scaling}]
$\avg{\delta\dot\omega} \lesssim \sigma$ is smaller than this
variance, which leads to the bound on $\ell$,
\begin{equation}
\label{eq:ell-bound-est-pericenter}
\ell^{2+2\wp} \lesssim
\frac{|\sigma|}{\avg{\dot{\omega}}}
\frac{ GmG\mu R_{1}^{\wp}R_{2}^{\wp} }{(4s+1)!! C_{1}^{r}C_{2}^{r}}
\left[
\frac{(Gm)^{2}}{R_{1}R_{2}}
\right]^{s}
v^{2(1-2s)} \,.
\end{equation}
Here we have set $t=s$ since all known binary systems are two NSs, and
therefore they will have the same type of scalar hair.  These scaling
estimates of bounds are presented in Fig.~\ref{fig:est-bounds-all}.

\subsubsection{Parameterization of Scalar Radiation}
\label{sec:param-scal-rad}

A binary system where one or both bodies sources scalar hair will emit
scalar radiation.  This in turn will affect the rate of inspiral of
the binary and thus change the gravitational waveform.  The scalar
radiation may be found in terms of derivatives of the binary's combined
scalar moments~\cite{Will:1996zj},
\begin{align}
\label{eq:theta-FZ-NZ-int}
\theta = \sum_{q=0}^{\infty} \frac{(-)^{q}}{q!}
\left(
\frac{1}{r}\mu^{Q}_{\txt{bin}}
\right)_{,Q}\,,
\end{align}
where $r$ is the distance to the binary, and where
\begin{align}
\label{eq:multipole-source-integral}
\mu^Q_{\txt{bin}}(u) = \int_{\mathcal{M}} \tau_{\txt{eff}}(u,\mathbf{x}) x^{Q} d^{3}x\,,
\end{align}
with the point-particle effective source is given by the
superposition
\begin{equation}
\label{eq:tau-eff}
\tau_{\txt{eff}} = (-)^{s}\mu_{1}^{S} \pd_{S} \delta^{(3)}(\mathbf{x}-\mathbf{x}_{1}) + 
(-)^{t}\mu_{2}^{T} \pd_{T} \delta^{(3)}(\mathbf{x}-\mathbf{x}_{2}) \,.
\end{equation}
Stein and Yagi find the $q$th binary moment as
\begin{equation}
\label{eq:mu-bin-from-BHs}
\mu_{\txt{bin}}^{Q}=
\begin{cases}
\frac{q!}{s!}\mu_{1}^{(k_{1}\cdots k_{s}}x_{1}^{k_{s+1}\cdots
 k_{q})} \plusonetwo & q\ge s\,, \\
0 & \text{otherwise.}
\end{cases}
\end{equation}

Now for each $q$ in Eq.~\eqref{eq:theta-FZ-NZ-int} one must compute
$(\mu^{Q}_{\txt{bin}}/r)_{,Q}$.  It is not immediately clear which
radiative binary moment will dominate this sum.  Therefore SY simply
parameterized this as $\ell_{\txt{rad}}$, so that the radiative
solution may be approximated as
\begin{equation}
\label{eq:theta-rad}
\theta_{\txt{rad}} \approx \frac{1}{r} \frac{n^{W}}{w!}
\ {}^{(w)}\!\mu_{\txt{bin}}^{W}(u)\,,
\end{equation}
where $w=|W|=\ell_{\txt{rad}}$, ${}^{(q)}f = (\pd/\pd t)^{q} f$, and
$n^{i}$ is the radial outward unit vector at some field point, when
the binary is taken at the origin of coordinates.

We know that $\ell_{\txt{rad}}$ must be greater than or equal to
$\min(\ell_{1},\ell_{2})$, the smaller of the two individual bodies'
scalar multipole moments (since all binary moments of lower valence
will vanish).  It is possible that the dominant radiative multipole
moment is simply $\ell_{\txt{rad}} = \min(\ell_{1},\ell_{2})$.
However, the computation of ${}^{(w)}\mu_{\txt{bin}}^{W}$ for
$w = \min(\ell_{1},\ell_{2})$ does not contain any time derivatives of
the separation vector between the two bodies.  The $w$th derivative of
this quantity depends on changes to the internal structure of the
bodies, or on the precession of their spins.

On the other hand, for $w = 1 +\min(\ell_{1},\ell_{2})$,
${}^{(w)}\mu_{\txt{bin}}^{W}$ will indeed involve a time derivative of
the separation vector, will thus vary on an orbital timescale, and is
thus expected to be important.  The binary moment for this $w$ is
\begin{equation}
\label{eq:mu-bin-w-equal}
\mu_{\txt{bin}}^{aS} = w x_{1}^{(a}\mu_{1}^{S)} \plusonetwo.
\end{equation}
This sum may contain only one term if e.g.~$\ell_{1}\ne \ell_{2}$
if the binary is a BH-NS system.  For a circular orbit, SY find the
relevant time derivative to be
\begin{equation}
{}^{(w)}\mu_{\txt{bin}}^{aS} =
\frac{w}{(Gm)^{s}}
\begin{cases}
(-)^{(1+s)/2} \mu_{\txt{red}}^{(S} n_{12}^{a)} v^{3s+1} \,, & |\epsilon|=1 \\
(-)^{s/2} \ \ \quad \mu_{\txt{red}}^{(S} v_{12}^{a)} v^{3s}  \,, & |\epsilon|=0
\end{cases}
\end{equation}
where $\ell_{\txt{rad}}=w=1+s=1+\min(\ell_{1},\ell_{2})$, and where SY have
defined a certain combination
\begin{equation}
\label{eq:mu-red-def}
\mu_{\txt{red}}^{S} \equiv \left[
\frac{m_{2}}{m}\mu_{1}^{S} -
\frac{m_{1}}{m}\mu_{2}^{S}
\right]\,.
\end{equation}

By combining this calculation with the estimates of
Eqs.~\eqref{eq:even-mu-Q-general} 
and~\eqref{eq:odd-mu-Q-general-Shat}, it is possible to estimate the
scaling of $\mu_{\txt{bin}}^{W}$ and its time derivatives.  This
scaling is controlled by the difference $s-\wp$.  Stein and Yagi
present this scaling estimate for the four cases $s-\wp=-1,0,+1,+2$.

This quantity appears in the energy flux due to scalar radiation,
which SY find to be
\begin{equation}
\dot{E}^{(\vartheta)} = - \frac{1}{(w!)^{2}} \frac{4\pi}{2w+1}
\delta_{(VW)} \avg{ {}^{(w+1)}\!\mu_{\txt{bin}}^{V}
  {}^{(w+1)}\!\mu_{\txt{bin}}^{W} }\,,
\end{equation}
where $|W|=|V|=w$.  Using the above assumptions, this result becomes
\begin{equation}
\label{eq:Edot-theta}
\dot{E}^{(\vartheta)} = - \frac{1}{(s!)^{2}} \frac{4\pi}{2w+1}
\frac{1}{(Gm)^{2w}}\mu_{\txt{red}}^{S}\mu_{\txt{red}}^{T}
\delta_{(abST)}
\begin{cases}
\avg{ n_{12}^{a} n_{12}^{b} v^{6s+8} }\,, & |\epsilon|=0 \\
\avg{ v_{12}^{a} v_{12}^{b} v^{6s+6} }\,, & |\epsilon|=1
\end{cases}
\end{equation}
where $|S|=|T|$ and $w=1+s$.

This is easily seen to be of $+(3s-1)$~pN order relative to the
leading GR energy flux due to quadrupole radiation:
\begin{equation}
\frac{\dot{E}^{(\theta)}}{\dot{E}^{\txt{GW}}} \sim \frac{5\pi^{2}}{(2w+1)(s!)^{2}}
\frac{\mpl^{2}}{\mu^{2}}
\frac{|\mu_{\txt{red}}^{S}|^{2}}{(Gm)^{2s}}
v^{6s-2}\,.
\end{equation}
Using the scaling estimates, which recall are controlled by $s-\wp$,
SY derive that for $|\epsilon|=0$ theories:
\newcommand{\thisLHS}{\frac{\dot{E}^{(\vartheta)}}{\dot{E}^{\txt{GW}}}}
\newcommand{\thisfactor}[1]{\frac{ #1 v^{6s-2}}{(2w+1)(s!)^2} \left( \frac{\ell}{Gm} \right)^{2+2\wp}}
\begin{subequations}
\label{eq:E-dot-theta-ratio}
\begin{align}
\label{eq:E-dot-theta-ratio-s-p--1}
\thisLHS &\sim \thisfactor{\eta^{-4}}
\frac{\delta m^{2}}{m ^{2}} & (s-\wp=-1) \\
\label{eq:E-dot-theta-ratio-s-p-0}
\thisLHS &\sim \thisfactor{\eta^{-2}}
\frac{\delta m^{2}}{m^{2}} & (s-\wp=0) \\
\label{eq:E-dot-theta-ratio-s-p-1}
\thisLHS &\sim \thisfactor{}
& (s-\wp=+1) \\
\label{eq:E-dot-theta-ratio-s-p-2}
\thisLHS &\sim \thisfactor{}
\frac{\delta m^{2}}{m^{2}}\,, & (s-\wp=+2)
\end{align}
\end{subequations}
where $\eta=m_{1}m_{2}/m^{2}=\mu/m$ is the symmetric mass ratio.
These scaling estimates agree with previously-derived results in
specific theories such as EdGB (where $s-\wp=-1$ for two BHS) and dCS
(where $s-\wp=0$ for two BHs; though note this theory has
$|\epsilon|=1$)~\cite{Yagi:2011xp}.

\subsubsection{Connection to PPE and Estimates of GW Bounds}
\label{sec:conn-ppe-estim}
There are four physical effects which SY identify that correct the
inspiral and thus the GW signal.  These effects are:
\begin{enumerate}
\item The conservative scalar pole-pole interaction, which modifies
  the binding energy, and the relation between $r_{12}$ and the
  orbital frequency $\omega$ (the ``Kepler'' relation).  This was
  discussed in Sec.~\ref{sec:binary-scal-inter}.
\item A conservative effect from the correction to the metric
  multipole moments of the bodies.  For example, a shift in the
  bodies' metric quadrupoles from the GR values will shift the metric
  monopole-quadrupole interaction energy from the GR value.
\item The dissipative effect of the energy lost due to scalar
  radiation.
\item A dissipative correction to how much energy is radiated in
  GWs.
\end{enumerate}
Effects (i) and (iii) were discussed above in
Secs.~\ref{sec:binary-scal-inter} and~\ref{sec:param-scal-rad}.
Each effect may contribute to the gravitational waveform, though they
may all enter at different post-Newtonian orders; therefore we do not
know, a priori, which effect will dominate the correction to the GR
waveform.  In EdGB, for a BH-BH system, effect (iii) is a relative
$-1$PN effect and thus dominates~\cite{Yagi:2011xp}.  Meanwhile, in dCS,
all of these corrections enter the waveform at the same order: +2pN
relative to the leading GR phase~\cite{Yagi:2013mbt,Yagi:2012vf}.

Since Ref.\cite{Stein:2013wza} focused only on the scalar interaction,
they choose to simply parameterize effects (ii) and (iv) for the
purposes of computing waveforms and mapping to the PPE framework 
explained in Sec.~\ref{sec:parameterized-GW}.

To parameterize effect (ii), SY wrote the shift to the binding energy
due to the metric deformation as
\begin{equation}
\label{E-deform}
\delta E_\txt{bind}^\txt{def} = \frac{C_\txt{def}}{r^{1+n_\txt{def}}}\,.
\end{equation}
If this is the leading conservative non-GR correction, then
$C_{\txt{def}}=E_{\txt{GR}}A_{\ppE}$ and
$1+n_{\txt{def}}=p_{\ppE}$~\cite{Chatziioannou:2012rf}.  If there is
another effect at this same post-Newtonian order, then $C_{\txt{def}}$
adds linearly to $E_{\txt{GR}}A_{\ppE}$.  Otherwise, this contribution is
sub-dominant.

To parameterize effect (iv), SY write the correction in the energy
flux due to the correction in GWs as
\begin{equation}
\dot{E}^{(h)} = C_h \left( \frac{G m}{r_{12}} \right)^{5+n_h}\,.
\end{equation}
Again if this is the leading dissipative non-GR correction, then this
corresponds to $C_{h}=\dot{E}_{\txt{GR}} B_\ppE$ and
$5+n_{h}=q_\ppE$~\cite{Chatziioannou:2012rf}.  If there is another effect
at the same order, then $C_{h}$ adds linearly to
$\dot{E}_{\txt{GR}}B_\ppE$.  Otherwise, this contribution is sub-dominant.

Having all effects (i-iv) parameterized, SY are then able to go
through the standard post-Newtonian procedure to find the corrections
to the GW phase.  From the two conservative effects,
they find the combined binding energy in frequency space is
\begin{equation}
\label{eq:mod-energy}
E_{\txt{bind}}  \sim \mu (G m \omega)^{2/3}
\left[
1 +
\frac{1}{Gm\mu}\frac{|\mu_{1}\mu_{2}|}{(Gm)^{2s}}
(Gm\omega)^{4s/3}
+\frac{1}{Gm\mu}
\frac{C_{\txt{def}}}{(Gm)^{n_{\txt{def}}}} (Gm\omega)^{2n_{\txt{def}}/3}
\right] \,.
\end{equation}
Including these corrections and the two dissipative effects, they then
find the corrected energy flux in frequency space as
\begin{align}
\label{eq:mod-Edot}
\dot{E}  \sim \frac{\eta^2}{G} (G m \omega)^{10/3}
\Bigg[
1 &{}+
\frac{1}{Gm\mu}\frac{|\mu_{1}\mu_{2}|}{(Gm)^{2s}}
(Gm\omega)^{4s/3}
+\frac{1}{Gm\mu} \frac{C_{\txt{def}}}{(Gm)^{n_{\txt{def}}}}
(Gm\omega)^{2n_{\txt{def}}/3}\nn\\
&{}+ \frac{G}{\eta^{2}(Gm)^{2}}
  \frac{|\mu_{\txt{red}}|^{2}}{(Gm)^{2s}} (Gm\omega)^{2s-2/3}
+ \frac{G C_{h}}{\eta^{2}} (Gm\omega)^{2n_{h}/3}
\Bigg] \,.
\end{align}

By going through the stationary phase approximation, SY can then find
the correction to the gravitational waveform phase due to these 4 effects.  They
find
\begin{align}
\label{eq:mod-phase}
\Psi (f) &\sim  \frac{1}{\eta} (\pi G m f)^{-5/3} +
\frac{1}{\eta^{2} G\, m^{2}} \frac{|\mu_{1}\mu_{2}|}{(Gm)^{2s}}
(\pi G m f)^{(4s-5)/3} \nn \\
&\qquad \qquad+ \frac{1}{\eta^2 G\, m^{2}} \frac{C_\txt{def}}{(Gm)^{n_\txt{def} }}
(\pi G m f)^{(2n_\txt{def}-5)/3} \nn\\
&\qquad\qquad + \frac{1}{\eta^3 G\, m^{2}} \frac{|\mu_\txt{red}|^2}{(Gm)^{2s}}
(\pi G m f)^{(6s-7)/3} + \frac{G C_{h}}{\eta^3} (\pi G m f)^{(2n_h-5)/3}\,.
\end{align}
From this equation it is straightforward to read off the PPE
coefficients in Eq.~\eqref{eq:simplest-PPE}. 
The corresponding $b_{\ppE}$ parameters are
\begin{subequations}
    \label{eq:b-ppE-all}
  \begin{align}
    \label{eq:b-ppE-1}
    b^{(i)}_\ppE &= (4s-5)/3  \\
    \label{eq:b-ppE-2}
    b^{(ii)}_\ppE &= (2n_\txt{def}-5)/3 \\
    \label{eq:b-ppE-3}
    b^{(iii)}_\ppE &= (6s-7)/3 \\
    \label{eq:b-ppE-4}
    b^{(iv)}_\ppE &= (2n_h-5)/3 \,.
  \end{align}
\end{subequations}
The $\beta_{\ppE}$ coefficients for effects (ii) and (iv) are
given as
\begin{subequations}
  \label{eq:beta-ppE-all}
  \begin{align}
    \label{eq:beta-ppE-2}
    \beta^{(ii)}_{\ppE} &\sim \frac{1}{\eta^{1+2n_{\txt{def}}/5}G \, m^{2}}
    \frac{C_{\txt{def}}}{(Gm)^{n_{\txt{def}}}} \\
    \label{eq:beta-ppE-4}
    \beta^{(iv)}_{\ppE} &\sim \frac{G C_{h}}{\eta^{2+2n_{h}/5}} \,.
  \end{align}
For effect (i), in the case with $s=t$ and $|\epsilon|=0$, SY find
\begin{equation}
  \label{eq:beta-ppE-1}
  \beta^{(i)}_{\ppE} \sim \frac{\ell^{2}}{\eta^{1+4s/5}(G m)^{2}}
  \left(
\frac{\ell^{2}}{R_{1}R_{2}}
  \right)^{\wp}
  \left(
    \frac{R_{1}R_{2}}{(Gm)^{2}}
  \right)^{s} (C_{1}C_{2})^{r} \,.
\end{equation}
Finally, as seen in Eq.~\eqref{eq:E-dot-theta-ratio}, different
values of $s-\wp$ give different expressions for effect (iii).  For
$s-\wp=-1,0,+1,+2$, SY find
\begin{align}
  \label{eq:beta-ppE-3-s-p--1}
  \beta^{(iii)}_{\ppE} &\sim \frac{1}{\eta^{(12+6\wp)/5}}
  \left( \frac{\ell}{Gm} \right)^{2+2\wp}
  \left( \frac{\delta m}{m} \right)^{2}& (s-\wp=-1) \\
  \label{eq:beta-ppE-3-s-p-0}
  \beta^{(iii)}_{\ppE} &\sim \frac{1}{\eta^{(8+6\wp)/5}}
  \left( \frac{\ell}{Gm} \right)^{2+2\wp}
  \left( \frac{\delta m}{m} \right)^{2}& (s-\wp=0) \\
  \label{eq:beta-ppE-3-s-p-1}
  \beta^{(iii)}_{\ppE} &\sim \frac{1}{\eta^{(4+6\wp)/5}}
  \left( \frac{\ell}{Gm} \right)^{2+2\wp}
  & (s-\wp=+1)\\
  \label{eq:beta-ppE-3-s-p-2}
  \beta^{(iii)}_{\ppE} &\sim \frac{1}{\eta^{(10+6\wp)/5}}
  \left( \frac{\ell}{Gm} \right)^{2+2\wp}
  \left( \frac{\delta m}{m} \right)^{2}\,. & (s-\wp=+2)
\end{align}
\end{subequations}
This completes the mapping onto the PPE framework.  These results
agree with previous calculations within specific theories such as EdGB~\cite{Yagi:2011xp}
and dCS~\cite{Yagi:2011xp,Yagi:2012vf}.

Having mapped the effects of generic scalar interactions onto the PPE
framework, it is possible to reuse PPE parameter estimation studies to
determine how $\ell$ can be bounded in these theories.  Consider a GW
inspiral measurement with a given SNR, over a
certain frequency range $f_{\min}\le f \le f_{\max}$, which is found
to be consistent with the predictions of GR.  For such a measurement,
Cornish et~al.~\cite{Cornish:2011ys} established the estimated bound
\newcommand{\udiff}[1]{\Delta u^{b^{#1}}}
\begin{equation}
\label{eq:Cornish-beta-estimate}
  |\beta|\lesssim \frac{3}{\text{SNR} \udiff{}}\,,
\end{equation}
where
$\Delta u^{b}\equiv|(u_{\min})^{b_{\ppE}}-(u_{\max})^{b_{\ppE}}|$,
with the dimensionless frequency parameter
$u=\pi G\mathcal{M}f=\eta^{3/5}v^{3}$ (recall that
$\mathcal{M} = m \eta^{3/5}$ is the chirp mass of the inspiral).  This
can be immediately converted into an estimated bound on $\ell$.

Specifically, for effect (i), SY find the estimated bound
\begin{equation}
  \label{eq:ell-bound-beta-1}
  \ell^{2+2\wp} \lesssim \frac{3}{\text{SNR}\udiff{(i)}}
  \frac{\eta^{+1+4s/5}}{C_{1}^{r}C_{2}^{r}}
  (Gm)^{2}R_{1}^{\wp}R_{2}^{\wp}
  \left[
\frac{(Gm)^{2}}{R_{1}R_{2}}
  \right]^{s} \,.
\end{equation}
Similarly, SY derive inequalities for effect (iii), which are
controlled by the difference $s-\wp$.  For the four values previously
considered, we have
\begin{subequations}
  \label{eq:ell-bound-beta-3}
  \begin{align}
  \label{eq:ell-bound-beta-3-s-p--1}
    \ell^{2+2\wp} &\lesssim
    \frac{3 \eta^{(12+6\wp)/5}}{\text{SNR}\udiff{(iii)}}(Gm)^{2+2\wp}
    \left(
      \frac{m}{\delta m}
    \right)^{2} &(s-\wp=-1) \\
  \label{eq:ell-bound-beta-3-s-p-0}
    \ell^{2+2\wp} &\lesssim
    \frac{3 \eta^{(8+6\wp)/5}}{\text{SNR}\udiff{(iii)}}(Gm)^{2+2\wp}
    \left(
      \frac{m}{\delta m}
    \right)^{2} &(s-\wp=0) \\
  \label{eq:ell-bound-beta-3-s-p-1}
    \ell^{2+2\wp} &\lesssim
    \frac{3 \eta^{(4+6\wp)/5}}{\text{SNR}\udiff{(iii)}}(Gm)^{2+2\wp}
    &(s-\wp=+1) \\
  \label{eq:ell-bound-beta-3-s-p-2}
    \ell^{2+2\wp} &\lesssim
    \frac{3 \eta^{(10+6\wp)/5}}{\text{SNR}\udiff{(iii)}}(Gm)^{2+2\wp}
    \left(
      \frac{m}{\delta m}
    \right)^{2}\,. &(s-\wp=+2)
  \end{align}
\end{subequations}

\begin{figure}[tbp]
  \centering
  \includegraphics[height=10cm]{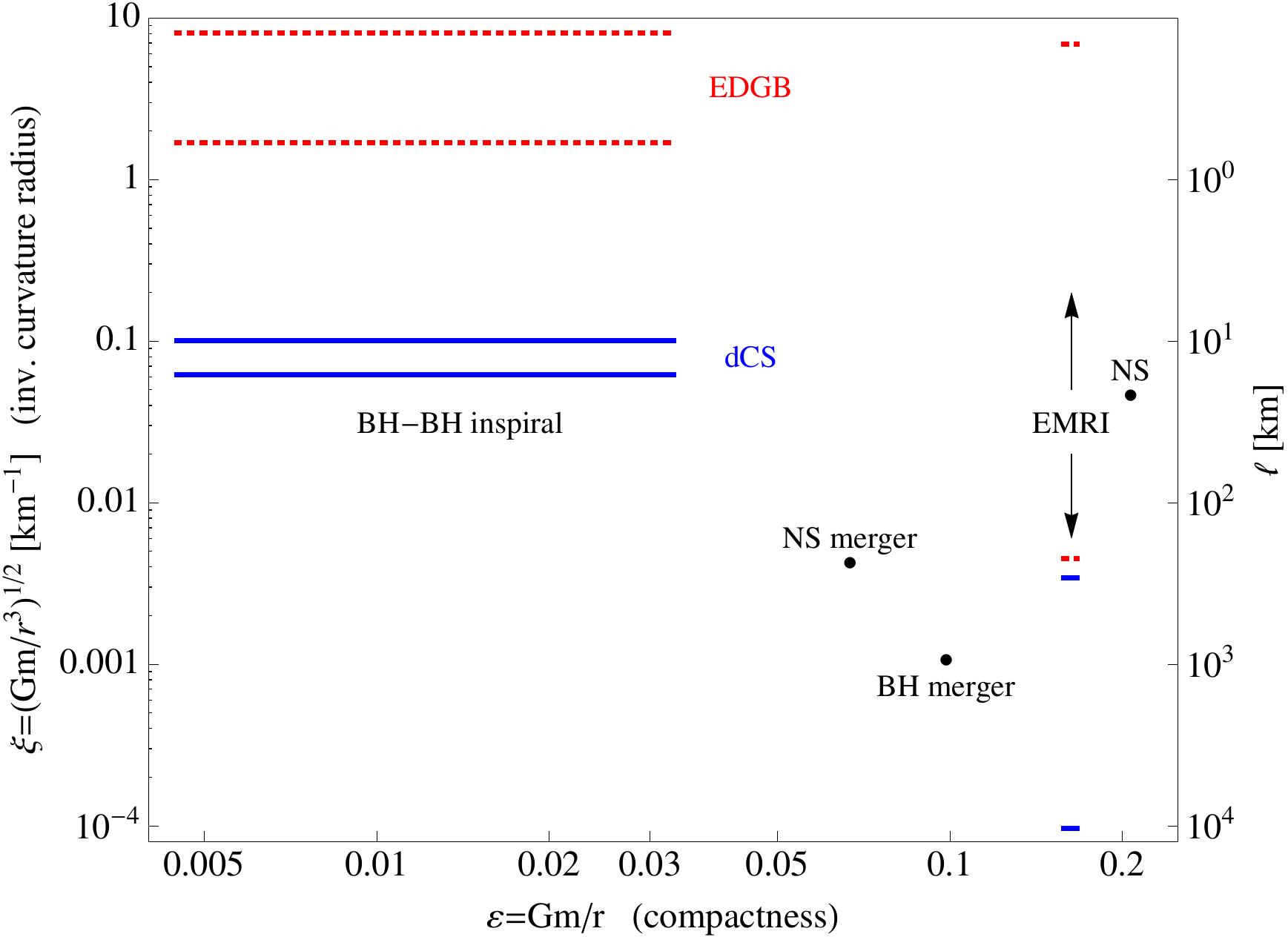}
  \caption{%
    Estimated bounds on $\ell$ from GW
    measurements, shown by the dashed region
    in Fig.~\ref{fig:est-bounds-all}.
    We assume a GW detection at SNR 30 that is consistent
    with GR. Solid
    (blue) lines are for dCS, while dotted (red) lines
    are for EdGB. Estimated bounds from a BH-BH inspiral in LIGO
    (an EMRI of a BH-BH binary in LISA)
    appear at left (right). The horizontal extent of such bounds
    represents the range of frequencies in band. 
    Figure from~\cite{Stein:2013wza}.
    \label{fig:est-bounds-beta}}
\end{figure}

Some estimated bounds based on these inequalities have been plotted in
Fig.~\ref{fig:est-bounds-beta}.  Stein and Yagi plot examples arising
from BH-BH inspirals in both dCS and EdGB.  For each example in each
theory, SY plot both bounds arising from effects (i) and (iii)
[respectively Eq.~\eqref{eq:ell-bound-beta-1} and
Eqs.~\eqref{eq:ell-bound-beta-3}]. For each theory, the two example
systems considered are a $10M_{\odot}$--$11M_{\odot}$ binary detected
in LIGO over the frequency range 20-400Hz; and a
$10M_{\odot}$--$10^{6}M_{\odot}$ binary detected in LISA over a
frequency range corresponding to 1 year of measurement terminating at
a plunge.  Both of these examples are assumed to be measured at an SNR
of 30, so that Eq.~\eqref{eq:Cornish-beta-estimate} gives
$|\beta|\lesssim 0.1/\udiff{}$.

Qualitatively, we can see in Fig.~\ref{fig:est-bounds-beta} that
stellar mass BHs yield better estimated bounds on $\ell$.  This makes
sense for theories like dCS and EdGB, which are higher-curvature
corrections to GR.  The curvature at the horizon of a BH goes as $\sim
1/M^{2}$, so lighter BHs are better probes of these
higher-curvature theories.

This analysis used dCS and EdGB as examples, in order to validate the
approach against well-studied theories.  The utility of this work is
that it can easily be reused for a broad class of theories which
include a long-ranged scalar field which is non-minimally coupled to
curvature.  Rather than redoing a whole suite of calculations in every
specific theory, one performs a matching calculation for compact
objects, and can then plug in parameters
$(|\epsilon|,d,r,\ell_{\BH},\ell_{\NS}, \mu^{S}_{\BH},
\mu^{S}_{\NS})$ to immediately find how well such a theory could be
constrained by pulsar timing or GW measurements.

\subsection{EMW Tests}
\label{sec:electr-wave-tests}

Let us now review BH based tests of GR with EMW observations~\cite{lrr-2008-9,Bambi:2015kza}.
We consider a BH with an accretion disk system.
We cover X-ray continuum spectrum, Fe line emission, BH shadow
and QPO in turn. 
We also comment on possible systematic errors in using these tests to probe GR.
For simplicity, we will neglect the light bending effect in the explanation below.
If one is to take such an effect into account,
 one needs to solve the null geodesic equations numerically using the ray-tracing
  algorithm~\cite{Psaltis:2010ww,Vincent:2011wz}.
Other possible tests that are not covered in this review include
e.g.~X-ray polarization~\cite{2012ApJ...754..133K,Liu:2015ibq},
stars~\cite{1999ApJ...514..388W,2008ApJ...674L..25W,Merritt:2009ex,Sadeghian:2011ub,2012ApJ...747....1L,Liu:2014uka} 
and hot spots~\cite{2014ApJ...787..152L,Li:2014coa,Liu:2014awa} around a black hole,
and BH jets~\cite{Bambi:2012ku,Bambi:2012zg}.
We refer the readers to e.g.~\cite{Bambi:2015kza} for more thorough review on these topics.

\subsubsection{Continuum Spectrum}
\label{sec:continuum}

A soft X-ray component of the generic spectrum of a stellar-mass BH 
with an accretion disk is interpreted as the thermal spectrum of an optically thick, 
geometrically thin disk. Such a disk is realized when the radiative 
efficiency is relatively large and the disk is described by the Novikov-Thorne
model~\cite{1973blho.conf..343N}. 
When the central object is the Kerr BH, 
the spectrum depends mainly on five parameters, the BH mass and spin, 
the inclination angle, the mass accretion rate and the distance from an 
observer. If one has independent measurements of the BH mass,
 distance and the inclination angle, 
 one can apply the continuum fitting method to determine the
 BH spin (and the accretion rate). 
Such a method can also be used to distinguish the Kerr BH and a
naked singularity~\cite{Kovacs:2010xm,Takahashi:2010pw}.
On the other hand, if the central object is not the Kerr BH, the spectrum 
depends on additional parameters~\cite{Harko:2009rp,Harko:2009gc,Harko:2010ua,
Bambi:2011jq,Bambi:2013sha,Vincent:2013uea,
Kong:2014wha,Bambi:2014sfa,Johannsen:2014wba} 
that may be related to coupling constants
in non-GR theories.
Current observational data have already been used to place constraints
on possible deviations from the Kerr BH~\cite{Bambi:2011jq,
Kong:2014wha,Bambi:2014sfa}.

The spectrum can be derived as follows.
For an optically thick, geometrically thin accretion disk around a BH, 
the time-averaged flux $\mathcal{F}(r)$ from
the disk surface is given via the conservation of energy by~\cite{1973blho.conf..343N,Bambi:2011jq}
\be
\mathcal{F}(r) = - \frac{\dot M}{4 \pi \sqrt{-\mathcal{G}}} \frac{\partial_r \Omega}{(E-\Omega L_z)^2} 
\int^r_{r_\mathrm{in}} (E-\Omega L_z) (\partial_r L_z) dr\,,
\ee
where $r_\mathrm{in}$ is the inner radius of the accretion disk, which is usually chosen
to be the ISCO radius,
$\dot M$ is the constant accretion rate that is determined from the mass 
conservation law and $\sqrt{-\mathcal{G}} \equiv \sqrt{N^2 g_{rr} g_{\phi\phi}}$ 
with $N$ representing the lapse function. 
The choice of the inner radius is motivated from e.g.~\cite{Steiner:2010kd} when the source
is in the high/soft state, though it is currently unclear whether such a choice is valid 
also in the low/hard state.
$E$, $L_z$ and $\Omega$
are the specific energy, axial component of the specific angular momentum
and the angular velocity for circular geodesics in the equatorial plane given by~\cite{Bambi:2011jq}
\ba
\label{eq:Omega}
\Omega &=& \frac{- \partial_r g_{t\phi} \pm \sqrt{(\partial_r g_{t\phi})^2-(\partial_r g_{tt}) (\partial_r g_{\phi\phi})}}
{\partial_r g_{\phi\phi}}\,, \\
E &=& - \frac{g_{tt} + g_{t\phi} \Omega}{\sqrt{-g_{tt}-2g_{t\phi} \Omega-g_{\phi\phi}\Omega^2}}\,, \quad
\label{eq:Lz}
L_z = \frac{g_{t\phi} + g_{\phi\phi} \Omega}{\sqrt{-g_{tt}-2 g_{t\phi} \Omega - g_{\phi\phi} \Omega^2}}\,,
\ea
where + (-) sign in $\Omega$ corresponds to the corotating (counterrotating) orbits.
%
Assuming a black body radiation, the effective temperature 
is determined from $\mathcal{F}$ by $\mathcal{F} = \sigma T^4$,
where $\sigma$ is the Stefan-Boltzmann constant. 
Neglecting the light bending effect, one finds the luminosity as
\be
L(\nu) = 8 \pi h \cos i \int^{r_\mathrm{out}}_{r_\mathrm{in}} \int^{2 \pi}_0 
\frac{\nu^3 \; \sqrt{-\mathcal{G}} }{\exp [h \nu_e/kT]-1} \; d\phi \; dr\,,
\ee
where $h$ is the Planck constant, $k$ is the Boltzmann constant,
$i$ is the inclination angle and 
$r_\mathrm{out}$ is the outer radius of the accretion disk.
$\nu$ and $\nu_e$ are the frequencies in the local rest frame of the observer
and the emitter respectively, which are related through the redshift factor $g$ by
\be
\label{eq:g}
g \equiv \frac{\nu}{\nu_e} = \frac{\sqrt{-g_{tt}-2 g_{t\phi} \Omega - g_{\phi\phi}\Omega^2}}
{1 + \Omega r \sin \phi \sin i}\,.
\ee
Such a factor encodes the frame dragging,
 the gravitational redshift and the special and 
 general relativistic effects of Doppler boost.

\subsubsection{Iron Line Emission}
\label{sec:iron}

Another way to test GR is to use the iron line emission from an accretion disk around
a stellar-mass or super-massive BH. Thermal photons are scattered by electrons
(the inverse Compton scattering) in a hot corona above the disk, which produces 
a power-law component in the X-ray spectrum. Such photons irradiate the disk and
 produces spectral lines by fluorescence, the strongest of which is the K$\alpha$
 iron line at 6.4keV (this becomes roughly 7.1keV for 
 the fully ionized iron line). This line is narrow in frequency intrinsically, but is broadened
 and skewed due to relativistic effects. The existence of such a line was detected in e.g.~\cite{1989MNRAS.238..729F,1995Natur.375..659T} 
 (see~\cite{Reynolds:2002np,2007ARA&A..45..441M} for reviews).
 
For a Kerr BH, the shape of the emission spectrum depends on the BH spin,
the inclination angle, the disk emissivity and the outer radius of the emission
region. In particular, the lower-energy tail depends on the inner radius of the 
disk, which is again commonly identified as the ISCO radius. Since such a radius
depends on the BH spin in units of the BH mass, one can measure the dimensionless BH spin
from the shape of the line profile even if the BH mass and distance 
are not measured. One can also use such a profile to test the BH geometry, as studied
in~\cite{Bambi:2013sha,Vincent:2013uea,Johannsen:2014wba,Johannsen:2012ng,Bambi:2012at,
Bambi:2013hza,Bambi:2013jda,Jiang:2014loa,Jiang:2015dla,Moore:2015bxa}.
The continuum spectrum method cannot be applied to an accretion disk around a 
super-massive BH because the spectrum lies in the ultra-violet (UV) band, which is difficult
to measure due to dust absorption, while the iron line method can be applied to
both stellar-mass and super-massive BHs.

The iron line emission profile can be calculated as follows. 
In the celestial coordinates $(\alpha, \beta)$ in the observer's
plane in the sky, the observed flux at energy $E$ is given by~\cite{Bambi:2012at,Moore:2015bxa}
\be
\label{eq:F-alpha-beta}
\mathcal{F} (E) = \frac{1}{d^2} \int \int g^3 I_e (E_e) d\alpha d\beta\,,
\ee
where $d$ is the distance between the observer and the BH and the redshift 
factor $g$ is given by Eq.~(\ref{eq:g}).  $I_e(E_e)$
is the emitted specific intensity at the emitted energy $E_e$ given by~\cite{Bambi:2012at,Moore:2015bxa}
\be
I_e (E_e) = r^{-q} \delta (E_e - E_{\mathrm{K} \alpha}) 
= r^{-q} g \delta (E - g E_{\mathrm{K} \alpha})\,, 
\ee
where $E_{\mathrm{K} \alpha} = 7.1$~keV for fully ionized iron lines and $q (>0)$
is the emissivity index. 
Neglecting the light-bending effect, which is a good approximation for emission from 
a small inclination (face-on) disk, one can convert the integral 
over $(\alpha,\beta)$ in Eq.~(\ref{eq:F-alpha-beta}) to that over $(r,\phi)$ as~\cite{Bambi:2012tg,Moore:2015bxa}
\be
\mathcal{F} (E) = \frac{\cos i}{d^2} \int_0^{2\pi} \int_{r_\ISCO}^{r_\mathrm{out}} 
g^4 r^{-q} \delta (E - g E_{\mathrm{K} \alpha}) \sqrt{-\mathcal{G}} \, dr \, d\phi\,,
\ee
where $i$ is the inclination angle and $r_{\mathrm{out}}$ is the outer radius of the 
emission region in the disk.

\subsubsection{BH Shadow}

VLBI at (sub-)millimeter wavelengths
will allow us to directly image BHs in the near future using
e.g.~Event Horizon Telescope.
Due to strong gravity in the vicinity of a BH, some light rays emitted from
an accretion disk terminates inside the BH event horizon and never reaches
an observer at infinity. This, in turn, produces a dark region in the image, which
is called a BH shadow.
Such a shadow within GR has been calculated by many 
authors~\cite{Falcke:1999pj,Takahashi:2004xh,Takahashi:2005hy,
Huang:2007us,Broderick:2008qf,Bambi:2008jg,Hioki:2009na,
Bambi:2010hf,Takahashi:2011dr,Lu:2014zja,Cunha:2015yba}.
Since the shape of a shadow depends strongly on the BH geometry, 
one can in principle measure not only the BH spin but also a possible deviation
from the Kerr BH~\cite{Vincent:2013uea,Broderick:2005at,Johannsen:2010ru,Amarilla:2010zq,Amarilla:2011fx,
Bambi:2011yz,Amarilla:2013sj,Li:2013jra,
Loeb:2013lfa,Atamurotov:2013dpa,Wei:2013kza,Atamurotov:2013sca,Johannsen:2015qca,
Tsukamoto:2014tja,Bambi:2014mla,Psaltis:2014mca,Wei:2015dua,Moffat:2015kva,Ghasemi-Nodehi:2015raa}, 
which can be used to test GR.

The boundary of the BH shadow (a photon ring) is determined from an unstable photon orbit
with a constant radius. Such a boundary for the Kerr BH
 can be calculated as follows~\cite{Hioki:2009na}. 
The existence of the Carter constant $\mathcal{Q}$ allows one to separate the null geodesic equations. 
In particular, the radial component of the equations in the Boyer-Lindquist coordinate is given by
\be
\rho^2 \frac{dr}{d\lambda} = \pm \sqrt{\mathcal{R}}\,,  \quad \mathcal{R} \equiv (r^2 + a^2 - a\xi)^2
 - \Delta \mathcal{I}\,,
\ee
where $\rho^2 \equiv r^2 + a^2 \cos^2 \theta$, $\Delta \equiv r^2 - 2 M r + a^2$, $\lambda$ is the 
affine parameter and $M$ and $a$ are the BH mass and Kerr parameter respectively.
$\mathcal{I}$ is a function of $\tilde \xi \equiv L_z/E$ and $\tilde \eta \equiv \mathcal{Q}/E$ given by
$\mathcal{I} \equiv \tilde  \eta + (a - \tilde \xi)^2$ with $E$ and $L_z$ representing the specific energy and 
the axial component of the specific angular momentum of a photon orbit.
The conditions for the unstable photon orbit are given by
\be
\mathcal{R} = 0\,, \quad \frac{d\mathcal{R}}{dr} = 0\,,
\ee
which can be solved for $\tilde \xi$ and $\tilde \eta$ to yield\footnote{%
$\tilde \xi$ on the equatorial plane is related to the real part of the 
QNM frequency of the central BH within the WKB 
approximation~\cite{Mashhoon:1985cya,Cardoso:2008bp}. 
}
\be
\tilde \xi = \frac{r^2 + a^2}{a} - 2 \frac{r \Delta}{a (r - M)}\,, \quad 
\tilde \eta = - \frac{r^3 [r (r-3M)^2-4a^2 M]}{a^2 (r-M)^2}\,.
\ee
The photon ring in the celestial coordinate $(\alpha,\beta)$ that an observer sees is given by
\ba
\alpha = - \lim_{r\to \infty} \frac{r p^{(\phi)}}{p^{(t)}} = -\tilde \xi \csc i\,, \quad 
\beta = \lim_{r\to\infty} \frac{r p^{(\theta)}}{p^{(t)}}  = \sqrt{\tilde \eta + a^2 \cos^2 i - \tilde \xi^2 \cot^2 i}\,, \nn \\
\ea
where $i$ is the inclination angle and $(p^{(t)},p^{(r)},p^{(\theta)},p^{(\phi)})$ are the tetrad components of
the photon momentum in reference frames that are locally nonspinning.

\begin{figure}[htb]
\begin{center}
\includegraphics[width=7.5cm,clip=true]{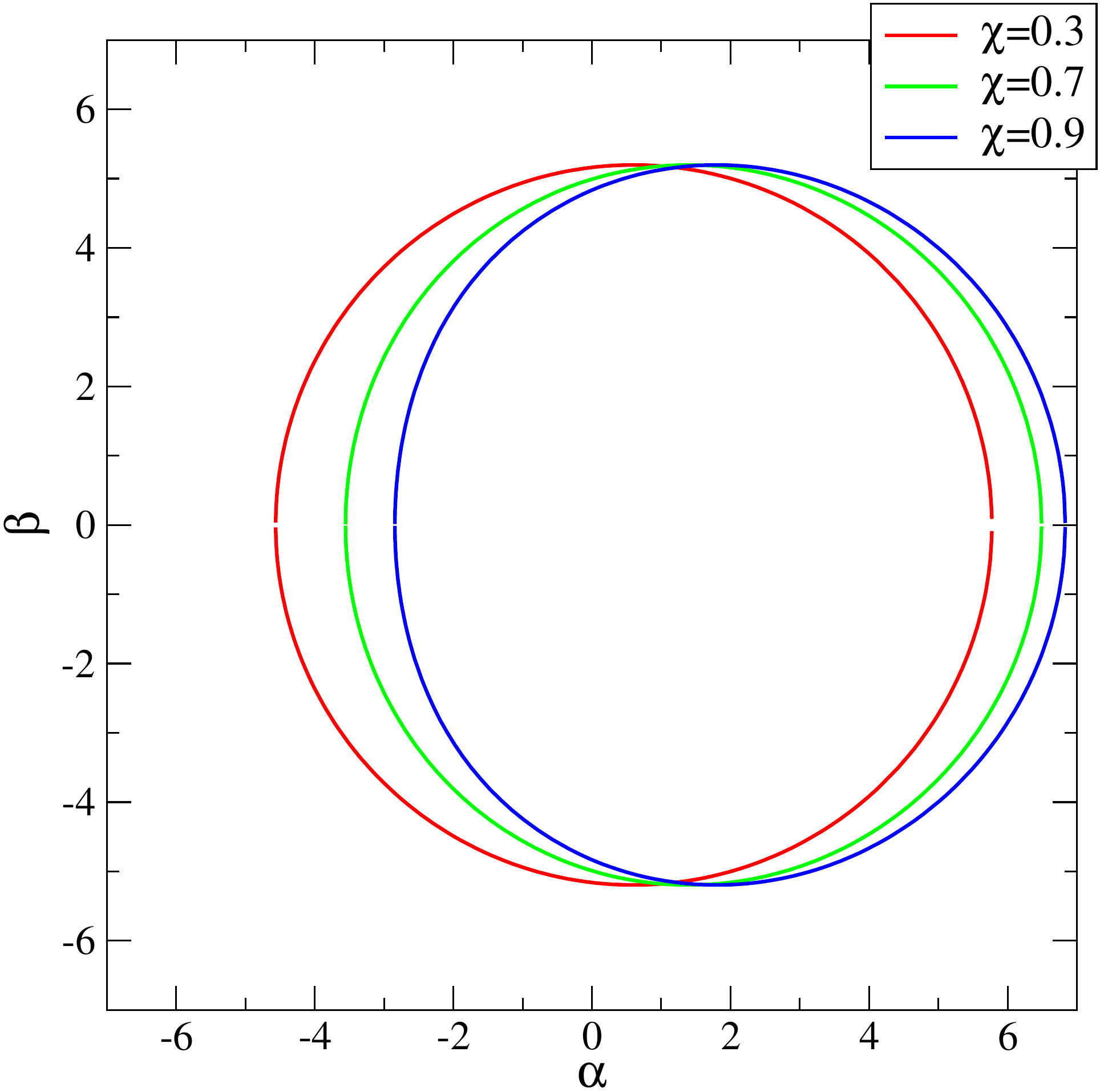}
\caption{\label{fig:photon-ring}
  (Color online) Photon ring of a Kerr BH with an inclination angle of
  $\pi/2$ and with various dimensionless spin parameter $\chi$.
  Observe how the shape changes as one varies $\chi$.
}
\end{center}
\end{figure}

Figure~\ref{fig:photon-ring} presents the photon ring for the Kerr BH with $i = \pi/2$ and various spin parameters
$\chi = a/M$. Observe that the center of the ring shifts to positive $\alpha$ and the deviation in the shape from
a circle becomes larger as one increases the spin. This shows that one can extract the information of the BH spin
from the shape of the BH shadow provided that one can break the degeneracy between the spin and inclination.
BH spacetimes in non-GR theories in general do not admit the Carter constant. 
(See e.g.~\cite{Yagi:2012ya} for a concrete example).
In such cases, one cannot separate the null geodesic equations and one needs to solve them numerically 
 using the ray-tracing algorithm~\cite{Psaltis:2010ww,Vincent:2011wz}.

\subsubsection{Quasi-periodic Oscillation}
\label{sec:QPO}

Observations of LMXBs show peaked features in the power 
density spectra called QPOs 
(see Refs.~\cite{2006csxs.book..157M,2006csxs.book...39V} for reviews).
In particular, high frequency QPOs from a BH-accretion disk system are observed
in commensurable pairs with the ratio 3:2~\cite{2001A&A...374L..19A, 2006csxs.book..157M}.
Although the origin of such QPOs is unclear, many models exist and they usually
relate QPO frequencies to fundamental oscillation frequencies of a test particle orbiting
a central object, namely the Keplerian and epicyclic frequencies. 

Such fundamental frequencies can be calculated as follows. 
First, from the normalization condition $u^\mu u_\mu = -1$ of the 4-velocity 
$u^\mu$ of a test particle, one finds
\be
g_{rr} \dot{r}^2 + g_{\theta\theta} \dot{\theta}^2 = V_\mrm{eff}\,, \quad 
V_\mrm{eff} \equiv \frac{E^2 g_{\phi\phi} 
+ 2 E L_z g_{t\phi} + L_z^2 g_{tt}}{g_{t\phi}^2 - g_{tt} g_{\phi\phi}}-1\,,
\ee
where the over dots refer to derivatives with respect to an affine parameter. 
The conditions for a circular orbit in the equatorial plane, $V_\mrm{eff}=0$,
$\partial_r V_\mrm{eff}=0$ and $\partial_\theta V_\mrm{eff}=0$ give 
$\Omega$, $E$ and $L_z$ shown in Eqs.~(\ref{eq:Omega}) and~(\ref{eq:Lz}).  
The Keplerian frequency is given by $\nu_\phi = \Omega / (2 \pi)$.
Next, let us consider small perturbations along the radial ($\delta_r$) and 
polar ($\delta_\theta$) directions from a fiducial orbit with $r=r_0$ and
$\theta = \pi/2$. Linear order perturbations are given by the solutions to
\be
\label{eq:epicyclic}
\frac{d^2 \delta_r}{dt^2} + \Omega_r^2 \delta_r = 0\,, \quad 
\frac{d^2 \delta_\theta}{dt^2} + \Omega_\theta^2 \delta_\theta = 0\,,
\ee
where 
\be
\Omega_r^2 = - \frac{1}{2 g_{rr} (u^t)^2} \frac{\partial^2 V_\mrm{eff}}{\partial r^2}\,, \quad 
\Omega_\theta^2 = - \frac{1}{2 g_{\theta\theta} (u^t)^2} \frac{\partial^2 V_\mrm{eff}}{\partial \theta^2}\,.
\ee
The radial and vertical epicyclic frequencies are given by $\nu_r = \Omega_r/(2\pi)$ 
and $\nu_\theta = \Omega_\theta/(2 \pi)$ respectively. For the Kerr BH, the fundamental 
frequencies are given by~\cite{Aliev:1980hz,1986Ap&SS.124..137A}
\ba
\nu_\phi &=& \frac{1}{2\pi} \, \frac{M^{1/2}}{r^{3/2} \pm a M^{1/2}}\,, \\
\nu_r &=& \nu_\phi \left( 1 - \frac{6M}{r} \pm \frac{8a M^{1/2}}{r^{3/2}} - \frac{3 a^2}{r^2} \right)\,, \\
\nu_\theta &=& \nu_\phi \left( 1 \mp \frac{4a M^{1/2}}{r^{3/2}} + \frac{3a^2}{r^2} \right)\,,
\ea
where the upper (lower) sign corresponds to a corotating (counterrotating) orbit.

We now review various QPO models. We classify them into three classes~\cite{Torok:2011qy}, 
(i) kinematic, (ii) resonant and (iii) diskoseismic models. We summarize the upper and lower
 kHz QPO frequencies in each model in the kinematic and resonant models in Table~\ref{table-QPO}.
 These QPO models were applied to probe possible deviations from the Kerr BH 
 in~\cite{Vincent:2013uea,Johannsen:2010bi,Bambi:2012pa,Aliev:2012rj,Bambi:2013fea,Maselli:2014fca}.

\emph{Kinematic Models}: 
These models consider hot spots or blobs orbiting in the accretion disks
around the central compact objects as the origin of QPOs.
One example is 
the \emph{relativistic precession} model~\cite{Stella:1997tc,Stella:1998mq,Morsink:1998mg},
which was first proposed to explain the QPOs for NSs in LMXBs and 
was applied to those for stellar-mass BHs in~\cite{Stella:1999sj}. 
The higher and lower kHz QPOs in such a model are given by
$\nu_{\phi}$ and $\nu_\phi - \nu_r$, where the latter corresponds to the 
periastron precession frequency.
Such a model also explains the low-frequency QPO as $\nu_\phi - \nu_\theta$,
which corresponds to the nodal precession frequency.
These triplets QPOs were observed in 
GRO J1655-40~\cite{Strohmayer:2001yn,Casella:2005vy,Motta:2012sy} 
with the Rossi X-ray Timing Explorer (RXTE).
The mass derived from such a QPO model~\cite{Motta:2013wga} is consistent with that
from optical/near-infrared spectro-photometric observations~\cite{Beer:2001cg}, while the
spin measurement~\cite{Motta:2013wga} is inconsistent with that 
from continuum spectrum fitting~\cite{Shafee:2005ef}.
Another kinematic model is the \emph{tidal disruption} model~\cite{2008A&A...487..527C,2009A&A...496..307K}, 
in which QPOs are caused by a blob of inhomogeneity inside the accretion disk being tidally stretched.

\fulltable{\label{table-QPO} Upper and lower kHz QPO frequencies in various models. The 
first two models correspond to the kinematic models while other models correspond to the resonant models.
(See also Table~1 of~\cite{Torok:2011qy}.) $\nu_\phi$ is the Keplerian frequency while $\nu_r$ and $\nu_\theta$
are the radial and vertical epicyclic frequencies respectively. $\nu_+$ and $\nu_-$ are defined by
$\nu_+ = \nu_\theta + \nu_r$ and $\nu_- = \nu_\theta-\nu_r$.
\vspace{1em}
}
\hline\hline
Models & References & lower frequency   &  upper frequency \\ \hline
relativistic precession & \cite{Stella:1997tc,Stella:1998mq,Morsink:1998mg} & $\nu_\phi - \nu_r$  & $\nu_\phi$  \\ 
tidal disruption & \cite{2008A&A...487..527C,2009A&A...496..307K} & $\nu_\phi $  & $\nu_\phi + \nu_r$  \\  \hline
parametric resonance & \cite{2003PASJ...55..467A,2005A&A...436....1T} & $\nu_r$  & $\nu_\theta$  \\ 
forced resonance & \cite{2005A&A...436....1T} & $\nu_-$  & $\nu_\theta$  \\ 
 & \cite{2005A&A...436....1T} & $\nu_\theta$  & $\nu_+$  \\ 
Keplerian resonance & \cite{2005A&A...436....1T} & $\nu_r$  & $\nu_\phi$  \\ 
& \cite{2005A&A...436....1T} & $2 \nu_r$  & $\nu_\phi$  \\ 
& \cite{2005A&A...436....1T} & $\nu_\phi$  & $3 \nu_r$  \\ 
warped disk oscillation & \cite{2001PASJ...53....1K,2004PASJ...56..559K,2004PASJ...56..905K,2005PASJ...57..699K,2008PASJ...60..111K} & $2 (\nu_\phi - \nu_r)$  & $2 \nu_\phi - \nu_r$  \\ 
non-axisymmetric disc-oscillation  & \cite{2004ApJ...617L..45B} & $\nu_\phi - \nu_r$  & $\nu_\theta$  \\ 
& \cite{2010ApJ...714..748T} & $\nu_\phi - \nu_r$  & $2 \nu_\phi - \nu_\theta$ \\ 
\hline\hline
\endfulltable

\emph{Resonant Models}:
Commensurability of QPO frequencies arises more naturally in the resonant models. 
Resonance models~\cite{Aliev:1980hz,2001A&A...374L..19A,2003PASJ...55..467A,2005A&A...436....1T} 
promote Eq.~(\ref{eq:epicyclic})
by introducing nonlinear couplings $F_r$ and $F_\theta$ between two epicyclic modes on the right hand side:
\be
\label{eq:epicyclic2}
\frac{d^2 \delta_r}{dt^2} + \Omega_r^2 \delta_r = \Omega_r^2 F_r \left( \delta_r, \delta_\theta, \dot \delta_r, \dot \delta_\theta \right)\,, \quad 
\frac{d^2 \delta_\theta}{dt^2} + \Omega_\theta^2 \delta_\theta = \Omega_\theta^2 F_\theta \left( \delta_r, \delta_\theta, \dot \delta_r, \dot \delta_\theta \right)\,,
\ee
where the dot refers to the time derivative. 
The \emph{parametric resonance} model~\cite{2003PASJ...55..467A,2005A&A...436....1T}  is 
given by $F_r=0$ and $F_\theta = - \delta_r \delta_\theta$. 
In such a case, the equation for $\delta_\theta$ becomes the Mathieu equation and 
the ratio between $\nu_r$ and $\nu_\theta$ is given by $\nu_r/\nu_\theta = 2/n$,
where $n$ is a positive integer. For the Kerr background, $\nu_\theta > \nu_r$, and hence
$n=3$ gives the dominant contribution. $n=1$ and $n=2$ modes may arise in non-Kerr
backgrounds. 
The \emph{forced resonance} model~\cite{2005A&A...436....1T} corresponds to $F_r = 0$ and 
$F_\theta = \delta_r/\Omega_\theta^2 + \bar F_\theta (\delta_\theta)$ 
where $\bar F_\theta$ represents the nonlinear term in $\delta_\theta$.
Such a nonlinear term gives resonant solutions for $\delta_\theta$, with the simplest
 frequencies being $\nu_+ = \nu_\theta + \nu_r$ and $\nu_- = \nu_\theta-\nu_r$.
The \emph{Keplerian}~\cite{2005A&A...436....1T} resonance model predicts couplings 
between Keplerian frequency and an
epicyclic frequency, although such a model maybe more unrealistic than the parametric or forced resonance
model. Kato~\cite{2001PASJ...53....1K,2004PASJ...56..559K,2004PASJ...56..905K,
2005PASJ...57..699K,2008PASJ...60..111K} 
proposed the \emph{warped disk oscillation} models, where QPOs
are caused by several resonances in a deformed disk with somewhat exotic geometry.
Other models include the \emph{non-axisymmetric disc-oscillation} models 
proposed in~\cite{2004ApJ...617L..45B,2010ApJ...714..748T},
where the frequencies of oscillation modes of a disk is similar to frequencies predicted 
in the relativistic precession model. Although the resonant models can naturally explain
commensurability of QPO frequencies, the underlying mechanisms are unclear in most of 
the models.

\emph{Discoseismic Models}:
These models predict that QPO frequencies correspond to those of the 
fundamental diskoseismic modes~\cite{1980PASJ...32..377K,1987PASJ...39..457O,
1992ApJ...393..697N,Perez:1996ti,2001ApJ...548..335S,2001ApJ...559L..25W,
2002ApJ...567.1043O,2008ApJ...680.1319S}, 
namely the g-modes, c-modes and p-modes.
The g-modes are inertial gravity wave modes that occur at frequencies close to
the maximum radial epicyclic frequency near the inner edge of the disc~\cite{Perez:1996ti}.
The c-modes are corrugation modes that exist only in disks that corotate with 
the central BH and whose frequency coincides with the Lense-Thirring frequency
at a certain radius close to the BH ISCO~\cite{2001ApJ...548..335S}. 
The p-modes are inertial-pressure 
oscillation modes that arises close to the edge of the accretion disc.
Unfortunately, magnetohydrodynamic simulations show that such modes typically
damp due to various instabilities~\cite{2009MNRAS.393..992T,
2009ApJ...690.1386F,2011MNRAS.410..399F}
 and do not reproduce the 3:2 QPOs.

\subsubsection{Systematics}
\label{sec:systematics}

We will now comment on possible systematic errors that may deteriorate tests of GR
with EMW observations. One major origin of such systematics is the uncertainty in 
the accretion disk model. For example, both the continuum spectrum and iron line methods assume
that the disk is geometrically thin and optically thick. However, recent quasar observations with
foreground microlensing are not fully consistent with the standard geometrically thin disk 
model~\cite{1990ApJ...358L..33W,1991ApJ...381L..39R,Pooley:2006rh,Dai:2009bp,
Morgan:2010xf,Blackburne:2010eq,JimenezVicente:2012wr,Jimenez-Vicente:2014lta}, and
the finite disk size may produce systematic errors~\cite{Miller:2014aaa}. 
The BH shadow observations are also affected by the uncertainty in the disk model.
Although the photon ring (Fig.~\ref{fig:photon-ring}) does not depend on the disk model, 
the intensity map of the image does depend on such a model.

The continuum spectrum method requires the disk inclination is known 
independently~\cite{Davis:2006cm}, and one way is to simply assume that 
the disk is aligned with the BH spin. 
On one hand, such an alignment is supported by calculations of binary 
population synthesis~\cite{2010ApJ...719L..79F}. However, 
the time scale for the Bardeen-Peterson effect~\cite{Bardeen:1975zz} 
to complete seems to be too long~\cite{Martin:2008wa}, and both 
observational data~\cite{Maccarone:2002ph} and theoretical calculations~\cite{Fragile:2000mf} 
suggest that tilted disks may exist. The observational data for 
XTE J1550-564~\cite{Steiner:2011vr} shows a good alignment, 
while those for GRO J1655-40~\cite{Hjellming:1995tv,Greene:2001wd} 
and V4641 Sgr~\cite{Orosz:2001dc} infer misalignment. The continuum 
spectrum method mentioned in Sec.~\ref{sec:continuum} assumed that the radiation 
is a blackbody, which is not completely true in nature, and one needs to correct 
the temperature by introducing a hardening factor (or a color factor). 
The correct computation of such a factor requires an accurate modeling of 
the disk atmosphere~\cite{Li:2004aq}, and significant progress has been 
made in e.g.~\cite{Davis:2004jf,Davis:2006cm}.

The iron line method requires the knowledge of the emissivity, which was taken as 
a simple power-law in Sec.~\ref{sec:iron}. However, this is clearly an approximation, 
and one may need a better modeling in order to reduce systematic errors. For example,
a cutoff in the emissivity may exist near ISCO, and its position is likely to affect
the measurement accuracy of the spin and deviations from 
Kerr~\cite{Noble:2011wa,Beckwith:2008pu,Noble:2008tm}.
Also, the 
assumption of taking the inner radius of the disk to be ISCO is not fully justified, as 
BHs are typically in the low/hard state. Furthermore, the thermal component of the 
disk may overlap with the low energy tail of the iron line, which makes the modeling 
more complicating. One may also need to account for Compton 
broadening~\cite{Bambi:2015kza}.

The QPO method is affected by the uncertainty in identifying the correct QPO model. 
Namely, one can carry out 
tests of GR only if the correct QPO model is known \textit{a priori}. Furthermore, current models
are incomplete, as they cannot explain e.g.~the energy dependence of QPOs~\cite{Miller:2014aaa}. 
They also cannot explain why some sources show QPOs with a 3:2 frequency ratio, 
while many sources exist with no high frequency QPOs.  

\section{Scalar-Tensor Theories}
\label{sec:scal-tens-theor}

We now look at specific alternative theories of gravity in turn. First, we will
focus on scalar-tensor theories.

\subsection{Basics}
\label{sec:ST-basics}

We here review basic properties of scalar-tensor theories,
BH solutions and current constraints from solar system and
binary pulsar observations.

\subsubsection{Theories}
\label{sec:ST-theories}


Scalar-tensor theories are one of the simplest and most well-studied
alternative theories of gravity~\cite{fujii}. One introduces one (or
more) additional scalar field to GR.  Such a scalar field can explain
the current accelerating expansion of the Universe~\cite{perrotta} or
inflation~\cite{steinhardt}, and it also arises from low energy
effective theory of string theory~\cite{polchinski1,polchinski2} as a
dilaton.

After an appropriate field redefinition, the most general
action for scalar-tensor theories with a single scalar field in
the Jordan frame, with at most second derivatives of the fields
up to total derivatives%
\footnote{The most general scalar-tensor theory with a single
  scalar field and second-order field equations is Horndeski's
  theory~\cite{Horndeski:1974wa}, whose action can be expressed in
  terms of Galileon interactions~\cite{Deffayet:2011gz}.
}
is given by~\cite{Bergmann:1968ve,Wagoner:1970vr}
\begin{equation}
S=\frac{1}{16\pi}\int d^4x \sqrt{-g} \left[ \vartheta R
                             - \frac{\omega(\vartheta)}{\vartheta}g^{\mu\nu}
                             \left( \partial_{\mu}\vartheta \right) \left( \partial_{\nu}\vartheta \right)
                             + V(\vartheta) +16\pi \mathcal{L}_{\MAT}(\Psi) \right]\,.
\end{equation}
Here, $g_{\mu\nu}$, $g$ and $R$ are the metric, its determinant and
the Ricci scalar in the Jordan frame, $\vartheta$ and $V$ represent
the scalar field and its potential, while $\mathcal{L}_{\MAT}$
corresponds to the matter Lagrangian with $\Psi$ representing the
matter Lagrangian density.  The coefficient $\omega(\vartheta)$ in
front of the kinetic term of the scalar field is a function of
$\vartheta$ that characterizes the theories. When
$\omega(\vartheta)= \omega_\BD$ and $V(\vartheta) = 0$ with
$\omega_\BD$ a constant called the Brans-Dicke (BD) parameter, 
the action reduces to
that of massless BD theory~\cite{brans}.  The inverse of $\omega_\BD$
characterizes the strength of a coupling between the scalar and matter
field.  The theory reduces to GR in the limit
$\omega_\BD \rightarrow \infty$. On the other hand, if one assumes
$V (\vartheta) = (1/2) m_s \vartheta^2$, the theory reduces to massive
BD theory~\cite{alsing,Berti:2012bp} with $m_s$ representing the mass
of the scalar field.  The scalar-tensor theories with multiple scalar
fields are studied in~\cite{Damour:1992we}.

Other well-studied scalar-tensor theories include quasi BD theory
 proposed by
Damour and Esposito-Far\`ese~\cite{Damour:1993hw,Damour:1996ke}.  Let
us first move to the Einstein frame by redefining the scalar field as
$\vartheta = \vartheta (\varphi)$ and carrying out a conformal
transformation on the metric as
$g_{\mu\nu} \to \tilde g_{\mu\nu} \equiv A (\varphi)^{-2} g_{\mu \nu}$
with $A(\varphi) = \vartheta^{-1/2}$.  In such a frame, the action is
given by
\begin{equation}
S=\frac{1}{16\pi}\int d^4x \sqrt{- \tilde g} \left\{ \tilde R
                             -2 \tilde g^{\mu\nu}
                             ( \partial_{\mu} \varphi ) ( \partial_{\nu} \varphi )
                             + U(\varphi) +16\pi \mathcal{L}_{\MAT}
                             \left[ \Psi, A(\varphi) \tilde g_{\mu\nu} \right] \right\}\,,
\end{equation}
where $\tilde g$ and $\tilde R$ are the determinant of the metric and
the Ricci scalar in the Einstein frame while
$U (\varphi) = A(\varphi)^4 V(\vartheta)$ is the potential for the
scalar field in this frame.  Notice that the scalar field is now
minimally coupled to the metric. The relation between
$\omega(\vartheta)$ and $A(\varphi)$ is given by~\cite{Will:2014kxa}
\begin{equation}
\omega (\vartheta) = \frac{1-3\, \alpha(\varphi)^2}{2 \, \alpha(\varphi)^2}\,,
\quad \alpha(\varphi) \equiv \frac{d\ln A(\varphi)}{d \varphi}\,.
\end{equation}
Damour and Esposito-Far\`ese~\cite{Damour:1993hw,Damour:1996ke} set
$U=0$ and expanded $\alpha (\varphi)$ around a constant $\varphi_0$ at
spatial infinity as
\begin{equation}
\alpha (\varphi) = \alpha_0 + \beta_0 (\varphi - \varphi_0) + \mathcal{O}\left[ (\varphi - \varphi_0)^2 \right]\,,
\end{equation}
where $\alpha_0$ and $\beta_0$ are constants.
In fact, having two constants $\alpha_0$ and $\varphi_0$ are 
redundant as they both enter 
only in the constant term up to the above truncation. 
For example, one can choose $\varphi_0 = \alpha_0/\beta_0$
such that the constant term vanishes and 
$\alpha (\varphi) = \beta_0 \varphi + \mathcal{O}\left[ (\varphi - \varphi_0)^2 \right]$~\cite{Palenzuela:2013hsa,
Barausse:2012da}.

\subsubsection{BH Solutions}
\label{sec:ST-bh-solutions}

We now briefly describe BH solutions in scalar-tensor theories (see
Sec.~3.1 of~\cite{Berti:2015itd} for a recent review on this topic).
As in GR, no-hair theorem exists for stationary, isolated BHs
in vacuum in scalar-tensor theories with real, time-independent
scalars~\cite{hawking-no-hair,1971ApJ...166L..35T,
  1970CMaPh..19..276C,Heusler:1995qj,Sotiriou:2011dz}.  This is
because in the Einstein frame in vacuum, the scalar-tensor theory
reduces to GR with a minimally coupled scalar field.  Such an analysis
has recently been extended to stationary BHs with a single,
real, time-dependent scalar field in~\cite{Graham:2014ina}.  On the
other hand, Jacobson~\cite{Jacobson:1999vr} perturbatively showed that
a BH can acquire a scalar hair if both the metric and the scalar field
are time dependent.  Regarding BH solutions with a complex scalar
field, Pena and Sudarsky~\cite{Pena:1997cy} showed that a spherically
symmetric BH solution does not exist, but BH solutions with a
rotating configuration have been constructed numerically
in~\cite{Herdeiro:2014goa,Herdeiro:2015gia,Herdeiro:2015tia} 
(see~\cite{Herdeiro:2015waa} for a recent review).  
If a BH is surrounded by matter, Cardoso
\et~\cite{Cardoso:2013fwa,Cardoso:2013opa} showed that it can
spontaneously acquire a scalar hair due to a tachyon instability,
similar to spontaneous scalarization for neutron
stars~\cite{Damour:1993hw,Damour:1996ke}.

In Horndeski's theory, Hui and Nicolis~\cite{Hui:2012qt}
extended Refs.~\cite{hawking-no-hair,Sotiriou:2011dz} and claimed that if
the scalar field enjoys a shift symmetry, a static, spherically
symmetric, asymptotically flat BH in vacuum cannot have a scalar hair.
Sotiriou and Zhou~\cite{Sotiriou:2013qea,Sotiriou:2014pfa} pointed out
that an exceptional case exists in this claim, which is when a scalar
field is linearly coupled to the Gauss-Bonnet invariant density.  The static,
spherically symmetric, asymptotically flat BH solution in such a
theory were previously constructed
in~\cite{Mignemi:1992nt,Yunes:2011we}.  Babichev and Charmousis
extended Ref.~\cite{Babichev:2013cya} to Horndeski's theory and showed that
BHs can acquire a scalar hair with a time-dependent scalar field.
Linear perturbations of static, spherically symmetric BHs in
Horndeski's theory in both even and odd parity sectors and necessary
conditions for their stability were studied
in~\cite{Kobayashi:2012kh,Kobayashi:2014wsa}.

\subsubsection{Curent Constraints}
\label{sec:ST-curent-constraints}

We here review current constraints on scalar-tensor theories.  A
strong bound on such theories has been placed by the Cassini satellite
while on its way to Saturn~\cite{cassini}.
In the PN gauge, the metric suitable for the solar system is given
by~\cite{TEGP,Will:2014kxa}
\begin{equation}
ds^2 = -(1-2 \Phi )dt^2 + (1-2\gamma \Phi)\delta_{ij} dx^i dx^j + \mathcal{O}(\epsilon^{3/2})\,,
\end{equation}
where the dimensionless parameter $\epsilon$ is defined via
$\epsilon \equiv M_{\odot}/r$ with $r$ representing the typical length
scale and 
$\Phi = -M_{\odot}/r + \mathcal{O}(\epsilon^2)$
is the gravitational potential.  Cassini placed a
bound on $\gamma$ by measuring the time delay that occurs when the
light from the satellite is deflected as it passes near the Sun
(Shapiro time delay).  Cassini found that $|\gamma -1|$ must be within
$\sim 10^{-5}$, which corresponds to a bound on the massless BD theory
as $\omega_\BD > 4\times10^4$.  The bound on massive BD theory using
the Shapiro time delay measurement is derived in~\cite{peri,alsing}.

Another strong bound comes from binary pulsar observations.  Since the
dipolar radiation is suppressed for NS/NS binaries, stronger bounds
have been placed from the orbital decay rate measurement of NS/white dwarf (WD)
binaries, such as PSR J1141-6545~\cite{bhat}, PSR
J1738+0333~\cite{Freire:2012mg} and PSR J0348+0432~\cite{2.01NS}.  For
example PSR J1738+0333~\cite{Freire:2012mg} places a bound on massless
BD theory that is comparable to the solar system bound.  Such binary
pulsar constraints place stronger bounds on some parameter region in
quasi BD theory than the solar system one.  The bound
on the massive BD theory from such observations have been derived
in~\cite{alsing}.

\subsection{GW Tests}
\label{sec:ST-gw-tests}

We now explain future projected constraints on scalar-tensor theories
with GW observations.

\subsubsection{Massless BD Theory}
\label{sec:massless-bd-theory}

Let us first review how one can derive corrections to GWs emitted from
a compact binary in massless BD theory. Will~\cite{will1977} derived
the scalar field in a region far from the binary.  He found that the
perturbation to the scalar field $\tilde \vartheta$ from the background
 is given by
\begin{equation}
\tilde{\vartheta}=-\frac{4}{3+2\omega_\BD} \eta \frac{m}{r}\mathcal{S}\,\bm{n}\cdot\bm{v}
+ \mathcal{O} \left( \frac{m^2}{r^2} \right)\,,
\label{phi-tilde}
\end{equation}
where $r$ is a distance from the binary to the field point, $\bm{v}$
is the velocity of the binary constituents and $\bm{n}$ is the unit
vector from the binary to the field point. 
We recall that $\eta$ is the symmetric mass ratio and $m$ is the total mass of the binary.
$\mathcal{S}$ is given by
\begin{equation}
\mathcal{S} \equiv s_1 - s_2\,, \quad s_a\equiv -\left[ \frac{\partial(\ln m_a)}{\partial(\ln G)} \right]_\infty\,.
\end{equation}
Here, $s$ is the sensitivity which is related to the scalar charge and
the subscript $\infty$ shows that the quantity inside the square brackets is
to be evaluated at spatial infinity.  Notice that in BD theory, the
gravitational constant $G$ depends on the scalar field.  $s$ is
roughly given by the object's compactness, so $s \sim 0.2$ for NSs and
$s = 0.5$ for BHs.  Will and Zaglauer~\cite{zaglauer} then calculated
the energy flux emitted from a binary.  Keeping to leading order in
the BD correction in terms of PN order, such a flux is given by
\begin{equation}
\frac{dE_{\mathrm{GW}}}{dt}=-\frac{32}{5} \left\langle\frac{\eta^2 m^4}{r_{12}^4}v^2
                                            \left[ 1+\frac{5}{48}\mathcal{S}^2\bar{\omega} v^{-2}  + \mathcal{O} \left( v^2
                                            \right) \right]\right\rangle\,,
\label{power-rad-bd}
\end{equation}
where $r_{12}$ is the binary
separation, $\bar \omega \equiv 1/\omega_\BD$ and the angular brackets
represent the orbital average.
Based on the energy flux above, Will~\cite{will1994} 
derived the gravitational waveform of a compact binary in this
theory. The phase in Fourier space is given by that in the PPE waveform in Eq.~\eqref{eq:simplest-PPE}
with
\begin{equation}
\beta_\ppE = -\frac{5}{3584} \eta^{2/5} \mathcal{S}^2\bar{\omega}\,, \qquad b_\ppE = -\frac{7}{3}\,.
 \label{eq:bd-gw-phase}
\end{equation}
This BD correction in the waveform phase is of ``$-1$PN'' order relative
to the leading GR contribution.

Regarding higher PN corrections, Mirshekari and
Will~\cite{Mirshekari:2013vb} derived the equations of motion of a
compact binary in scalar-tensor theories with a single scalar field
and a vanishing potential up to 2.5PN order.  They showed that the
radiation reaction enters at 1.5PN order due to the scalar dipolar
radiation, which is 1PN order lower than that in GR due to quadrupolar
radiation.  For BH binaries, such equations of motion are exactly the
same as those in GR to 2.5PN order.  Yunes \et~\cite{Yunes:2011aa}
showed that the equations of motion for BH EMRIs are the same as those
in GR to all PN orders in the small mass ratio limit. This can be
interpreted as a generalization of the BH no-hair theorem in scalar-tensor
theories to a BH binary system.  The gravitational and scalar
radiation are calculated to 2PN~\cite{Lang:2013fna} and to
1.5PN~\cite{Lang:2014osa} orders respectively.  Then,
Lang~\cite{Lang:2014osa} derived the total energy flux carried away to
infinity to 1PN order.

We now review proposed constraints on massless BD theory from future
GW observations with NS/BH binaries.  The results are summarized in
Tables~\ref{table-bd-previous1} and~\ref{table-bd-previous2}.  The
bound on $\omega_\BD$ using GWs from compact binaries was first
studied by Eardley~\cite{eardley}.  
Will~\cite{will1994} derived the BD
correction to the gravitational waveform phase from compact binaries
as mentioned above and carried out a matches filtering analysis 
with Adv.~LIGO.  Scharre
and Will~\cite{scharre} and Will and Yunes~\cite{willyunes} performed
a similar analysis using LISA.  In particular, Will and Yunes improved
previous work by studying how the bounds on $\omega_\BD$ depend on
LISA's various noise sources, such as position and acceleration
noises.  They also studied how the bounds change as a function of the
LISA's arm lengths.  Berti \textit{et al.}~\cite{bertibuonanno}
calculated the measurement accuracy in $\omega_\BD$ with LISA, taking
the effect of spin-orbit coupling into account.  They also carried out
a Monte Carlo simulation, where they randomly distributed $10^4$
sources over the sky, and derived an averaged bound on $\omega_\BD$.
They found that LISA can place the proposed bound of
$\omega_\BD>10799$ on average with a 1yr observation of GWs emitted
from a (1.4+1000)M$_{\odot}$ BH/NS binary with an SNR of $\sqrt{200}$.

\begin{figure}[htb]
\begin{center}
\includegraphics[width=7.5cm,clip=true]{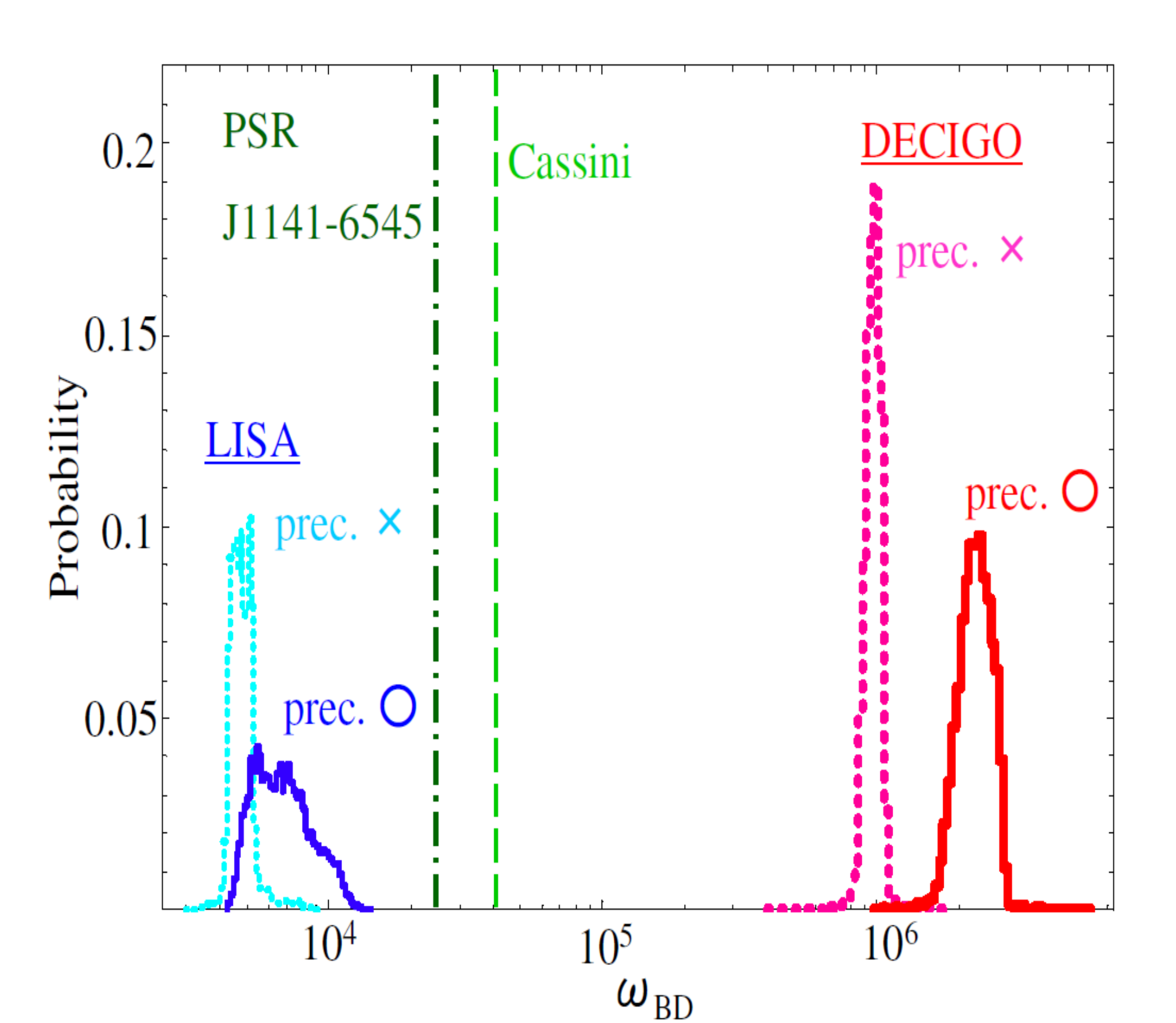}
\caption{\label{fig:BD} (Color online)
  Histograms showing probability distributions of the lower bound of
  $\omega_\BD$ obtained in~\cite{yagiLISA,yagiDECIGO}.  The authors
  carried out Monte Carlo simulations of $10^4$ BH/BH binaries for a
  circular orbit with and without spin precessions in massless BD
  theory.  The mass and distance of binaries are chosen as
  $(1.4+10^3)M_{\odot}$ for LISA and $(1.4+10)M_{\odot}$ for
  DECIGO/BBO, both with an SNR of $\sqrt{200}$.  The light blue
  (magenta) dotted and blue (red) solid histograms
  show the probability distribution using LISA (DECIGO/BBO) without and
  with precessions respectively.  The dashed and dotted-dashed vertical
  lines are the bounds from the Cassini satellite~\cite{cassini} and
  PSR J1141-6545~\cite{bhat} respectively.  Observe that DECIGO/BBO
  places a stronger constraint than the current bounds by more than an
  order of magnitude.  Observe also that the effect of spin precession
  is not so significant in constraining massless BD theory. 
  This figure is taken from~\cite{yagiLISA}. }
\end{center}
\end{figure}

Yagi and Tanaka~\cite{yagiLISA} extended previous analyses by carrying
out a Monte Carlo simulation to estimate proposed bounds on massless BD
theory using LISA, taking both the spin precession and eccentricity of
a binary system into account.  They found that the inclusion of
eccentricity weakens the bound by 4--5 times than the one without
including eccentricity. This is because both $\omega_\BD$ and
eccentricity enter at a negative PN order in the gravitational
waveform phase relative to the leading GR term, 
which results in a strong correlation between these
two parameters. One can break the degeneracy by imposing prior
information on eccentricity.  Figure~\ref{fig:BD} presents the
probability distribution of the lower bound on $\omega_\BD$, assuming
LISA detects GW signals from (1.4+1000)M$_{\odot}$ BH/NS binaries with
a circular orbit and an SNR of $\sqrt{200}$. Observe that the bound
is weaker than the current solar system bound.
In~\cite{yagiDECIGO}, Yagi and Tanaka performed a similar analysis
using DECIGO/BBO.  The results are also shown in Fig.~\ref{fig:BD},
where the authors assumed that DECIGO/BBO detects GW signals from
(1.4+10)M$_{\odot}$ BH/NS binaries with a circular orbit and an SNR of
$\sqrt{200}$. Observe that the proposed constraints using DECIGO/BBO
are more than one order of magnitude stronger than the current bound.
DECIGO/BBO performs better than LISA for three reasons. First, the
number of GW cycles is larger since the GW frequency is higher, which
allows one to perform more accurate measurement on parameters for a
fixed SNR. Second, the velocity of binary constituents at 1yr before
coalescence is smaller, which makes the ``$-1$PN'' dipolar radiation
effect larger. Third, the effective frequency range is larger, which
again allows one to measure parameters more accurately. DECIGO/BBO can
place even stronger constraints by taking the advantage of the fact that
the expected detection rate of NS/BH binaries is $\sim 10^4$.  Such a
large detection rate makes the constraint on $\omega_\BD$ to be
$\omega_\BD > 3.77 \times 10^8$, which is indeed four orders of
magnitude stronger than the solar system bound!

Recently, Arun and Pai~\cite{Arun:2013bp} and Yagi~\cite{Yagi:2013du}
extended~\cite{bertibuonanno} by calculating the upper bound on
$\omega_\BD$ using detectors other than LISA or DECIGO/BBO.
Reference~\cite{Arun:2013bp} found that the bound with
second-generation GW interferometers, such as Adv.~LIGO, is 400 times
weaker than the current bound, while the bound with third generation
interferometers, such as ET, can be slightly larger than the current
one.  Reference~\cite{Yagi:2013du} found that the bound with ASTROD-GW
is comparable to that with LISA.  The bound with eLISA is also
comparable to the LISA one~\cite{Arun:2013bp,Yagi:2013du}.

\fulltable{\label{table-bd-previous1} Summary of previous work on
  probing BD theory with GWs from NS/BH binaries.  The second column
  represents the maximum PN order that is taken into account.  The
  third, fourth and fifth columns present whether the spin-orbital
  coupling $\beta$, spin precession and eccentricity are included
  respectively.  The sixth column shows whether each work considers a
  detection from multiple sources.  The seventh column describes a
  type of analyses being performed (either the pattern-averaged (PA)
  or Monte Carlo (MC).)
 This table is taken from~\cite{Yagi:2013du}.
}
 \hline\hline
 Reference & PN & $\beta$ &  prec. & ecc. & multi. & analy. 
 \\ \hline
Will (1994)~\cite{will1994} &1.5  & $\times$ & $\times$ & $\times$ & $\times$ & PA  
\\
Damour \& Esposito-Far\`ese (1998)~\cite{damour-GW-ST} & 1.5  & $\times$ & $\times$ & $\times$ & $\times$ & PA 
\\
Scharre \& Will (2002)~\cite{scharre} & 1.5   & $\times$ & $\times$ & $\times$ & $\times$ & PA 
\\
Will \& Yunes (2004)~\cite{willyunes} & 1.5  & $\times$ & $\times$ & $\times$ & $\times$ & PA  
\\
Berti \et~(2005)~\cite{bertibuonanno} & 2 & $\checkmark$  & $\times$ & $\times$ & $\times$ & MC 
\\
Yagi \& Tanaka (2010)~\cite{yagiLISA} & 2 & $\checkmark$ & $\checkmark$ & $\checkmark$  & $\times$ & MC 
\\
Yagi \& Tanaka (2011)~\cite{yagiDECIGO} & 2 & $\checkmark$ & $\checkmark$ & $\checkmark$ & $\checkmark$ & MC  
\\
Arun \& Pai~(2013)~\cite{Arun:2013bp} & 2 & $\checkmark$  & $\times$ & $\times$ & $\times$ & PA 
\\
Yagi (2013)~\cite{Yagi:2013du} & 2 & $\checkmark$ & $\times$ & $\times$ & $\times$ & PA  
\\ \hline\hline
\endfulltable


\fulltable{\label{table-bd-previous2}
Summary of the proposed bounds on $\omega_\BD$ with various GW interferometers.
Numbers in brackets represent the total mass of a binary in a unit of $M_\odot$.
 This table is taken from~\cite{Yagi:2013du}.}
\hline\hline
 Reference & adv.~LIGO & ET &  LISA& eLISA  & DECIGO & ASTROD  \\ \hline
Will (1994)~\cite{will1994} & $2\times 10^3$ (3.7) & &  & & &    \\
Scharre \& Will (2002)~\cite{scharre} &  &  &   $2\times 10^5$ ($10^3$)   & & &  \\
Will \& Yunes (2004)~\cite{willyunes} & &  &   $2\times 10^5$ ($10^3$)   & & & \\
Berti \et~(2005)~\cite{bertibuonanno} & &  &   $10^4$ ($10^3$)   & & & \\
Yagi \& Tanaka (2010)~\cite{yagiLISA} & &  &   $7\times 10^3$ ($10^3$)   & & & \\
Yagi \& Tanaka (2011)~\cite{yagiDECIGO} & &  &  & & $4\times 10^8$ (11.4)  &   \\
Arun \& Pai~(2013)~\cite{Arun:2013bp} & $10^2$ (6.4) & $10^5$ (6.4)  &  & $10^4$ (400)  & &   \\
Yagi  (2013)~\cite{Yagi:2013du} & &  & $8\times 10^3$ ($10^3$) & \ $9\times 10^3$ ($10^3$) & &  $8\times 10^3$ ($10^3$)  \\ \hline\hline
\endfulltable

\subsubsection{Massive BD Theory}
\label{sec:massive-bd-theory}

Regarding proposed GW constraints on massive scalar-tensor theories,
Berti \et~\cite{Berti:2012bp} derived corrections to the gravitational
waveform emitted from compact binaries in massive BD theory.  They
showed that the quadrupolar and dipolar scalar radiation is produced
when $f >\bar m_s/(2 \pi)$ and $f >\bar m_s/(\pi)$ respectively, where
$\bar m_s \equiv m_s/\hbar$ is the reduced mass of the scalar field.
The correction to the gravitational waveform phase is given
by~\cite{Berti:2012bp}
\begin{eqnarray}
\delta \Psi (f) &=& \Psi_{0\mrm{PN}}(f) \left[ \zeta + \xi \Gamma^2 \nu \left( \frac{5}{462} x^{-3}
- \frac{1}{1632} \nu x^{-6} \right) \Theta (2 \pi f - \bar m_s)  \right. \nn \\
& & \left. - \xi \mathcal{S}^2 \left( \frac{5}{84} x^{-1}
- \frac{25}{1248} \nu x^{-4} \right) \Theta (\pi f - \bar m_s) \right]\,,
\end{eqnarray}
where $\nu \equiv \bar m_s^2 m^2$, $\Theta$ is the Heaviside function and
\begin{eqnarray}
\xi & \equiv & \frac{1}{2 + \omega_\BD}\,, \\
\Gamma &=& 1 - 2 \frac{s_1 m_1 + s_2 m_2}{m}\,, \\
\zeta &=& \frac{2}{3} \xi (s_1 + s_2 - 2 s_1 s_2) + \frac{1}{2} \xi
- \frac{1}{12} \xi \Gamma^2 \Theta(2 \pi f - \bar m_s)\,.
\end{eqnarray}
They then carried out a Fisher analysis and derived proposed bounds on
the BD parameter $\omega_\BD$ and $\bar m_s$ in massive BD theory with
Adv.~LIGO, ET and eLISA.  They assumed such interferometers detect GW
signals from a NS/BH binary and neglected the effect of BH spins. 
In order to have dipolar radiation present within e.g.~Adv.~LIGO 
frequency band, the scalar field mass needs to be smaller than $\mathcal{O}(10^{-13})$eV.
The authors
found that space-borne GW interferometers place more stringent
constraints than the ground-based ones because the target frequency is
lower for the former and GW frequencies produced from binaries in
massive BD theory is proportional to $\bar m_s$.  An eLISA measurement
of GWs with an SNR of 10 can place constraints on the theory that are
stronger than current bounds from Cassini and Lunar Laser Ranging
experiments and binary pulsar observations.

An interesting fact regarding binary evolution in massive
scalar-tensor theories was revealed in~\cite{Cardoso:2011xi}.  In GR,
binary separation shrinks due to gravitational radiation.  On the
other hand, when a compact object (other than a black hole) orbits around a rotating 
super-massive black hole (SMBH) in
massive scalar-tensor theories, orbits can float at a same separation
 (\emph{floating orbits})
due to superradiance.  Namely, if the central
BH is rotating sufficiently fast, an orbiting body excites
superradiant modes of the scalar field, which can fuel the orbiting
body energy at the same rate as that radiated to infinity (and to the
horizon of the central BH) due to gravitational radiation. If the
central BH is orbiting slowly, the scalar radiation due to
superradiance that goes into the horizon becomes positive, which
enhances the inspiral rate of the system (\emph{sinking orbits}).

Yunes \et~\cite{Yunes:2011aa} constructed gravitational waveform in
massive BD theory for binaries with both floating and sinking
orbits. Regrading the former, they found that such an effect can
produce a large dephasing when carrying out a matched filtering
analysis using the GR inspiral template. This means that if a GW
signal from an EMRI consisting of a SMBH and a NS
 is detected with LISA, one can place a stringent
constraint on $\omega_\BD$ for scalar field masses that produce
supperradiant resonances at frequencies that can be detected with
LISA. Moreover, the timescale for an orbiting body to go through the
``floating'' resonance becomes larger than the Hubble timescale for
binaries with frequencies that are lower than the lowest frequency
limit of LISA.  This means that once the EMRI signal is detected, one
can also rule out the parameter region that produces floating orbit at
a frequency \emph{lower} than the detected one.  Based on these facts,
the authors in~\cite{Yunes:2011aa} found that LISA can place
constraints that are more than 10 orders of magnitude stronger than
the current bound on the theory!  
Although this analysis does not consider systematic errors, it is 
unlikely that systematics due to e.g.~astrophysical matter around 
EMRIs and uncertainties in the waveform modeling compensate 
such a large effect due to floating orbits.
Regarding GW signals from
binaries with a sinking orbit, they found that such effect is small
and one cannot place meaningful constraints with LISA from such
systems.

\subsubsection{Quasi BD Theory}
\label{sec:damo-espos-farese}

Gravitational and scalar radiation from compact binaries in quasi BD theory 
was derived in~\cite{damour-GW-ST}. In this
theory, stars can spontaneously scalarize once their binding energy
exceeds the threshold~\cite{Damour:1993hw,Damour:1996ke}
(\emph{spontaneous scalarization}).  Such a phenomenon can be
understood in analogy with second-order phase transitions, such as
ferromagnetism.  Such scalarization allows one to place stringent
constraints on the theory from binary pulsar
observations~\cite{bhat,Freire:2012mg,2.01NS,Wex:2014nva,Berti:2015itd}.
Similar scalarization~\cite{Barausse:2012da,
  Palenzuela:2013hsa,Shibata:2013pra,Taniguchi:2014fqa,Sampson:2014qqa}
occurs in a compact binary when the separation shrinks and the binding
energy of the binary exceeds the threshold (\emph{dynamical
  scalarization}), or when a star acquires a scalar charge and
scalarizes its companion (\emph{induced scalarization}).

Regarding the evolution of BH binaries in quasi BD theory, Healy
\et~\cite{Healy:2011ef} performed a numerical relativity simulation of
late inspiral and merger of BH binaries in an inhomogeneous scalar
field. In particular, they assumed that binaries are inside a scalar
bubble. They calculated the scalar dipole radiation and found that the
GWs are indistinguishable from those in GR unless an external
mechanism induces dynamics in the scalar field. Alternatively, if one
allows the scalar field dynamics, the scalar field bubble can collapse
and accretes onto BHs, which makes the BH mass larger and changes the
binary evolution and GWs from those in GR.

\subsubsection{Generic Scalar-tensor Theories}

Regarding generic scalar-tensor theories, Berti
\et~\cite{Berti:2013gfa} carried out a different simulation of a
BH binary evolution by relaxing the asymptotic flatness condition
on the metric, which is motivated from
cosmology~\cite{Sahni:1999qe,Hu:2000ke}. They studied a non-spinning
BH binary inspiral with a circular orbit in a constant scalar field
gradient.  Such a background scalar field leads to scalar dipole
radiation from the binary, which can be understood from the miracle
hair growth of a BH scalar hair found by
Jacobson~\cite{Jacobson:1999vr} under a time-dependent situation.
Unfortunately, such scalar radiation is too small to be detected in
the near future for realistic values of the scalar field gradient.

\subsection{EMW Tests}
\label{sec:ST-electr-wave-tests}

\subsubsection{Quasi BD Theory}

Regarding EMW tests of scalar-tensor theories with a BH, Liu
\et~\cite{Liu:2014uka} recently derived a proposed bounds on
quasi BD theory from a BH/pulsar binary.  They assumed
that future radio telescopes, such as Five-hundred-metric Aperture
Spherical radio Telescope (FAST)~\cite{Nan:2011um} and the Square
Kilometre Array (SKA)~\cite{Carilli:2004nx}, will detect signals from a
pulsar orbiting a stellar-mass BH.  They carried out a mock data
simulations and estimated the measurement accuracy of post-Keplerian
parameters. In particular, they applied the measurement accuracy of
the orbital decay rate to constrain quasi BD
theory. They assumed a system with the mass $(1.4,10)M_\odot$, the
orbital period $P_b= 5$days, the eccentricity $e = 0.8$ and the BH
spin $\chi =0$.  They found that a 5yr observation with FAST or SKA
can place constraints that are stronger than the current bound on the
coupling parameters $\alpha_0$ and $\beta_0$ in the theory.  For
example, the bound on $\alpha_0$ (which is related to the BD
parameter) can be increased by more than one order of magnitude.  Such
a bound from BH/pulsar binaries can even be stronger than the proposed
bounds from \emph{future} solar system experiments, such as GAIA, in a
certain parameter space of the theory.

\subsubsection{Generic Scalar-tensor Theories}

Although a stationary BH in scalar-tensor theories does not possess a
scalar hair~\cite{Graham:2014ina}, a non-stationary BH can acquire
such a hair. For example, Jacobson~\cite{Jacobson:1999vr} showed this
perturbatively, where he assumed that the timescale in which the
scalar field varies is much larger than $ G M/c^3$ with $M$
representing the BH mass, and kept to the leading order in this small
timescale. Such a treatment reduces to considering a time-dependent
scalar field under the Kerr background. Jacobson added a term $\bar \mu \, t$
to a static scalar field under the Kerr background with $\bar \mu$
representing a constant.  Such a term may arise from cosmological
evolution or a BH slowly moving in an asymptotic spatial gradient in
the scalar field.  Imposing regularity at the horizon, he found that
the asymptotic behavior of the scalar field around spatial infinity is
given by
\begin{equation}
\vartheta = \vartheta_0 + \bar \mu t - \frac{2 \bar \mu M^2 (1 + \sqrt{1-\chi^2})}{r} + \mathcal{O}\left( \frac{M^2}{r^2} \right)\,,
\end{equation}
where $\vartheta_0$ is a constant that represents the scalar field at
spatial infinity while $\chi \equiv J/M^2$ with $J$
representing the magnitude of the spin angular momentum of the BH. The
coefficient of the $1/r$ term in the above equation corresponds to the
scalar hair. Notice that such a scalar field vanishes when $\bar \mu =0$,
which clearly shows that such a scalar hair arises due to the time
dependence of the scalar field. 
A related solution was found in shift-symmetric 
Horndeski theories in~\cite{Babichev:2013cya}.
Horbatsch and
Burgess~\cite{Horbatsch:2011ye} derived a constraint on $\bar \mu$ from
the orbital decay measurement of a SMBH binary candidate OJ
287~\cite{2008Natur.452..851V}.  Such a system shows a periodic burst
that occurs when the smaller BH passes through the accretion disk of
the larger one.  They derived the dipolar radiation flux due to the
emergent scalar hair above and found a bound as
$|\bar \mu^{-1}| > 1.4 \times 10^6$~s.

\section{Massive Gravity Theories}
\label{sec:mass-grav-theor}

\subsection{Basics}
\label{sec:MG-basics}

\subsubsection{Theories}
\label{sec:MG-theories}

Massive gravity theories are a simple extension of GR, where one
introduces a finite mass to the graviton.  Many different kinds of
massive gravity theories exist (see e.g.\ Refs.~\cite{rubakov,
  Hinterbichler:2011tt, deRham:2014zqa} for recent reviews).
Originally, Fierz and Pauli~\cite{fierz} proposed a Lorentz-invariant
massive gravity model by introducing a mass term of the graviton in
the Einstein-Hilbert action.  However, such a theory shows a
pathological feature that the theory does not approach GR at 
linear order in perturbation in the massless limit.  This is known
as vDVZ discontinuity~\cite{vdv,z} and
originates from the fact that the helicity-0 component of the graviton
does not decouple from matter.  Although such a discontinuity seems to
contradict solar system experiments, Vainshtein showed that one
recovers GR in the massless limit by including the nonlinear
contribution~\cite{vainshtein}.
He pointed out that the linear approximation already breaks down at a
distance much longer than the Schwarzschild radius (the so-called
Vainshtein radius) in massive gravity.
References~\cite{dvali,nicolis} showed that such a mechanism indeed
works in the DGP braneworld model.  Although the Fierz-Pauli theory
is free of ghost modes in a flat background, it suffers from the
Boulware-Deser ghost in a curved background due to the presence of a
helicity-0 mode of the graviton~\cite{boulware}.  Such a pathology
occurs irrespective of how one generalizes the Fierz-Pauli theory to a
curved background~\cite{Creminelli:2005qk}.

Regarding other types of massive gravity theories,
Rubakov~\cite{rubakov2} and Dubovsky~\cite{dubovsky} proposed a
Lorentz violating massive gravity theory that evades pathologies
related to the Bouleware-Deser instability.  Although the helicity-0
mode is present in this theory, such a mode is screened thanks to the
Vainshtein mechanism.  Such a Lorentz-violating massive gravity theory
has recently been extended so that the theory becomes UV complete by
introducing vector fields~\cite{Blas:2014ira}.  Chamseddine and
Mukhanov proposed a new massive gravity model inspired by the Higgs
mechanism~\cite{chamseddine}.  They introduced four scalar fields with
a global Lorentz symmetry and when such a symmetry is spontaneously
broken, the graviton absorbs scalar degrees of freedom and acquires a
finite mass.  Although Vainshtein mechanism seems to work also in this
theory~\cite{alberte}, the existence of ghost modes cannot be
avoided~\cite{chamseddine2}.

Recently, de Rham \et~\cite{derham1,derham2} generalized the
Fierz-Pauli theory in a nonlinear way and proposed \textit{nonlinear
  massive gravity} (or \emph{ghost-free massive gravity}) under a flat
reference metric.  Reference~\cite{derham2} showed that the
Hamiltonian constraint, which is necessary to kill the Boulware-Deser
ghost modes, exists up to fourth order in nonlinearities in the
unitary gauge.  Hassan and Rosen~\cite{hassan-generic} generalized the
theory to a generic reference metric.  The Hamiltonian constraint was
shown to exist in all orders in the unitary gauge with (i)
flat~\cite{hassan-flat}, (ii) generic but
non-dynamical~\cite{hassan-curv-nondyn}, and (iii) generic and
dynamical (bi-gravity)~\cite{hassan-bi} reference
metrics.  Furthermore,
Refs.~\cite{hassan-secondary,derham-stuckel,derham-helicity} showed
the existence of the secondary constraint in both nonlinear massive
and bimetric gravities.  This concludes that the Boulware-Deser ghost
is absent to all orders in nonlinearities in these massive
gravity theories.  The extension of nonlinear massive gravity includes (i)
mass-varying~\cite{Huang:2012pe,Huang:2013mha} and (ii) quasi-dilaton
massive gravity~\cite{DeFelice:2013tsa,DeFelice:2013dua}.  The former
promotes the mass as a function of an external scalar field while the
latter rescales the reference metric globally using the quasi-dilation
scalar field so that the theory acquires a global rescaling symmetry.
The absence of the Boulware-Deser ghosts in these extended theories
were shown in~\cite{Huang:2012pe,Huang:2013mha,Mukohyama:2013raa}.

Regarding BH solutions in nonlinear massive gravity and bi-gravity, the Kerr spacetime 
is also a solution to the modified field equations in
some of these theories\footnote{%
In nonlinear massive gravity, the two metrics cannot both be in a diagonal form~
\cite{Deffayet:2011rh}.
For example, one can use the Eddington-Finkelstein coordinates instead of
the Schwarzschild coordinates for finding a static and spherically symmetric solution. 
}~\cite{Babichev:2014tfa}.  However, the
Schwarzschild BH suffers from an instability against monopole
fluctuations if the graviton mass is smaller than a
threshold~\cite{Babichev:2013una,Brito:2013wya,Brito:2013yxa}.  On the
other hand, the Kerr BH (with two metrics differing by an overall constant)
suffers from superradiant
instabilities, whose time scale is orders of magnitude shorter
than that for the spin-0 and spin-1 fields~\cite{Brito:2013wya}.  These studies led the authors
in~\cite{Brito:2013xaa} to find numerically a new non-rotating,
asymptotically flat BH solution in the theory.
We refer the readers to recent reviews~\cite{Volkov:2013roa,
Tasinato:2013rza,Volkov:2014ooa,Babichev:2015xha} for more details on 
BH solutions and their stability analyses in these theories.

\subsubsection{Current Constraints}
\label{sec:MG-current-constraints}

A robust constraint on the mass of the graviton is obtained from solar
system experiments.  In massive gravity theories, the gravitational
potential is given by a Yukawa-type form.  Then, the acceleration of a
planet in massive gravity theories relative to that in GR is given by
\begin{equation}
\frac{a_\MG}{a_\GR} = 1-\frac{1}{2} \left( \frac{r}{\lambda_g} \right)^2
+ \mathcal{O}\left[ \left( \frac{r}{\lambda_g} \right)^3 \right]\,,
\end{equation}
where $a_\MG$ and $a_\GR$ represent the massive gravity and GR
contribution to the acceleration and $\lambda_g \equiv h/(m_g c)$ is
the graviton Compton length.  The observation of the Kepler's third
law of Mars places a constraint as~\cite{talmadge}
\begin{equation}
\lambda_g > 2.8 \times 10^{17} \mathrm{cm}\,.
\end{equation}

Other bounds on the mass of the graviton exist that are stronger than
the solar system one above but are less robust due to theoretical
uncertainties.  Such bounds include those from bound clusters and
tidal interactions between galaxies~\cite{goldhaber} and weak
lensing~\cite{choudhury}.  Reference~\cite{sjors} derived a new bound
on the mass of the graviton in nonlinear massive gravity from galactic
lensing and velocity dispersion.

One can also place a bound from binary pulsar observations.  For
example, Ref.~\cite{finnsutton} calculated modifications to the energy
flux emitted from a binary in a Fierz-Pauli type massive gravity and
obtained a bound using orbital decay rate observations of PSR B1913+16
and PSR B1534+12.  Such a constraint turned out to be two orders of
magnitude weaker than the solar system bound.  
Reference~\cite{Jimenez:2015bwa} derived the bound on the propagation speed of the graviton
in (beyond) Horndeski theories. If one assumes that such a bound is also applicable
to constrain the graviton mass using Eq.~\eqref{vg} below, one finds 
$\lambda_g \gtrsim 3 \times 10^{15}$cm. Such a bound is comparable to 
the one in~\cite{finnsutton}.
These current
constraints are summarized in
Table~\ref{table-mg-current}.

\fulltable{\label{table-mg-current}
Current bounds on $\lambda_g$~\cite{bertisesana}. 
$h_0 = H_0/(100 \mrm{km/s/Mpc})$
with $H_0$ representing the current Hubble constant.
}
\hline\hline
& Solar system & Clusters & Weak lensing & Galaxies & Binary pulsars\\ \hline
$\lambda_g (\mrm{cm})$ & $2.8\times 10^{17}$~\cite{talmadge} & $6.2\times 10^{24} h_0$~\cite{goldhaber}  & $1.8\times 10^{27}$~\cite{choudhury} & $10^{26}$~\cite{sjors} & $1.6\times 10^{15}$~\cite{finnsutton} \\ \hline\hline
\endfulltable

\subsection{GW Tests}
\label{sec:MG-gw-tests}

\subsubsection{GW Signals Alone}
\label{sec:gw-signals-alone}

\fulltable{\label{table-mg-previous1} Summary of previous work on
  constraining massive gravity theories with GW observations alone.
  The second column shows the maximum PN order considered.  (Berti
  \et~(2011) is left blank because they do not specify this
  information in the paper.)  The third, fourth, fifth and sixth
  columns indicate whether they include the effect of the spin-orbit
  coupling $\beta$, higher harmonics (HH), precession and eccentricity
  respectively.  The seventh column presents whether they consider the
  inspiral phase only or all of the inspiral, merger and ringdown
  (IMR) phases in the evolution of a binary.  The eighth column shows
  whether they assume a detection of GW signals from multiple sources.
  The ninth column describes whether they carry out a
  model-independent (MI) calculation.  The last column represents a
  type of analyses they performed (the pattern-averaged (PA), Monte
  Carlo (MC) or Bayesian (B).)  This table is taken
  from~\cite{Yagi:2013du}.  }
\hline\hline
 Reference & PN & $\beta$ & HH & prec. & ecc. & IMR & multi. & MI & analy. \\ \hline\hline
Will (1998)~\cite{will1998} &1.5 & $\times$ & $\times$ & $\times$ & $\times$ & $\times$ & $\times$ & $\times$ & PA   \\
Will \& Yunes (2004)~\cite{willyunes} & 1.5 & $\times$ & $\times$ & $\times$ & $\times$ & $\times$ & $\times$ & $\times$ & PA  \\
Berti \et~(2005)~\cite{bertibuonanno} & 2 & $\checkmark$ & $\times$ & $\times$ & $\times$ & $\times$ & $\times$ & $\times$ & MC   \\
Arun \& Will (2009)~\cite{arunwill} & 3.5 & $\checkmark$ & $\checkmark$ & $\times$ & $\times$ & $\times$ & $\times$ & $\times$ & PA   \\ 
Stavridis \& Will (2009)~\cite{stavridis} & 2 & $\checkmark$ & $\times$ & $\checkmark$ & $\times$ & $\times$ & $\times$ & $\times$ & MC   \\
Yagi \& Tanaka (2010)~\cite{yagiLISA} & 2 & $\checkmark$ & $\times$ & $\checkmark$ & $\checkmark$ & $\times$ & $\times$ & $\times$ & MC   \\
Yagi \& Tanaka (2011)~\cite{yagiDECIGO} & 2 & $\checkmark$ & $\times$ & $\checkmark$ & $\checkmark$ & $\times$ & $\times$ & $\times$ & MC   \\
Keppel \& Ajith (2010)~\cite{keppel} & 3.5 & $\times$ & $\times$ & $\times$ & $\times$ & $\checkmark$ & $\times$ & $\times$ & PA   \\
Del Pozzo \et~(2011)~\cite{delpozzo} & 2 & $\times$ & $\times$ & $\times$ & $\times$ & $\times$ & $\checkmark$ & $\times$ & B   \\
Cornish \et~(2011)~\cite{cornish-PPE} & 3.5 & $\times$ & $\times$ & $\times$ & $\times$ & $\times$ & $\times$ & $\checkmark$ & B   \\
Berti \et~(2011)~\cite{bertisesana} &  & $\times$ & $\times$ & $\times$ & $\times$ & $\times$ & $\checkmark$ & $\times$ & MC   \\
Huwyler \et~(2011)~\cite{Huwyler:2011iq} & 2 & $\checkmark$ & $\checkmark$ & $\checkmark$ & $\times$ & $\times$ & $\times$ & $\checkmark$ & MC   \\
Arun \& Pai (2013)~\cite{Arun:2013bp} & 3.5 & $\times$ & $\checkmark$ & $\times$ & $\times$ & $\times$ & $\times$ & $\times$ & PA   \\ 
Yagi (2013)~\cite{Yagi:2013du} & 2 & $\times$ & $\times$ & $\times$ & $\times$ & $\times$ & $\times$ & $\times$ & PA   \\ \hline \hline
\endfulltable

\fulltable{\label{table-mg-previous2}
  Summary of the proposed bounds on $\lambda_g$ with various GW
  interferometers.  The bounds are normalized by $10^{18}$cm for
  adv.~LIGO and ET, while $10^{21}$cm for LISA, eLISA, DECIGO and
  ASTROD-GW.  Numbers in brackets represent the total mass of a binary
  in a unit of $M_\odot$.  This table is taken
  from~\cite{Yagi:2013du}.  }
\hline\hline
 Reference & adv.~LIGO & ET & LISA & eLISA & DECIGO  & ASTROD \\ \hline\hline
Will (1998)~\cite{will1998} &0.6 (20)  &  & 7 ($2\times 10^7$) &  & &   \\
Will \& Yunes (2004)~\cite{willyunes} &   &  & 5 ($10^7$) &  & &  \\
Berti \et~(2005)~\cite{bertibuonanno} &  &   & 1 ($2\times 10^6$) &  & &    \\
Arun \& Will (2009)~\cite{arunwill} & 0.7 (60) & 10 (400) & 5 ($2\times 10^6$)  &  & &   \\ 
Stavridis \& Will (2009)~\cite{stavridis} &   &  & 7 ($2\times 10^7$) &  & &    \\
Yagi \& Tanaka (2010)~\cite{yagiLISA} &   &  & 3 ($1.1\times 10^7$) & & &   \\
Yagi \& Tanaka (2011)~\cite{yagiDECIGO} &  &  &  &  & 0.3 ($1.1\times 10^6$)  &  \\
Keppel \& Ajith (2010)~\cite{keppel} & 8 (360) & 70 (3000) & \ 60 ($4.8\times 10^7$) &  & &   \\
Del Pozzo \et~(2011)~\cite{delpozzo} & 0.5--2.5  &  &  &  & &   \\
Cornish \et~(2011)~\cite{cornish-PPE} & 0.9 (18--24)   &  & 4 ($4-5\times 10^6$) &  & &    \\
Berti \et~(2011)~\cite{bertisesana} &  &  & 6.5--7.5 & 3--5 & &    \\
Huwyler \et~(2011)~\cite{Huwyler:2011iq} &   &  & 7 ($1.3\times 10^7$) &  & &    \\
Arun \& Pai~(2013)~\cite{Arun:2013bp} &  &  &  &  0.1 & &   \\
Yagi  (2013)~\cite{Yagi:2013du} &  && 4 ($1.1\times 10^7$) &  1.3 ($1.1\times 10^7$)  & &  6 ($1.1\times 10^7$)  \\ \hline\hline
\endfulltable

In this subsection, we review proposed bounds with future GW
observations.  When the mass of the graviton is non-vanishing, the
group velocity of GWs is given by~\cite{will1998}
\begin{equation}
v_{\mathrm{g}}^2=1-\frac{1}{f^2\lambda_g^2}\,.
\label{vg}
\end{equation}
Such a deviation from the speed of light modifies the gravitational
waveform phase which can be mapped to that in the PPE waveform as~\cite{will1998}
\be
\label{eq:phase-mg2}
\beta_\ppE = - \frac{\pi^2 D \mc}{ \lambda_g^2 (1+z)}\,, \qquad b_\ppE = -1\,.
\ee
Here, $z$ represents the source redshift and $D$ corresponds to a
distance parameter defined in~\cite{will1998,bertibuonanno}. Notice
that the above correction to the phase is of 1PN
relative to the leading term in GR.
If the late-time acceleration of the Universe was due to some modifications
of GR that gives a mass to the graviton, one needs to require $\lambda_g$
to be the Hubble scale, namely $\lambda_g \sim 10^{28}$cm (or 
the graviton mass to be $\sim 10^{-33}$eV). We will see below that 
projected bounds from future GW observations with BHs are weaker than this.

Let us now briefly explain previous work on deriving proposed bounds
with GW observations of BH/BH binaries alone.  We summarize the
difference in computational methods and assumptions among each work in
Table~\ref{table-mg-previous1}, while Table~\ref{table-mg-previous2}
shows a summary of proposed constraints with various GW interferometers
with different binary systems.
Will~\cite{will1998} derived a bound with Adv.~LIGO and LISA by
carrying out a Fisher analysis, where he included $\lambda_g$ into a
parameter set that needs to be fit against the signal using the
matched filtering analysis.  Will and Yunes~\cite{willyunes} carried
out a similar analysis using the improved noise curve for LISA with a
pattern-averaged waveform for non-spinning binaries.  They studied how
the constraint depends on the LISA position and acceleration noises and
arm lengths.  Berti \textit{et al.}~\cite{bertibuonanno} performed
Monte Carlo simulations for spinning binaries, where the authors
included the spin-orbit coupling $\beta$ into the parameter set.  They
found a constraint $\lambda_g>1.33\times 10^{21}$cm, assuming that
LISA detects GW signals from a ($10^6+10^6$)M$_{\odot}$ BH/BH binary
at 3Gpc with an observational period of 1yr.  Observe that such a bound
is roughly four orders of magnitude stronger than the solar system
one.

Arun and Will~\cite{arunwill} and Arun and Pai~\cite{Arun:2013bp} took
the effect of higher harmonics in the waveform into account and
obtained a stronger bound for high-mass binaries than that obtained by
using the waveform with only the dominant harmonic.
Stavridis and Will~\cite{stavridis} included the effect of spin
precession by solving the precession equations numerically.  They found
that LISA can place $\lambda_g > 5\times 10^{21}$cm by observing GW
signals from a precessing binary with masses of $(10^6+10^7)M_\odot$
at 3Gpc.  Interestingly, such a constraint is comparable to that
obtained for non-spinning binaries.  This suggests that the precession
breaks the degeneracy between spins and other parameters including
$\lambda_g$.  Huwyler \et~\cite{Huwyler:2011iq} included both the spin
precession and higher harmonics effects.  Their bound on $\lambda_g$
is comparable to that in~\cite{stavridis}.  Such a fact suggests that
the effect of spin precession is more important than that of higher
harmonics when breaking the degeneracy between $\lambda_g$ and other
parameters.

\begin{figure}[htb]
\begin{center}
\includegraphics[width=7.5cm,clip=true]{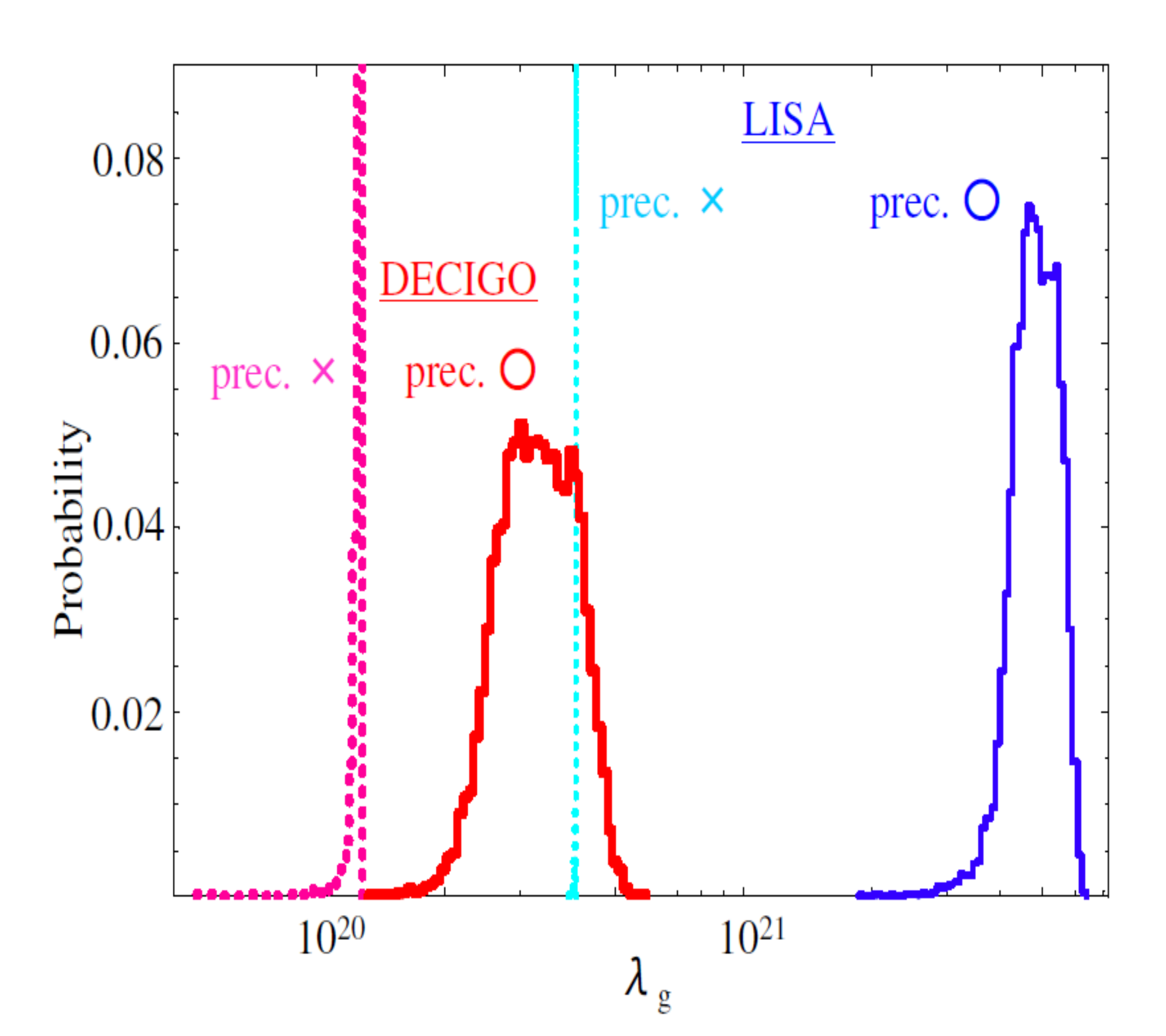}
\caption{\label{fig:massive} (Color online)\label{massive}
  Histograms showing probability distributions of the lower bound of
  $\lambda_g$ obtained in~\cite{yagiLISA,yagiDECIGO}.  The authors
  carried out Monte Carlo simulations of $10^4$ BH/BH binaries for a
  circular orbit with and without spin precessions in massive gravity
  theories. The mass and distance of binaries are chosen as
  $(10^7+10^6)M_{\odot}$ for LISA and $(10^6+10^5)M_{\odot}$ for
  DECIGO/BBO, both at 3Gpc.  The meaning of each histogram is the same
  as in Fig.~\ref{fig:BD}.  Observe that LISA places a stronger
  constraint than DECIGO/BBO.  Observe also that the precession breaks
  the degeneracy among parameters and improve the constraint by an
  order of magnitude.  This figure is taken from~\cite{yagiDECIGO}.
}
\end{center}
\end{figure}

Yagi and Tanaka~\cite{yagiLISA,yagiDECIGO} included the effects of
both spin precession and orbital eccentricity and derived proposed
constraints on $\lambda_g$ with LISA and DECIGO/BBO by carrying out
Monte Carlo simulations.  They worked in a
simple precession framework~\cite{apostolatos}, where the mass of binary
constituents are equal or one of the spins of the constituents is
zero.  One can approximately solve the precession equations analytically within such
assumptions.  Figure~\ref{fig:massive} presents the probability
distribution of the upper bound on $\lambda_g$ using LISA and
DECIGO/BBO with and without taking the effect of precession into
account.  The binary orbit is assumed to be circular.  Observe that
LISA places stronger constraint on $\lambda_g$ than DECIGO/BBO.  This
is because the source mass is larger for the former, which in turn
brings a larger modification in the waveform phase.  The histograms
have a sharp peak for non-precessing binaries.  This is because in
such a case, $\lambda_g$ is completely degenerate with other
parameters and the upper bound is determined purely from the prior
information on spin.  Observe that the spin precession enhances the
constraint by an order of magnitude.  This shows that the precession
breaks the degeneracy between $\lambda_g$ and other parameters,
consistent with~\cite{stavridis}.
Reference~\cite{yagiLISA,yagiDECIGO} found that the effect of the
eccentricity is small.  This is because the eccentricity and the
massive gravity correction enter in the phase at negative and positive PN
orders respectively, which indicates that the degeneracy between such
effects are very weak.  Yagi~\cite{Yagi:2013du} estimated a bound for
non-spinning binaries with eLISA and ASTROD-GW, and found that the
former (the latter) bound is slightly weaker (stronger) than the LISA
one.

All of the studies explained in the previous two paragraphs consider
just the inspiral phase in the evolution of a binary.  Keppel and
Ajith~\cite{keppel} used the phenomenological hybrid
waveform~\cite{ajith,ajith-spin} that includes not only the inspiral
phase but also the merger and ringdown phases.  They carried out a
pattern-averaged analysis and derived bounds as
$\lambda_g > 5.9 \times 10^{22}$cm and
$\lambda_g > 7.8 \times 10^{18}$cm using LISA and Adv.~LIGO
respectively.  Notice that the latter bound is a few times stronger
than the solar system one.  Although their analysis does not take spins
of binary constituents into account, the results in
Ref.~\cite{stavridis} imply that the bounds in~\cite{keppel} should be
comparable to those where one includes the effect of the spin
precession.

Del Pozzo \et~\cite{delpozzo} improved previous work using a Fisher
analysis by performing a Bayesian inference study.  They also took an
advantage of the fact that future GW interferometers may detect
signals from $\sim 50$ different binaries.  They found a new bound of
$\lambda_g > 2.6 \times 10^{18}$cm with second-generation GW
interferometers.  A similar Bayesian analysis was carried out by
Cornish \et~\cite{cornish-PPE} within a model-independent framework.
Their constraint is consistent with that in~\cite{delpozzo}.  Berti
\et~\cite{bertisesana} also took the effect of detecting GW signals
from multiple sources into account.  They carried out 1000
realizations of merger simulations and found that at each realization,
approximately 40 BH binary mergers exist whose GW signals can be
detected with LISA.  They derived a bound of $\lambda_g >
6.5$--$7.5
\times 10^{21}$cm with LISA and a slightly weaker bound with eLISA.

Mirshekhari \et~\cite{Mirshekari:2011yq} extended the above analyses
by considering a generic modification to the dispersion relation of
the graviton as
\begin{equation}
\label{eq:mod-disp}
E^2 = p^2 c^2 + m_g^2 c^4 + A p^{\alpha} c^{\alpha}\,,
\end{equation}
where $E$ and $p$ are the energy and linear momentum of the graviton
while $A$ and $\alpha$ are constants that characterize the
modification to the usual dispersion relation of the graviton.  Such a
modified relation includes the relation predicted in Doubly Special
Relativity~\cite{AmelinoCamelia:2000ge,Magueijo:2001cr,
  AmelinoCamelia:2002wr,AmelinoCamelia:2010pd}, extra dimension
theories~\cite{Sefiedgar:2010we}, Ho\v{r}ava-Lifshitz
gravity~\cite{Horava:2008ih,Horava:2009uw} and theories with
non-commutative geometries~\cite{Garattini:2011kp}.  They derived a
modification to the waveform phase and found that the last term in
Eq.~(\ref{eq:mod-disp}) enters at $(1+3\alpha/2)$PN order relative to
the leading.  They carried out a Fisher analysis for non-spinning
binaries and derived constraints on $A$ for different $\alpha$ using
Adv.~LIGO, ET and eLISA. 
For example, Adv.~LIGO can typically probe the length scale of  
$\mathcal{O}(10^{-11}\mrm{cm})$ or larger, which is much larger than
the Planck scale of $\mathcal{O}(10^{-33}\mrm{cm})$.
Although corrections to the graviton dispersion relation are 
predicted to be typically
Planck suppressed, this may not always be the case.
For example, using perturbative quantum field theories,
Collins \textit{et al}.~\cite{Collins:2004bp,Collins:2006bw} suggested 
that such corrections may
not be Planck suppressed if one takes renormalization into account.

De Felice \et~\cite{DeFelice:2013nba} pointed out an interesting fact
that gravitons of physical and reference metrics oscillate 
in the ghost-free bi-gravity model, just like neutrinos.  Narikawa
\et~\cite{Narikawa:2014fua} constructed gravitational waveforms of
non-spinning compact binaries in this theory and found that a
characteristic peak exists in the amplitude at a certain GW frequency
that depends on the parameters in the theory.  They studied the
detectability of such a modified effect in gravitational waveform with
second-generation ground-based detectors using a simple Bayesian
hypothesis testing derived in~\cite{Vallisneri:2012qq}.  They also
carried out a Fisher analysis to study how well such interferometers
can place constraints on the parameters in the theory including the
effective mass of the graviton.

\subsubsection{Coincident Tests with EMW Signals}
\label{sec:coinc-tests-with}

Let us now review proposed constraints on $\lambda_g$ when not only GW
signals but their electromagnetic counterparts are also detected.
Kocsis \et~\cite{kocsis-mg} focused on a SMBH binary coalescence and
proposed to use the correlation between GW and EMW signals.  They
derived a proposed bound of $\lambda_g > 2.8 \times 10^{20}$cm
assuming that GWs and EMWs are emitted simultaneously and
 the timing uncertainty of GWs to be the inverse of the GW frequency
at ISCO.  Notice that such a constraint is three orders of magnitude
more stringent than the solar system bound.  However, such an analysis
suffers from uncertainties in the delay in the emission time of GWs
and EMWs, which introduce systematic errors in the bound that one
can obtain.  They pointed out that one can reduce such systematic
errors if one can identify the variability in EMW signals before
coalescence and relate that to the orbital period.

Hazboun and Larson~\cite{Hazboun:2013pea} derived proposed constraints
on the difference between the propagation speed of the photon and
graviton by detecting EMW and GW signals simultaneously in three
situations, (i) isolated pulsars with Adv.~LIGO, (ii) ultracompact
binaries with LISA, and (iii) supermassive BH binaries with pulsar
timing arrays.  The second and third situations are interesting in
terms of BH based tests of GR.  Although it is not clear how BH
binaries can emit EMWs, such a case might be possible if a gas,
such as an accretion disk, exists around them. 
Indeed, some known
supermassive BH binaries like OJ 287 indeed show periodic variation in
EMW brightness.  By comparing the phase of the GW and EMW
signals, they found that LISA and pulsar timing arrays may be able to
place bounds on the graviton mass that is one and four orders of
magnitude stronger than the solar system bound respectively.

Nishizawa and Nakamura~\cite{Nishizawa:2014zna} estimated how strong a
coincident detection of GW and EMW signals from short gamma-ray bursts
(sGRBs) can place constraints on the propagation speed of the
graviton.  The expected event rate of such a coincident detection with
second-generation ground-based interferometers is $\sim 1$/yr for a
NS/BH binary as the origin of a sGRB.  Comparing the difference in the
arrival time between gravitons and photons, they derived the bound as
$ |\delta_g| < \Delta \tau_\inter/T_0$,
where $\delta_g \equiv (c - v_g)/c$ is the fractional difference in
the propagation speed of the graviton and photon, $T_0$ is the
propagation time duration of the photon and $\Delta \tau_\inter$ is
the uncertainty in the intrinsic time delay between GW and EM
emissions.  With a typical intrinsic time delay of $10$s, which is estimated
from numerical calculations of long GRBs, the authors found that a
coincident detection of GW and EMW signals from sGRBs can typically
constrain $\delta_g$ to $\sim 10^{-15}$. This, in turn, leads to a
constraint on $\lambda_g$ that is comparable to that from binary
pulsar observations in~\cite{finnsutton}, but weaker than the solar
system bound and that from GW observations alone.  Since the
constraint on $\delta_g$ is generic, one can also apply such a bound
to constrain e.g.~Lorentz-violating theories of gravity such as
Einstein-\AE{}ther and Ho\v{r}ava-Lifshitz
gravity~\cite{Hansen:2014ewa}.

\section{Quadratic Gravity}
\label{sec:modif-quadr-grav}

\subsection{Basics}
\label{sec:quad-basics}

Quadratic gravity is an alternative theory of gravity where the Einstein-Hilbert action
acquires a correction that depends on the Riemann curvature tensor at quadratic order 
coupled to a dynamical, long-ranged scalar field. Such a theory is \emph{metric} in the sense that
the matter field is only coupled to the metric, 
while the scalar field indirectly couples to matter through curvature.
Matter is \emph{universally} coupled to the metric, and hence, the weak equivalence principle holds. 

Such a theory is motivated from at least two aspects. The first one is 
the bottom-up, EFT motivation, where at low energies, GR can acquire corrections
that are prescribed by expanding the action in curvature. Thus, the leading correction term in the action
would be at quadratic order in curvature. 
The second motivation is the top-down, high-energy one, where fundamental theories of quantum gravity,
such as string theory~\cite{polchinski1,polchinski2} and loop quantum gravity~\cite{Rovelli:2008zza,Rovelli:2010bf}, 
predict corrections to GR in which scalar fields, such as dilatons
and axions, are coupled to curvature squared scalars~\cite{Gross:1986mw,Metsaev:1987zx,Mignemi:1992nt,
Mignemi:1993ce,Kleihaus:2011tg,Taveras:2008yf,Alexander:2008wi,Maeda:2009uy}.

The quadratic gravity action is thus given by~\cite{Yunes:2011we,Yagi:2011xp}
\begin{equation}
  \label{eq:S-full-action}
  S = S_{\EH} + S_{\MAT} + S_{\vartheta} + S_{q}\,,
\end{equation}
where 
\begin{align}
S_{\EH} &= \kappa \int d^4x \sqrt{-g}  R
\label{eq:EH-action}
\end{align}
is the Einstein-Hilbert action with $\kappa \equiv 1/(16\pi)$ while $S_{\MAT}$ is the action for the matter field.
$S_{\vartheta}$ is the action for the canonical scalar field $\vartheta$ defined by
\begin{equation}
  \label{eq:S-vartheta}
  S_{\vartheta} = -\frac{1}{2} \int d^{4}x\sqrt{-g}
  \left[(\cd_{a}\vartheta)(\cd^{a}\vartheta) + 2U(\vartheta)\right]\,,
\end{equation}
with $U(\vartheta)$ representing the scalar field potential.
$S_q$ contains the coupling of the scalar field to the quadratic curvature.
In the EFT viewpoint, derivative interactions are higher operator order and 
one expects them to be suppressed relative to non-derivative 
interactions~\cite{Yagi:2015oca}.
Restricting to non-derivative interactions only, the most generic form of $S_q$
is given by 
\be
\label{eq:q-action}
 S_{q} ={} \int d^4x \sqrt{-g} \; \left[f_1(\vartheta) R^2
+ f_2(\vartheta) R_{ab} R^{ab} + f_3(\vartheta) R_{abcd} R^{abcd}
+ f_4(\vartheta) R_{abcd} {}^{*}R^{abcd}   \right]\,,
\ee
where $f_i(\vartheta)$ with $i=(1,2,3,4)$ are arbitrary functions of $\vartheta$ and
 the (left) dual Riemann tensor is defined by
\begin{equation}
  \label{eq:dual-riem-def}
  {}^{*}\! R^{ab}{}_{cd} = \frac{1}{2}\epsilon^{abef}R_{efcd}\,,
\end{equation}
with $\epsilon^{abcd}$ representing the Levi-Civita tensor.

In this review, we consider a subclass of the most generic form in Eq.~\eqref{eq:q-action}, where we set 
$f_i(\vartheta) = \alpha_i g(\vartheta)$ with $\alpha_i$ representing coupling constants and $g(\vartheta)$
an arbitrary function of $\vartheta$, namely  
\be
S_{q} = \int d^{4}x\sqrt{-g} \  g(\vartheta) \left[\alpha_{1} R^{2} + \alpha_{2} R_{ab} R^{ab} 
+ \alpha_{3} R_{abcd} R^{abcd} + \alpha_{4} \, {}^{*}R_{abcd} R^{abcd} \right]\,.
\label{eq:quad-action-we-focus-on}
\ee
Notice that $\alpha_i$ have a unit of length squared in geometric units 
since $\vartheta$ is dimensionless.
This subclass of quadratic gravity contains an interesting examples of theories including
\begin{itemize} 
\item \emph{EdGB}\footnote{EdGB theory was originally introduced in
the Jordan frame~\cite{Metsaev:1987zx,Maeda:2009uy}, 
where $e^{-\gamma \vartheta}$ is coupled to the Gauss-Bonnet invariant density. 
If one then moves to the Einstein frame via a conformal transformation, 
one finds $g(\vartheta) = e^{-\gamma \vartheta}$ and 
$(\alpha_1,\alpha_2,\alpha_3,\alpha_4) = (1, -4,1,0)\alpha_\EDGB$  in Eq.~\eqref{eq:quad-action-we-focus-on} 
with higher order terms in curvature tensors and (the derivative of) the scalar field. EdGB that 
we refer to in this review corresponds to truncated EdGB in~\cite{Maeda:2009uy} 
where one neglects such higher order terms.}~\cite{Metsaev:1987zx,Maeda:2009uy}:
$g(\vartheta) = e^{-\gamma \vartheta}$ and 
$(\alpha_1,\alpha_2,\alpha_3,\alpha_4) = (1, -4,1,0)\alpha_\EDGB$ with $\gamma$ and 
$\alpha_\EDGB$ representing coupling constants.
The current strongest bound on $\alpha_\EDGB$ comes from the existence of a BH solution as
$\sqrt{|\alpha_\EDGB|} \leq 1.4$km~\cite{Kanti:1995vq,Pani:2009wy}.
\item \emph{dCS}~\cite{Jackiw:2003pm,Smith:2007jm,Alexander:2009tp}: $g(\vartheta) = \vartheta$ and 
$(\alpha_1,\alpha_2,\alpha_3,\alpha_4) = (0, 0,0,-1/4)\alpha_\CS$ with $\alpha_\CS$ 
representing a coupling constant.
The strongest bound comes from the measurement of the frame-dragging effect of Earth 
using Gravity Probe B and LAGEOS satellites~\cite{AliHaimoud:2011fw} 
and from table-top experiments~\cite{Yagi:2012ya} as
$\sqrt{|\alpha_\CS|} \leq \mathcal{O}(10^8)$km.
\item \emph{decoupled quadratic gravity (dQG)~\cite{Yunes:2011we,Yagi:2011xp}}: $g(\vartheta) =  \vartheta$. 
Such a theory corresponds to 
Taylor expanding $g(\vartheta)$ in Eq.~\eqref{eq:quad-action-we-focus-on} around $\vartheta=0$ and
keep only up to linear order in the scalar field coupling. One also neglects the term that is independent of $\vartheta$.
This is motivated from a fact that such a term with either the Gauss-Bonnet combination 
($(\alpha_1,\alpha_2,\alpha_3,\alpha_4) = (1, -4,1,0)\alpha_\EDGB$)
or the CS combination 
($(\alpha_1,\alpha_2,\alpha_3,\alpha_4) = (0, 0,0,-1/4)\alpha_\CS$) becomes a total derivative
and does not contribute to the field equations. Notice that one recovers dCS gravity when one takes
the CS combination of the coupling constants in dQG.
The current strongest bound on $\alpha_3$ (or $\alpha_\EDGB$ in the decoupled EdGB) 
comes from the orbital decay rate of the BH LMXB
A0620-00~\cite{johannsen1} as $\sqrt{|\alpha_3|} \leq 1.9$km (or $\sqrt{|\alpha_\EDGB|} \leq 1.9$km)~\cite{kent-LMXB}.
\end{itemize}
Since we truncate the action to quadratic order in curvature, it is natural to work
within the small coupling approximation, where one takes the coupling constants to be small
and keep only the leading order corrections.
Such a procedure ensures that the field equations are of second order, and thus, makes the theory
well-posed.\footnote{This has been explicitly proven for dCS gravity in~\cite{Delsate:2014hba}.}
From here on, we set $U(\vartheta)=0$ for simplicity.

\subsection{BH Solutions}
\label{sec:quad-bh-solutions}

In this subsection, we review static, slowly-rotating and rapidly-rotating BH solutions
in quadratic gravity, in particular EdGB, dCS and dQG. We also review the current
status of the BH stability analysis in each theory.

\subsubsection{EdGB}
\label{sec:EdGB-BH}

Let us first review BH solutions in EdGB gravity.
Static BH solutions in EdGB were first derived in~\cite{Mignemi:1992nt,Mignemi:1993ce} analytically
within the small coupling approximation 
(and was later rederived in~\cite{Yunes:2011we,Sotiriou:2014pfa}). 
If one relaxes the small coupling approximation
and treat the theory as exact, then one needs to solve the field equations numerically, 
as done in~\cite{Kanti:1995vq,Torii:1996yi,Alexeev:1996vs} for static solutions, 
in~\cite{Pani:2009wy} for slowly-rotating solutions and 
in~\cite{Kleihaus:2011tg} for rapidly-rotating solutions. 
Reference~\cite{Kanti:1995vq} showed that the BH solutions only exist if $\alpha_\EDGB$
is smaller than a threshold that depends on the horizon radius and also 
the scalar field at the horizon, 
which places the current strongest bound on the theory. 

The stability of a non-rotating BH in EdGB gravity was studied
in~\cite{Kanti:1997br} under radial perturbations, which are a special
case of polar perturbations.  The authors derived a
Schr\"o{}dinger-type master perturbation equation and
solved the eigen value problem.  Later, Cardoso and
Pani~\cite{Pani:2009wy} studied the stability under axial
perturbations. A stability against axial perturbations is easier to
tackle than that against polar perturbations because the scalar field
perturbation does not appear in the axial perturbation equations.
Both calculations suggest that a non-rotating BH is stable in EdGB
gravity.  Such analyses are possible because the polar and axial
perturbations decouple as in GR.

\subsubsection{dCS}
\label{sec:dCS-BH}

We now review BH solutions in dCS gravity within the small coupling approximation.
Since spherically symmetric spacetimes do not break parity, non-rotating BH solutions
in dCS is exactly the same as the Schwarzschild solution. BH solutions
acquire dCS corrections when one includes spins. Slowly rotating BH solutions to
first order in spin were derived 
in~\cite{Yunes:2009hc,Konno:2009kg}, where the scalar field and the metric are 
given by
\ba
\label{eq:CS-BH-scalar}
\vartheta &=& \frac{5\alpha_\CS}{8} \chi \frac{\cos\theta}{r^2} \left( 1+\frac{2}{7}\frac{M}{r} + \frac{18}{5} \frac{M^2}{r^2} \right) \nn \\
&=& - \frac{\mu_\CS^i n^i}{r^2} + \mathcal{O}\left( \frac{M^3}{r^3} \right)\,, \qquad \mu_\CS^i  \equiv - \frac{5}{8} \alpha_\CS \chi^i\,, \\
\label{eq:CS-BH-metric}
ds^{2} &=& ds^{2}_{K} + \frac{5}{4}\zeta_\CS M \chi \frac{M^4}{r^4} \left( 1+\frac{12}{7}\frac{M}{r} + \frac{27}{10} \frac{M^2}{r^2} \right) \sin^2 \theta dt d\phi\,.
\ea
Here, $ds^{2}_{K}$ is the line element for Kerr given by Eq.~\eqref{eq:kerr-metric} 
while $M$ and $\chi^i$
corresponds to the BH mass and the dimensionless spin angular momentum vector respectively.
$n^i$ is the unit vector from the BH to the field point and $\chi^i n_i = \chi \cos \theta$,
while $\mu_\CS^i$ is the BH scalar dipole charge in dCS gravity.
Such a scalar dipole charge is important because it sources the dominant scalar radiation
emitted from a BH binary, as we will see in Sec.~\ref{sec:quad-gw-tests}.  
$\zeta_\CS$ is the dimensionless coupling constant defined via 
$\zeta_\CS \equiv \alpha_\CS^2/(\kappa M^4)$.
Such a modification in the metric shifts the location of ISCO.

Such calculations were extended to second order in spin in~\cite{Yagi:2012ya} 
using a BH perturbation scheme~\cite{Sago:2002fe}, treating a deformation due to spin
as a perturbation from Kerr. The scalar field is the same as that to linear order in spin, but
the metric acquires a CS correction in the even-parity sector, namely the $(t,t)$, $(r,r)$, $(\theta,\theta)$
and $(\phi,\phi)$ components.
Such modifications to the even-parity sector render CS corrections to the quadrupole moment and
 the location of the event horizon,
ergo region and ISCO.
These modifications, in particular the one to the quadrupole moment given by 
$Q = Q_K[1 -(201/1792) \, \zeta_\CS]$ 
with $Q_K$ representing the GR Kerr quadrupole moment, are crucial in modeling
gravitational waveforms from BH binaries, as we will see in Sec.~\ref{sec:quad-gw-tests}.
Although the Kerr solution and the dCS BH solution to linear order in spin
are both of Petrov type D, the dCS solution to second order in spin is of type I.
This fact leads to the absence of a second-rank Killing tensor or a Carter-like constant.

\begin{figure}[htb]
\begin{center}
\includegraphics[width=8.5cm,clip=true]{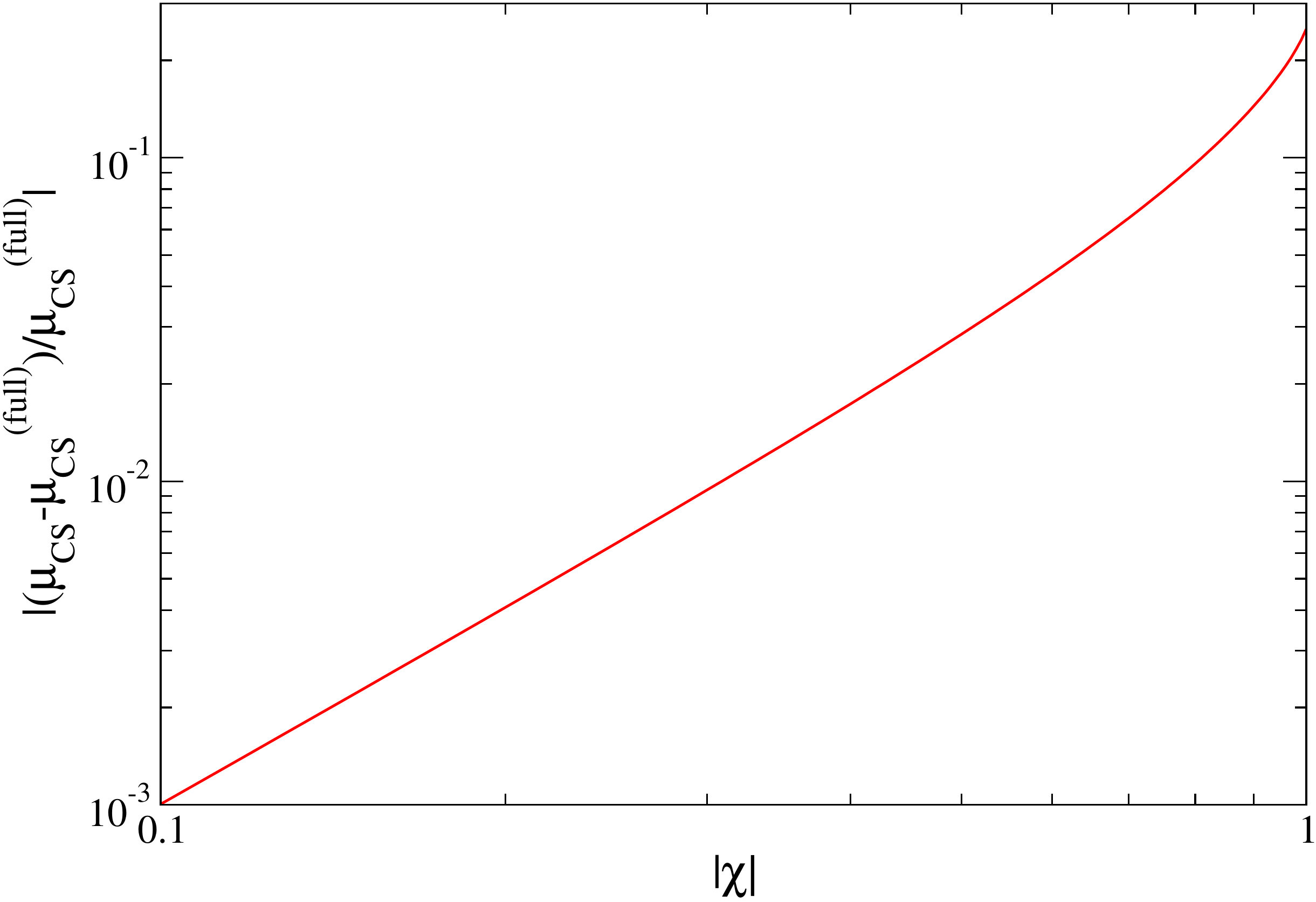}
\caption{\label{fig:slow-rot-vs-full} (Color online)
Absolute fractional difference between the (magnitude of the) dipole charge 
to full order in spin 
$\mu^{(\mathrm{full})}_\CS$ and that to 
leading order in spin $\mu_\CS$ in dCS.
Observe that the fractional difference is within 10$\%$ for $|\chi| < 0.8$.
}
\end{center}
\end{figure}

Regarding rapidly rotating BH solutions, Ref.~\cite{Yagi:2012vf}
derived the scalar dipole charge valid to all orders in spin by using
the separability of the scalar wave equation, solving the $\ell =1$
mode of the scalar equation of motion, and extracting the leading (in
powers of $r^{-1}$) behavior at spatial infinity. The authors found
the magnitude of the dipole as
\be
\mu^{\mathrm{(full)}}_\CS = \alpha_\CS \frac{2+2\chi^4-2\sqrt{1-\chi^2}-\chi^2(3-2\sqrt{1-\chi^2})}{2 \chi^3}\,.
\ee
Figure~\ref{fig:slow-rot-vs-full} presents the fractional difference between
$\mu_\CS^{\mathrm{(full)}}$ and $\mu_\CS$.
Observe that the leading order in spin charge is valid within 10\% when $|\chi| < 0.8$.

In the decoupling limit, the scalar wave equation on the Kerr
background separates.  Each $\ell$ mode's ordinary differential equation
(ODE) may be solved using
variation of parameters, and the formal solution written in terms of
quadrature.  This has been presented in~\cite{Konno:2014qua,
  Stein:2014wza, Stein:2014xba}.  To be more explicit, we write
\begin{equation}
  \vartheta = \sum_{\ell=0}^{\infty} P_{\ell}(\cos\theta) \vartheta_{\ell}(r)
\end{equation}
in Boyer-Lindquist coordinates, where $P_{\ell}(\cos\theta)$ are
Legendre polynomials.  Then we find separated ODEs for each $\ell$
mode,
\begin{equation}
  \label{eq:CS-radial-equation}
  \left[
\frac{\pd}{\pd r} \left(\Delta \frac{\pd}{\pd r}\right)  - \ell(\ell+1)
  \right] \vartheta_{\ell}(r) = S_{\ell}(r)\,,
\end{equation}
where $S_{\ell}(r)$ was first published in~\cite{Konno:2014qua}, and
then more compactly in terms of hypergeometric functions
in~\cite{Stein:2014wza}. 
We recall that $\Delta$ is defined below Eq.~\eqref{eq:kerr-metric}.
Equation~\eqref{eq:CS-radial-equation} is
solved via variation of parameters using the homogeneous solutions,
which are simply Legendre functions of the first and second kind,
$P_{\ell}(\bar \eta)$ and $Q_{\ell}(\bar \eta)$, where
$\bar \eta\equiv(r-M)/\sqrt{M^{2}-a^{2}}$ is a shifted and rescaled radial
coordinate.  Thus the solution for $\vartheta_{\ell}(r)$ can be
written as
\begin{align}
  \label{eq:vartheta-l-ansatz}
  \vartheta_{\ell}(\bar \eta) = \vartheta^{+}_{\ell}(\bar \eta) P_{\ell}(\bar \eta) +
  \vartheta^{\infty}_{\ell}(\bar \eta) Q_{\ell}(\bar \eta)\,,
\end{align}
where we have the two quadratures (due to the simplicity of the
Wronskian $\mathcal{W}[P_{\ell},Q_{\ell}]$)
\begin{align}
  \vartheta^{+}_{\ell}(\bar \eta) &=
  \int_{\eta}^{\infty} S_{\ell}(r') Q_{\ell}(\bar \eta') d\bar \eta' \,, &
  \vartheta^{\infty}_{\ell}(\bar \eta) &=
  \int_{1}^{\eta} S_{\ell}(r') P_{\ell}(\bar \eta') d\bar \eta' \,.
\end{align}
Here the constants of integration are fixed by regularity at the
horizon and vanishing at infinity, which means that
$\vartheta_{\ell}^{+}$ must vanish at infinity, while
$\vartheta_{\ell}^{\infty}$ must vanish at the horizon ($\bar \eta=1$).

\begin{figure}[tbp]
  \centering
  \includegraphics[height=8cm]{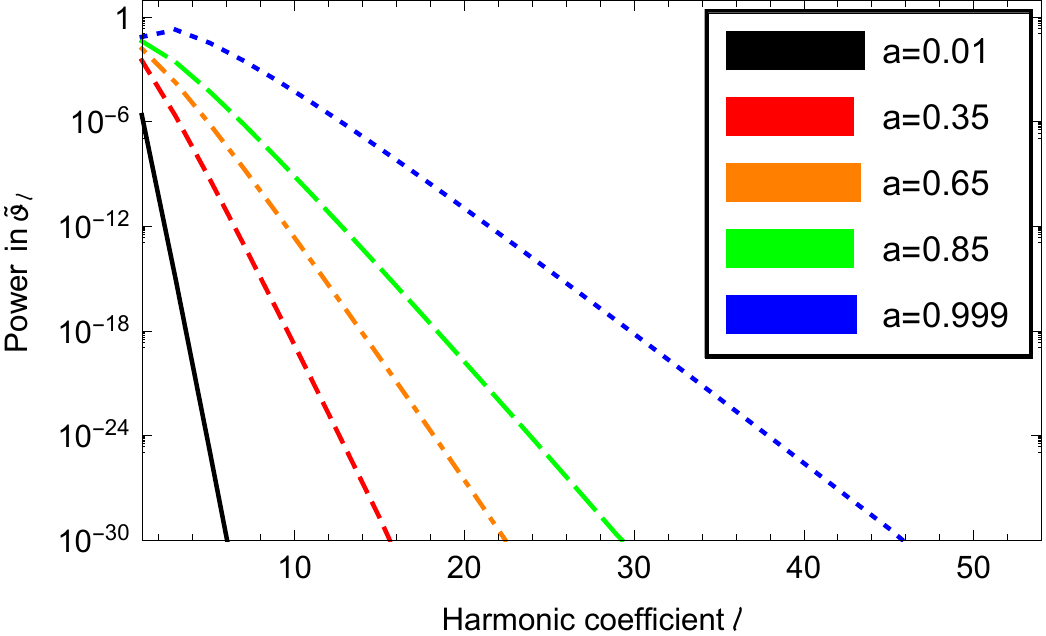}
  \caption{
    (Color online)
   Exponential convergence of smooth solutions for $\vartheta$ and $h$
     as a decomposition in Legendre
    polynomials $P_{\ell}(\cos\theta)$.
    The vertical axis represents the $L^{2}$ norm of
    $\vartheta_{\ell}(x)$.
    At low spin, the convergence is more rapid, and one only needs to keep fewer
    coefficients, while more coefficients must be kept as one increases spin.
    We only plot the odd coefficients of $\vartheta$.
    $h$ follows the same trend.
    Figure from~\cite{Stein:2014xba}.
  }
  \label{fig:power-theta}
\end{figure}

\begin{figure}[tbp]
  \centering
  \includegraphics[height=10cm]{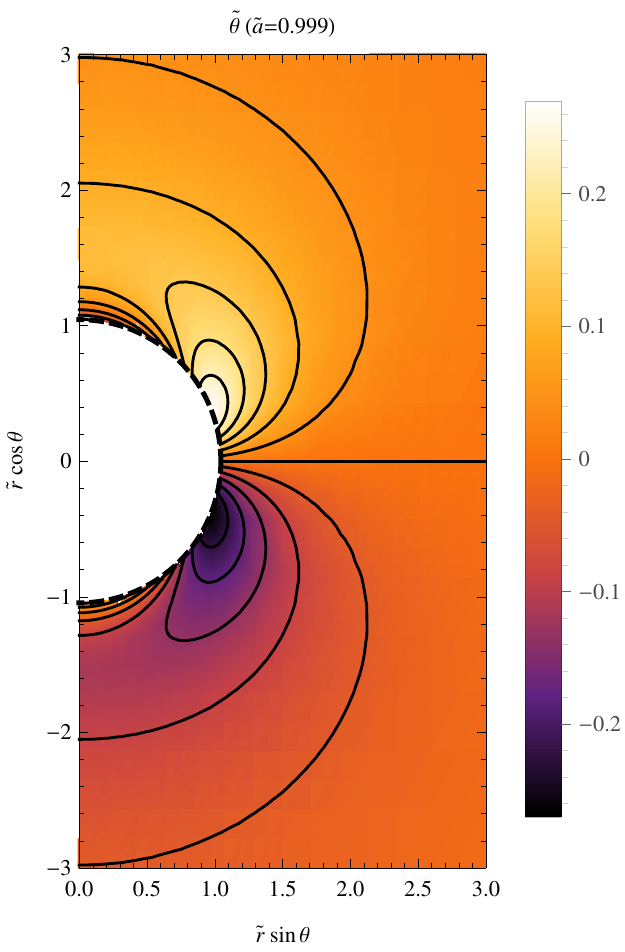}
  \caption{
    (Color online)
    Profile of solution for $\vartheta$ in a longitudinal
    ($\phi=\text{const.}$) section of the space, at high spin
    ($a/M=0.999$). The rotation axis is the left edge of the
    figure.  Color represents the value of the field and contours 
    of constant field value are spaced linearly.
    The dashed line corresponds to the horizon.
    Figure from~\cite{Stein:2014xba}.
  }
  \label{fig:theta-profile}
\end{figure}

The first few $\ell$ modes may individually be integrated in closed
form (a closed-form expression for general $\ell$ has not been
presented in the literature to date).  However, it is straightforward
to find $\vartheta_{\ell}$ numerically.  Konno and
Takahashi~\cite{Konno:2014qua} integrated the radial ODEs as an
initial value problem.  Meanwhile, Stein~\cite{Stein:2014xba} solved
these equations using a global pseudo-spectral method.  Either way,
since the source is $C^{\infty}$, there is exponentially decreasing
power with increasing $\ell$ number (see Fig.~\ref{fig:power-theta}),
as shown in~\cite{Stein:2014xba}.  Therefore, the solution is
faithfully captured by truncating in $\ell$.  At a fixed fractional
truncation error, the number of required $\ell$ modes increases with
increasing rotation $a$.  At rapid rotation, the solution displays
interesting multipolar structure as seen in Fig.~\ref{fig:theta-profile}


This solution for $\vartheta$ is the $\mathcal{O}(\alpha_{\CS})$
correction to the Kerr solution.  To this order, the metric is not
affected.  The metric is first corrected at order
$\mathcal{O}(\alpha_{\CS}^{2})$, as seen in
Eq.~\eqref{eq:CS-BH-metric} (recall that
$\zeta_{\CS}\propto\alpha_{\CS}^{2}$).  As mentioned before,
Eq.~\eqref{eq:CS-BH-metric} is only the $\mathcal{O}(a^{2})$
correction to the metric.  Finding the full metric to all orders in
$a$ is still an open problem.  However, some progress was made by
Stein~\cite{Stein:2014xba}. He noted that the trace of the metric's
equation of motion, in Lorenz gauge, is simply the scalar wave operator acting on the
trace of the metric deformation $h\equiv g^{ab}h_{ab}$, with a
different source term:
\begin{equation}
  \label{eq:eom-trace}
  \kappa\square h
  = - (\cd^{a}\vartheta)(\cd_{a}\vartheta)\,.
\end{equation}
Here the source term depends on the solution for $\vartheta$ to
all orders in $a$.  This is the same differential operator which acts
on $\vartheta$ in its equation of motion, so again separation of
variables will give radial ODEs of the form of
Eq.~\eqref{eq:CS-radial-equation}, acting on $h_{\ell}$ and with
different source terms.  This was also solved numerically
in~\cite{Stein:2014xba}.

Reference~\cite{Stein:2014xba} used this numerical solution to find
the regime of validity of the weak-coupling perturbation theory.
Specifically, for the perturbation theory to be valid, we should have
that the size of $h$ is under perturbative control, where $h$ is the
$\mathcal{O}(\alpha_{\CS}^{2})$ perturbation away from the Kerr
metric.  If we examine the perturbation to the volume element,
\begin{equation}
  \sqrt{-g} = \sqrt{-g^{\GR}} (1+ {\textstyle\frac{1}{2}}h) +
  \mathcal{O}(\alpha_{\CS}^{3})\,,
\end{equation}
we see that a condition on the trace, $|h|<1$ throughout the
spacetime, can serve to define the regime of validity of the
perturbative scheme.  The delineation between perturbation theory
being valid and invalid is presented in Fig.~1
of~\cite{Stein:2014xba}.

Regarding BH stability analyses in non-dynamical CS gravity, where
one sets $S_\vartheta =0$ in Eq.~\eqref{eq:S-full-action}, Yunes and
Sopuerta~\cite{Yunes:2007ss} studied the BH perturbation under the
Schwarzschild background and found that, unlike in GR, polar and axial
modes are generically coupled.  Due to the Pontryagin constraint,
generic BH oscillations are disallowed in this theory.  In particular,
a single parity oscillation (purely polar or axial oscillation) does
not exist.  The authors also looked at a stability analysis in dCS
gravity, but the situation remained similar and in particular, the
axial oscillations are severely suppressed.  Cardoso and
Gualtieri~\cite{Cardoso:2009pk} showed that the polar and axial modes
decouple when the background scalar field vanishes, and the scalar
field perturbation only couples to the axial perturbations.  Molina
\et~\cite{Molina:2010fb} carried out detailed numerical calculations
and showed that the Schwarzschild BH in dCS gravity is stable.
Motohashi and Suyama~\cite{motohashi2} carried out a perturbation
study under a static and spherically symmetric spacetime in both
non-dynamical and dynamical CS gravity without assuming that the
background scalar field vanishes. They found that the ghost modes are
present unless such a background scalar field vanishes.
However, the ghost modes only appear at a very high wavenumber, as
shown by~\cite{Dyda:2012rj}, and this wavenumber is outside of the
regime of validity of dCS when treated as an EFT.

Linear stability analysis of a BH in dCS within the high
frequency (namely, geometric optics or WKB) approximation has been
carried out in~\cite{garfinkle}.  The authors considered perturbations in 
dCS gravity and
showed how the propagation speed of GWs acquires a dCS correction
under the Schwarzschild background (see Sec.~\ref{sec:dec-QG-BH} 
below for a more detailed 
explanation of the BH linear stability analysis within the WKB
approximation).

\subsubsection{dQG}
\label{sec:dec-QG-BH}

We now describe BH solutions in dQG within the small coupling approximation. 
Static solutions were found in~\cite{Yunes:2011we}.
The authors found that the scalar field only depends on $\alpha_3$ as
\be
\vartheta =  \frac{2\alpha_3}{M r} \left( 1 + \frac{M}{r} + \frac{4}{3} \frac{M^2}{r^2} \right)
 =  \frac{\mu_\DQG}{r} + \mathcal{O}\left( \frac{M^2}{r^2} \right)\,, \qquad \mu_\DQG \equiv \frac{2 \alpha_3}{M}\,,
\label{eq:BH-scalar-dQG}
\ee
where $\mu_\DQG$ is the BH scalar monopole charge in dQG.
This is because within the small coupling approximation, the scalar field is source by 
the curvature scalar under the Kerr background and therefore $R^2 = 0 = R_{\mu\nu} R^{\mu\nu}$,
while $R_{abcd} {}^{*}R^{abcd}=0$ due to spherically symmetry. 
Thus, the scalar field is only sourced by a term proportional to 
$\alpha_3 R_{\mu\nu\rho\sigma} R^{\mu\nu\rho\sigma}$.
The metric acquires a dQG correction to the $(t,t)$ and $(r,r)$ components, which further modifies
the location of the ICSO and binding energy of a binary.

Slowly-rotating BH solutions in dQG were found in~\cite{Pani:2011gy} to linear order in spin
and in~\cite{Ayzenberg:2014aka} to quadratic order in spin. As in dCS, at linear order in spin,
the scalar field is proportional to $\alpha_4$ and
the metric acquires modifications to the odd-parity sector, which further modifies the location of
the ISCO.
At second order in spin, the scalar field is proportional to $\alpha_3$ and the metric acquires 
modifications to the even-parity sector, which further modifies the quadrupole moment and the location of
the ISCO, event horizon and ergo region. As in dCS gravity, dQG BH solutions are of Petrov type D to
linear order in spin while of type I to second order in spin.

Linear stability analysis of a BH in quadratic gravity within the WKB 
approximation has been
carried out in~\cite{Ayzenberg:2013wua}. The authors explicitly derived the propagation
speed of GWs under the slowly-rotating BH background.  When the scalar
field perturbation is larger than the metric one, the dispersion
relation of GWs in quadratic gravity is given
by~\cite{Ayzenberg:2013wua}
\begin{equation}
\label{eq:disp-rel-grav-QG}
(k^\mu k_\mu)^2 = 32 \xi_A S^{\mu \nu} S_{\mu \nu} = 32 \xi_A W^{\mu \nu} W_{\mu \nu}\,,
\end{equation}
where $A=3$ ($A=4$) when considering the even (odd) parity sector of
the theory, $\xi_A \equiv \alpha_A/\kappa$ 
 and $k^{\mu}$ represents the wave vector. $S_{\mu \nu}$ and
$W_{\mu \nu}$ are related to the Weyl tensor $C_{\mu \nu \rho \sigma}$
via
\begin{equation}
S_{\mu \nu} \equiv C_{\mu \rho \nu \sigma} k^{\rho} k^{\sigma}\,, \quad
W_{\mu \nu} \equiv {}^* C_{\mu \rho \nu \sigma} k^{\rho} k^{\sigma}
= \frac{1}{2} \epsilon_{\nu \sigma}{}^{\alpha \beta} C_{\mu \rho \alpha \beta} k^{\rho} k^{\sigma}\,,
\end{equation}
where $\epsilon_{\mu \nu \rho \sigma}$ is the Levi-Civita tensor.
Notice that the right hand side of Eq.~(\ref{eq:disp-rel-grav-QG}) is
already proportional to $\xi_A$, and hence one only needs to consider
the GR contribution to $S_{\mu \nu}$ and $W_{\mu \nu}$ if we only keep
to leading order in the coupling constant.  Observe that
Eq.~(\ref{eq:disp-rel-grav-QG}) reduces to $k^\mu k_\mu = 0$ in GR,
and hence GWs propagate at the speed of light irrespective of the
background spacetime.  From Eq.~(\ref{eq:disp-rel-grav-QG}), one finds
that the dispersion relation for GWs propagating in a radial direction
under any Petrov type D spacetime reduces to that in GR.

Let us now look at the GW dispersion relation under specific example
backgrounds.  Parameterizing the wave vector as
$k^\mu \equiv [\Omega, k_1, k_2, k_3]$, one finds the dispersion
relation of GWs under the non-spinning BH background
as~\cite{garfinkle,Ayzenberg:2013wua}
\begin{equation}
\label{eq:Omega-Schw}
\Omega = \Omega_\Schw \left[ 1 \pm \frac{12 M^3}{r^3} \zeta_A^{1/2}
\left( 1 - \frac{k_1^2}{\Omega_\Schw f^2} \right) \right]\,,
\end{equation}
with the GR dispersion relation given by
\begin{equation}
 \Omega_\Schw \equiv \pm \frac{1}{f} \sqrt{k_1^2 + f r^2 k_2^2 + f r^2 k_3^2 \sin^2 \theta}\,.
\end{equation}
Here, $M$ is the BH mass, $r$ is the distance from the BH and
$f \equiv 1 - 2 M/r$.  Observe that the quadratic gravity
correction in Eq.~(\ref{eq:Omega-Schw}) vanishes for the spherical
wavefront ($k_{2,3}=0$) and GWs propagate at the speed of light, which
is consistent with~\cite{motohashi2}.  On the other hand, the
quadratic gravity correction to the GW dispersion relation propagating
in the radial direction does not vanish under the \emph{spinning} BH
background, but such an effect decays rapidly at spatial infinity and
is quadratically proportional to the BH spin.  This means that it is
practically impossible to distinguish quadratic gravity and GR from
the GW propagation observed at infinity.  The situation is similar
when the metric perturbation is larger than the scalar one.




\subsection{GW Tests}
\label{sec:quad-gw-tests}

We now explain projected constraints on quadratic gravity with GWs
from BH binaries. 

\subsubsection{EdGB}

Reference~\cite{Yagi:2015oca} recently derived an order of magnitude estimate
on the future projected constraints
on EdGB gravity with GWs from BH binaries. The dominant correction to the waveform is due to the 
scalar dipole radiation, which is proportional to the square of the difference in the scalar monopole 
charges of the binary constituents
(see Sec.~\ref{sec:GW-dQG} for more details on how to derive scalar dipole radiation in quadratic gravity). 
Within the small coupling approximation, 
the scalar charge of a NS and a BH are
$\mathcal{O}(\alpha_\EDGB^2)$ and $\mathcal{O}(\alpha_\EDGB)$ respectively. Thus, when 
the coupling constant is smaller than the curvature length scale of a NS or a BH in a NS/BH binary, 
the NS scalar charge can be neglected. 
With an Adv.~LIGO observation of SNR=20, Cornish \textit{et al.}~\cite{cornish-PPE} found that the measurement accuracy
of the PPE parameter $\beta_\ppE$ for the dipole radiation is $\delta \beta_\ppE = 5 \times 10^{-4}$. 
Assuming that the BH scalar charge is roughly the same as that in dQG with the Gauss-Bonnet combination
of $(\alpha_1,\alpha_2,\alpha_3,\alpha_4) = (1,-4,1,0)\alpha_\EDGB$, one finds an approximate 
GW projected bound from a BH/NS binary with the mass $(M_\NS,M_\BH) = (1.4,5)M_\odot$ as
\begin{align}
\sqrt{|\alpha_{\EDGB}|}  \lesssim 3.0 \; {\rm{km}} \left(\frac{\delta \beta_{\ppE}}{5 \times 10^{-4}}\right)^{1/4} 
\left(\frac{M_{\BH}}{5 M_{\odot}} \right)\left(\frac{0.171}{\eta}\right)^{1/10}\,,
\label{eq:GB-GW-BH-NS}
\end{align}
where we recall that $\eta$ is the symmetric mass ratio.
Similarly, an approximate bound from a BH/BH binary with the mass $(m_1,m_2)=(10,5)M_\odot$ is derived as
\begin{align}
\sqrt{|\alpha_{\EDGB}|} \lesssim 3.4 \; {\rm{km}} \left(\frac{\delta \beta_{\ppE}}{5 \times 10^{-4}}\right)^{1/4} 
\left(\frac{m}{15 M_{\odot}}\right) 
\left(\frac{0.33}{\delta} \right)^{1/2}
\left(\frac{\eta}{0.22}\right)^{9/10}\,,
\label{eq:DDGB-BHBH-Const}
\end{align}
where $m$ is the total mass and $\delta \equiv (m_{1} - m_{2})/m$. Such bounds are comparable
to the current strongest bound from the existence of BHs~\cite{Kanti:1995vq,Pani:2009wy}.

\subsubsection{dCS}
\label{sec:dCS-GW}

GWs in theories that violate parity at the level of the field
equations are known to have an amplitude birefringent effect, where
the amplitude of one of the circular polarization modes is enhanced
while the other one is suppressed while they
propagate~\cite{Lue:1998mq,jackiw:2003:cmo,Satoh:2007gn,Satoh:2008ck}.
Alexander \et~\cite{Alexander:2007kv} considered such an effect on GWs
from compact binaries at a cosmological distance in (non-dynamical) CS gravity.  They
considered a metric perturbation around a flat
Friedmann-Robertson-Walker spacetime and derived the dispersion
relation for such a perturbation.  They then assumed that the CS
correction to the relation is small, namely, the scalar field evolves
on cosmological timescales. They derived the amplitude on
circularly-polarized gravitational waveform as
\begin{equation}
A_{\RL} = A_{\RL}^{(\GR)}  
\exp \left[ \lambda_{\RL} \frac{k(t)}{H_0} \bar  \xi (z) \right]\,, \qquad 
A_{\RL}^{(\GR)} \equiv (1 + \lambda_{\RL} \mu)^2 \frac{2 \mathcal{M}}{D_L} \left[ \frac{k(t) \mathcal{M}}{2} \right]^{2/3}\,.
\end{equation}
%
%
%
Here, the subscript R (L) refers to the right- (left-) handed
polarization, $\lambda_\R = +1$, $\lambda_\LL = -1$, $\mathcal{M}$ is
the chirp mass, $D_L$ represents the luminosity distance of the
source, $\mu$ is the cosine angle between the observer's line-of-sight
and the orbital angular momentum of the binary, $H_0$ is the current
Hubble constant, $k(t)$ is the instantaneous wave number of the wave
front that passes the GW interferometer at $t$ and $\bar \xi (z)$
encodes the accumulated CS correction to the amplitude due to the
birefringence that depends on the source redshift $z$. From this
equation, one finds
\begin{equation}
\frac{A_\R}{A_\LL} = \frac{1 + \mu}{1 - \mu} \exp \left[ \frac{2 k(t) \xi(z)}{H_0} \right]
\equiv \frac{1 + \tilde \mu}{1 - \tilde \mu}\,,
\end{equation}
where $\tilde \mu$ is the apparent cosine angle between the line-of-sight
and the orbital angular momentum that includes the CS correction.  The
authors in~\cite{Alexander:2007kv} carried out a Fisher analysis,
assuming that the amplitude parameters are completely decoupled from
the phase parameters, and found that if LISA can measure $\bar \xi$ to
$10^{-19}$ accuracy if it detects a GW signal from an edge-on,
equal-mass BH binary with $M = 10^6 M_\odot$ at $z=5$.

Yunes~\et~\cite{Yunes:2010yf} extended the above analysis by
considering a coincident detection of a short gamma-ray burst (sGRB)
and GWs from NS/NS or NS/BH binaries with second-generation
ground-based detectors such as Adv.~LIGO.  If the GW amplitude
birefringence due to parity-violation effect is present, the measured
luminosity distance assuming GR is correct is different from the
actual distance.  They assumed that the host galaxy of the sGRB source
can be identified so that one obtains its redshift.  By comparing the
luminosity distance estimated from such EMW observations with that
obtained from GW observations, one can carry out a consistency test of
gravity and constrain parity-violation effect in gravity.  The authors
found that one can place constraints that are typically two orders of
magnitude stronger than the solar system bound from the
LAGEOS satellites~\cite{Smith:2007jm} and are comparable to the binary
pulsar observations~\cite{AliHaimoud:2011bk} in non-dynamical CS gravity.
The authors also discuss that the measurement error of the distance from 
EMW observations should be subdominant as long as the source is not
too close (GW SNR < 108).

GWs from EMRIs in dCS gravity was first considered by Sopuerta and
Yunes~\cite{Sopuerta:2009iy}.  They used a slowly-rotating BH
solution to linear order in spin in this theory. They found that such
a BH admits a Carter-like constant that is constructed from the
second-rank Killing tensor.  They worked within the
\emph{semirelativistic} approximation~\cite{Ruffini:1981af}, where the
trajectory of a small compact object orbiting around the massive BH is
assumed to be a geodesic, and thus the effect of radiation reaction is
neglected (though the dCS correction to radiation reaction can be important
depending on masses and spins of binary constituents~\cite{Yagi:2011xp}). 
Gravitational radiation within such an approximation is
calculated from the usual GR multipolar decomposition of the radiative
field under the Minkowski background. The dCS correction to the
gravitational waveform comes from the difference in the massive BH
geometry from the Kerr one to linear order in spin. They derived a rough bound on the
characteristic length scale of the theory $\xi_\CS^{1/4} = \sqrt{|\alpha_\CS|}/\kappa^{1/4}$ 
with future
GW observations as
\begin{equation}
\label{eq:CS-bound-EMRI}
\xi_\CS^{1/4} < 2 \times 10^5 \, \mrm{km} \, \left( \frac{\Delta}{10^{-6}} \right)
 \left( \frac{M}{5 \times 10^5 M_\odot} \right)\,.
\end{equation}
Here, $M$ is the mass of the massive BH at the center and $\Delta$
is the accuracy to which one can distinguish the waveform phase in dCS
gravity from GR that depends on the SNR and the number of GW
cycles. Observe that the above proposed constraint is more than two
orders of magnitude stronger than the current bound.  The proposed
bound becomes even stronger if one can measure GW signals from
intermediate-mass ratio inspirals.

Canizares \et~\cite{Canizares:2012is} improved the above analysis in
two ways by (i) including the radiation reaction effects and (ii)
carrying out a parameter estimation study.  Regarding the first
extension, the authors took a hybrid approach~\cite{Gair:2005ih},
where one combines the PN approximations with the BH perturbation
results~\cite{Hughes:1999bq}.  They used the GR expression to take the
radiation reaction effect into account, and hence, the dCS
modification to the waveform again comes from the correction to the
background BH geometry.  Regarding the second extension, they carried
out a Fisher analysis and found the proposed bound with LISA to be
$\xi^{1/4} < 1.4 \times 10^4$km, with $M = 5 \times 10^5 M_\odot$,
which is 10 times stronger than the bound in
Eq.~(\ref{eq:CS-bound-EMRI}).

Unlike the above analyses that focused on EMRIs with a slowly-rotating
massive BH at the center, Pani \et~\cite{pani-DCS-EMRI} considered
EMRIs with non-spinning BHs. Although an isolated non-spinning BH
solution in dCS gravity is the same as that in GR, GWs from
non-spinning BH binaries still acquire dCS corrections from the
dissipative sector (radiation reaction). The authors solved a set of
three differential equations that govern the metric polar and axial
perturbations and the scalar field perturbation using Green's function
techniques.  Using these perturbations, they found that dCS
corrections to the gravitational and scalar radiation to spatial
infinity are of 6PN and 7PN orders relative to the GR leading
quadrupolar radiation. Interestingly, those to the horizon are larger
and of 5PN order relative to GR. They also showed that the dCS
correction to the number of GW cycles can exceed unity depending on
the mass of the binary constituents and the CS coupling parameter.
Here, the number of GW cycles exceeding unity
should be take as a necessary and not sufficient condition for the 
dCS effect to be detected, as one needs much more cycles due to
e.g.~correlations between parameters and systematics.

Reference~\cite{Yagi:2012vf} derived gravitational waveforms from slowly-rotating 
BH binaries with comparable masses in 
dCS gravity. Two corrections exist; conservative and dissipative. 
The former arises from the correction to the quadrupole moment of a BH solution as discussed in 
Sec.~\ref{sec:dCS-BH}, which enters at 2PN order relative to the leading Newtonian term in the waveform.
Such a BH solution also has a scalar dipole charge as in Eq.~\eqref{eq:CS-BH-scalar}, 
whose magnitude is given by $\mu_\CS = (5/8) \alpha_\CS \chi$ 
valid to leading order in spin~\cite{Yagi:2011xp}.
This scalar dipole charge induces a dipole-dipole interaction between two BHs, 
which also enters at 2PN order in the waveform.
Such an effect is similar to magnetic dipole interactions of two NSs~\cite{Ioka:2000yb}.
Regarding the dissipative corrections, Ref.~\cite{Yagi:2011xp} derived the energy flux correction
due to the scalar radiation that enters again at
2PN order.
Combining all of these modifications, the authors in~\cite{Yagi:2012vf} derived corrections to 
gravitational waveforms from BH binaries that can easily be mapped to the 
PPE parameters in the gravitational waveform in Eq.~\eqref{eq:simplest-PPE}.
%
Carrying out a Fisher analysis, Ref.~\cite{Yagi:2012vf} found that second-generation ground-based 
GW interferometers, such as Adv.~LIGO, may be able to place constraints that are six orders of magnitude
stronger than the current strongest bound from solar system~\cite{AliHaimoud:2011fw}
and table-top~\cite{Yagi:2012ya} experiments. 
  
\subsubsection{dQG}
\label{sec:GW-dQG}

We now move onto GW constraints on dQG. The correction to the energy flux emitted from a BH binary was
calculated in~\cite{Yagi:2011xp}. The authors perturbed the field equations around a flat background and 
derived the wave equations for the scalar field that is sourced by scalar charges in 
Eqs.~\eqref{eq:BH-scalar-dQG} and~\eqref{eq:CS-BH-scalar}.
Solving such wave equations in the near zone, one can reproduce the BH scalar field solution in
Eqs.~\eqref{eq:BH-scalar-dQG} and~\eqref{eq:CS-BH-scalar} to leading order in $M/r$.
One can then calculate the energy flux $\dot E^{(\vartheta)}$ of the scalar radiation emitted from a BH binary by solving such 
wave equations in the far zone. 

Let us first focus on the even-parity sector, where one finds the relative scalar energy flux from gravitational 
energy flux in GR as~\cite{Yagi:2011xp}
\be
\dot E^{(\vartheta)} = \dot E_\GR \left( 1 + \frac{5}{96} \zeta_3 \frac{\delta^2}{\eta^4} v^{-2} \right)\,,
\ee
where $\zeta_3 \equiv \alpha_3/(\kappa m^4)$, $v$ is the orbital velocity of the binary 
and we recall $\delta \equiv (m_1-m_2)/m$. Notice that such a correction is of $-1$PN order
relative to GR.
One can then calculate the correction to the gravitational waveform phase, which
can easily be mapped to the PPE waveform in Eq.~\eqref{eq:simplest-PPE} as~\cite{Yagi:2011xp}
\be
\beta_\ppE = -\frac{5}{7168} \zeta_3 \frac{\delta^2}{\eta^{18/5}}\,, \qquad b_\ppE = -\frac{7}{3}\,.
\ee
Using such a waveform and based on a Bayesian analysis in~\cite{cornish-PPE}, 
one obtains future projected constraints with Adv.~LIGO that are the same as 
Eqs.~\eqref{eq:GB-GW-BH-NS} and~\eqref{eq:DDGB-BHBH-Const} 
but replacing $\alpha_\EDGB$ with $\alpha_3$~\cite{Yagi:2011xp,Yagi:2015oca}. Approximate constraints with other
GW interferometers were calculated in~\cite{kent-LMXB} using a Fisher analysis.

Regarding the odd parity sector, one finds that if BHs are spinning, 
both the scalar and gravitational radiation gives 2PN 
correction to the waveform relative to GR, as already mentioned in Sec.~\ref{sec:dCS-GW}.
On the other hand, if BHs are non-spinning, 
the corrections to the energy flux are suppressed to 7PN (scalar radiation) and 6PN (gravitational radiation) order. 
The scalar energy flux obtained analytically within the PN approximation in~\cite{Yagi:2011xp} for a non-spinning BH binary
agrees beautifully with the numerical results in~\cite{pani-DCS-EMRI}.

\subsection{EMW Tests}
\label{sec:quad-electr-wave-tests}

We here review current and projected constraints on quadratic gravity with EMW observations
from a system containing a BH.

\subsubsection{EdGB}
\label{sec:edgb}

Regarding proposed constraints on EdGB gravity, Maselli
\et~\cite{Maselli:2014fca} calculated QPO frequencies using a
slowly-rotating BH solution to linear order in spin
constructed in~\cite{Pani:2009wy}. The authors adopted the
relativistic precession model~\cite{Stella:1997tc,Stella:1998mq}
(see Sec.~\ref{sec:QPO} for details).
This model was applied to a BH system GRO J1655-40, where three QPOs
were observed with the Rossi X-ray Timing Explorer
(RXTE)~\cite{Motta:2013wga} within $\sim 1\%$ accuracy.  In GR, such a
QPO triplet can be solved for the BH mass, spin and the QPO radius.
In order to constrain non-GR theories, one needs additional information
to constrain coupling parameters in such theories. The authors
in~\cite{Maselli:2014fca} assumed that future X-ray satellite LOFT
will detect \emph{two} QPO triplets from the same BH with different
QPO radii with the measurement accuracy that is 15 times higher than
that with RXTE. They calculated QPO triplets for a fiducial BH in EdGB
gravity with the dimensionless coupling parameter of
$\zeta_\EDGB (\equiv \alpha_\EDGB^2/(\kappa M^4)) = 0.1$ and recovered the BH
mass and spin assuming that GR is the correct theory.  One can
constrain the theory by checking the consistency on the recovered mass
and spin from each QPO triplets. They carried out a Monte Carlo
simulation, together with a $\chi^2$-test, and found that future X-ray
observations may be able to place constraints that are slightly
stronger than the current theoretical bound derived from the existence
of a BH solution.

\subsubsection{dCS}
\label{sec:dcs}

We now review EMW tests on dCS gravity. The dCS correction to the BH
shadows of slowly-rotating BHs to linear order in spin was studied
in~\cite{Amarilla:2010zq}.  As shown in~\cite{Sopuerta:2009iy}, a
Carter-like constant exists for such a BH solution. Thus, null
geodesic equations separate and one can follow the same procedure
in GR to calculate the dCS correction to the shadows. The authors
in~\cite{Amarilla:2010zq} showed that such shadows are indeed affected
by a non-vanishing coupling constant of the theory, but whether such
an effect can be measured with future observations remains unclear,
especially when one takes dCS gravity as an effective theory.  The
continuum X-ray spectrum from a geometrically thin, optically thick
accretion disk around a slowly rotating BH to linear order in spin in
dCS gravity was calculated in~\cite{Harko:2009kj}.  The authors showed that the dCS
correction to the spectrum can clearly be seen when the dimensionless
coupling parameter $\zeta_\CS (\equiv \xi_\CS/M^4)$ is of
$\mathcal{O}(10-100)$. We note that such a large coupling constant is
beyond the small coupling approximation if one wishes to treat the
theory as an effective theory.

Vincent~\cite{Vincent:2013uea} improved the above analyses by taking
the light bending effect into account by solving the null geodesic
equations for a slowly-rotating BH to linear order in spin in dCS
using the ray-tracing algorithm with the open-source code
GYOTO~\cite{Vincent:2011wz}.  The author also calculated the dCS
correction to the iron line emission and QPO frequencies, with the
epicyclic resonance~\cite{2001AcPPB..32.3605K,Abramowicz:2001bi} and
hot spot~\cite{2004ApJ...606.1098S} models for the latter.  The author
found that the typical dCS deviation from GR in these observables is
within $\sim \mathcal{O}(0.1\%)$, which makes it extremely difficult
to be measured with current or near-future observations.
Moore and Gair~\cite{Moore:2015bxa} recently calculated the iron line spectrum
from an accretion disk around a rotating BH to quadratic order in spin
in dCS gravity. They carried out a Bayesian parameter estimation study and
found that it would be difficult to place a meaningful constraint on the theory
with future observations. 

\subsubsection{dQG}

Let us now consider how one can constrain dQG with current EMW
observations.  As proved in~\cite{Yagi:2011xp,Yagi:2015oca}, ordinary stars such as
NSs do not possess monopole scalar hair.  This, in turn, means that
dipolar radiation is absent from a binary system with ordinary stars.
Therefore, a stringent constraint on the theory comes from a binary
system where at least one of the constituents is a BH.

One example is a BH-LMXB.  Based on the scalar radiation calculation
in~\cite{Yagi:2011xp}, Ref.~\cite{kent-LMXB} derived a constraint on
the theory from the upper bound on the orbital decay rate of
A0620-00~\cite{johannsen1}.
If one takes the Gauss-Bonnet combination of 
$(\alpha_1,\alpha_2,\alpha_3,\alpha_4) = (1,-4,1,0)\alpha_\GB$,
the bound is obtained as $\sqrt{|\alpha_\GB|} < 1.9$km. 
Such a constraint is six
orders of magnitude stronger than the solar system
bound~\cite{amendola}.  Although such a LMXB system has larger
astrophysical uncertainties than binary pulsars, the constraint
depends only weakly on such uncertainties.  Furthermore,
Ref.~\cite{kent-LMXB} showed that the unknown excess in the orbital
decay rate of XTE J1118+480~\cite{GonzalezHernandez:2011aa} compared
to the General-Relativity prediction can be explained by additional
scalar radiation in EdGB gravity. Interestingly, such an excess cannot
be explained with additional radiation in most of other alternative
theories of gravity as they have already been constrained strongly
from solar system experiments and binary pulsar observations.

Another example is to consider a BH/pulsar
binary.  
Based on~\cite{Liu:2014uka}, Ref.~\cite{Yagi:2015oca}
derived a proposed constraint on the theory with a BH/pulsar
system. The authors found the future projected constraint for the Gauss-Bonnet combination
of the coupling constants as
\begin{equation}
\sqrt{|\alpha_\GB|} < 0.065 \, \mrm{km} \left( \frac{M}{10M_\odot} \right) \left( \frac{\Delta \dot P}{10^{-3}} \right)^{1/4}
\left( \frac{v}{10^{-3}} \right)^{1/2}\,,
\end{equation}
where $\Delta \dot P$ is the measurement accuracy of the orbital decay
rate.  Such a constraint is indeed more than one order of magnitude
stronger than the LMXB bound.

\section{Large Extra Dimension}
\label{sec:large-extra-dimens}

\subsection{Basics}
\label{sec:large-extra-basics}

String theory predicts that our universe has more than four
dimensions~\cite{polchinski1,polchinski2} with extra dimensions being
compactified in a certain way.  One well-known and simple example of
such a compactification is the Kaluza-Klein compactification.  Particle
physics experiments place a strong bound on the size of the extra
dimension $\ell$ as $\ell \le 10^{-16}$cm.  Arkani-Hamed \textit{et
  al}.~\cite{arkani1,arkani2} proposed a braneworld model (the ADD
model), where the authors embedded a tension-less brane (which we live
on) in a flat and compact bulk spacetime. They also assumed that
ordinary matter is localized on the brane.  Then, only gravitons can
propagate through the bulk.  The size of extra dimensions can be
relatively large in the ADD model since the constraint on the gravity
sector is not as strong as that on the matter sector.
Moreover, the ADD model provides a novel explanation on the hierarchy
problem between the Planck scale and the electroweak scale.  Below, we
describe in detail a different type of braneworld models proposed by
Randall and Sundrum~\cite{randall1,randall2}.

\begin{figure}[htb]
\begin{center}
\includegraphics[width=11.5cm,clip=true]{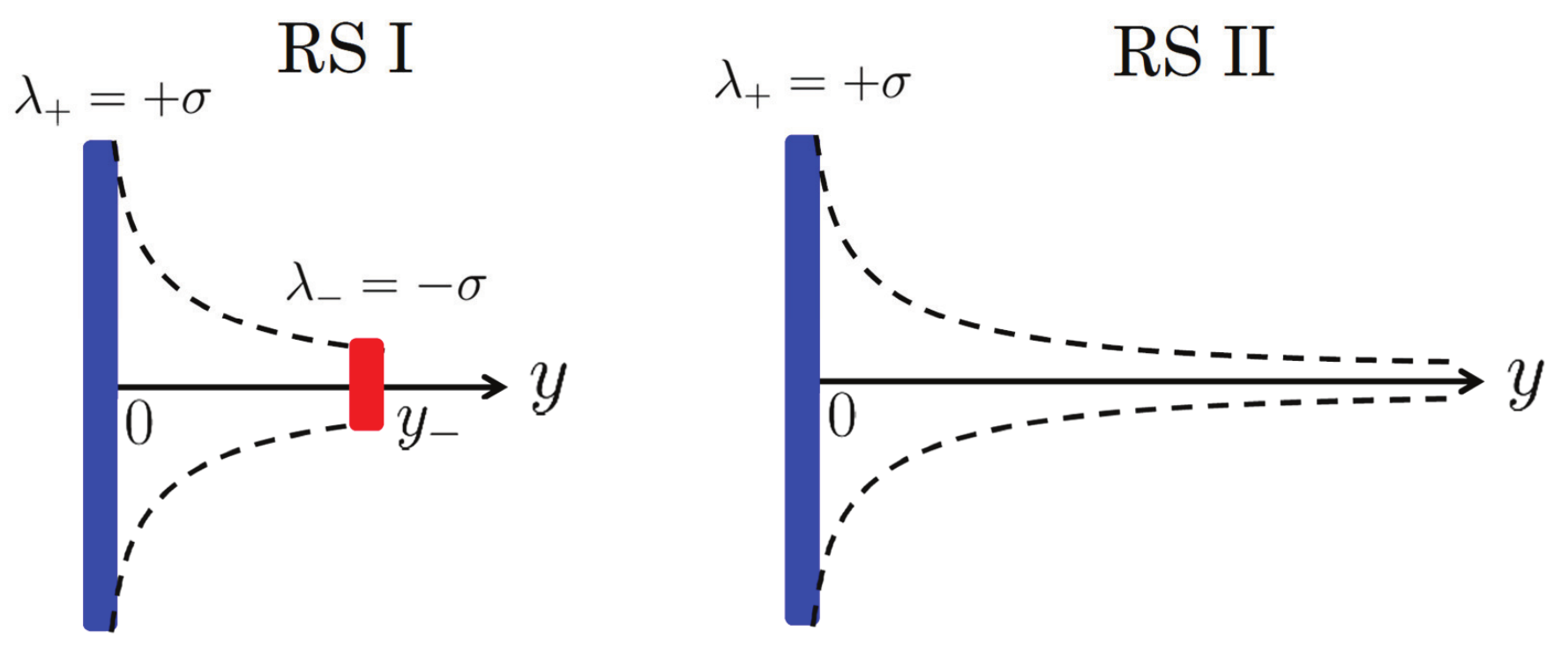}
\caption{\label{fig:brane} (Color online)
  Schematic picture of the RS I (left) and RS II (right) braneworld
  models, with $y$ representing the coordinate for the extra
  dimension. The bulk spacetime is AdS.  The RS I model has two branes
  with their tension represented by $\lambda_\pm$ and we live on the
  one with a negative tension.  The RS II model has one brane (with a
  positive tension).
}
\end{center}
\end{figure}

\subsubsection{RS braneworld Models}
\label{sec:rs-braneworld-models}

First, we explain the first RS braneworld model (the RS I model).  Let
us consider a (4+1)-dimensional theory with two
branes~\cite{randall1}, one with a positive tension at $y=y_+=0$ and
the other with a negative tension at $y=y_-$, where $y$ represents the
coordinate of the extra dimension (see Fig.~\ref{fig:brane}).  We
impose the $Z_2$ symmetry at $y=0$.  The action is given
by\footnote{In this section, we do not set the gravitational constant
to unity.}~\cite{randall1,tanahashi}
\begin{eqnarray}
S &=& \frac{1}{16\pi G_5} \int d^5x \sqrt{-g} (R-2\Lambda_5) \nn \\
& & - \int dy d^4x \sqrt{-\gamma} \left\{ (\lambda_+ + L_{m+})\delta (y) + (\lambda_- + L_{m-})\delta (y-y_-) \right\}\,.
\end{eqnarray}
Here, $G_5$ and $\Lambda_5$ are the five dimensional gravitational and
cosmological constant respectively, $\gamma_{MN}$ represents the four
dimensional induced metric on the $y=$const. surface with the indices
$(M,N)$ running from 0 to 4, $\lambda_{\pm}$ are tensions and
$L_{m\pm}$ are the matter Lagrangians for the positive and negative
tension branes respectively.
%
%
%
Imposing the flat brane ansatz for the metric as 
\begin{equation}
\label{eq:ansatz-RS-I}
ds^2 = a^2(y) \eta_{\mu\nu} dx^{\mu} dx^\nu + dy^2\,,
\end{equation}
with $a(y)$ representing the warp factor,
%
%
one solves the modified Einstein equations and finds
%
%
\begin{equation}
a(y) = e^{-|y|/\ell}\,, \quad
\lambda_+ = -\lambda_- = \frac{3}{4\pi G_5 \ell} \equiv \sigma\,.
\end{equation}
Here, $\ell$ denotes the AdS curvature radius and is related to
$\Lambda_5$ via $\Lambda_5= -6/\ell^2$.

One of the interesting feature about this model is that the hierarchy
problem can be explained naturally.
Assuming that we live on the negative-tension brane, one finds
\begin{equation}
\frac{1}{G_4} = \frac{\ell}{G_5} (e^{2y_-/\ell}-1)\,.
\end{equation}
Therefore, by choosing $y_-/\ell \sim 37$, we obtain the
four-dimensional Planck mass $M_4 = 10^{19}$GeV with the fundamental
five-dimensional Planck mass $M_5 \sim \ell^{-1} \sim 1$TeV.  (Here,
we set $\hbar=1$.)


Next, let us explain the second model~\cite{randall2} (the RS II
model), where we take the negative tension brane to infinity.
Therefore only one brane exists in this model and the size of the extra
dimension is infinite (see Fig.~\ref{fig:brane}).  We assume that we
live on the positive tension brane.  Although this model cannot solve
the hierarchy problem, the remarkable feature is that the Newtonian
gravity is reproduced in the low energy limit even though it has an
infinite size of extra dimension~\cite{garriga}.
Perturbing the metric ansatz in Eq.~(\ref{eq:ansatz-RS-I}) and solving
the modified Einstein equations, one can show that all the KK modes of
gravitons are suppressed and the gravity is localized on the brane.  The
gravitational potential is given by~\cite{randall2,garriga}
\begin{equation}
\label{pot}
V(r) = V_\mathrm{N} (r) \left( 1 + \frac{2}{3} \frac{\ell^2}{r^2} \right)\,,
\end{equation}
where $V_\mathrm{N} (r)$ represents the Newtonian potential.  Observe
that the above equation reduces to the Newtonian potential when
$\ell \ll r$.  Current table-top experiments place a constraint
as~\cite{adelberger}
\begin{equation}
\ell \leq 14 \mu \mrm{m}\,.
\end{equation}
%

\subsubsection{Emparan-Fabbri-Kaloper-Tanaka Conjecture}

Next, we describe an important conjecture on a brane-localized black
hole in RS-II braneworld model based on the AdS/CFT correspondence.  Such
a correspondence states that the gravity in the $AdS_5 \times S^5$
spacetime can be interpreted as the four-dimensional $\mathcal{N}=4$
$U(N)$ super Yang-Mills theory on the AdS
boundary~\cite{maldacena,aharony}.  This correspondence is an example
of the ``holographic'' principle~\cite{'tHooft:1993gx,susskind}.
When such a correspondence is applied to the RS-II braneworld
model~\cite{hawking-brane}, one finds that the five-dimensional RS-II
gravity corresponds to the four-dimensional CFT on the brane
interacting with the four-dimensional gravity and matter.
Such a correspondence is called the \textit{bulk/brane correspondence}
and many supporting evidence exists as discussed
in~\cite{garrigasasaki,duff,anderson1,anderson2,shiromizu1,
  shiromizu2,tanaka-friedmann,grisa}.

Regarding the CFT point of view, BHs evaporate via Hawking
radiation~\cite{hawking-rad}.  The BH temperature is given by
$T=\kappa_{g}/2\pi$ (under the unit $\hbar=1$) where $\kappa_{g}$ is the
surface gravity of the BH and $\kappa_{g} = (4G_4 M)^{-1}$ for the
Schwarzschild BH.  The radiated flux can be approximated by the black
body radiation as $F\propto T^4$.  Since the area of the BH event
horizon is given by
\begin{equation}
A^{(4\mrm{D})}=4\pi (2G_4 M)^2\,,
\label{area-4D}
\end{equation}
the BH mass loss rate is given by
$\dot{M} = FA^{(4\mrm{D})} = - (15360 \pi G_4^2 M^2)^{-1}$.
Therefore, in GR, the evaporation time scale becomes
$\tau^{(\mrm{GR})} = M/\dot{M} = 2.1\times 10^{67} (M/M_\odot)^3 \
\mrm{yr}$.
On the other hand, CFT has a huge number of degrees of freedom of
$N^2 \approx \pi \ell^2/G_4 =2.36 \times 10^{60} (\ell/14\mrm{\mu
  m})^2$~\cite{aharony}.
Thus, the Hawking radiation of a BH on a brane is enhanced compared to that in
GR by a factor $\sim N^2$
and becomes~\cite{emparan}
\begin{equation}
\dot{M} \propto \frac{\ell^2}{G_4^3 M^2}\,.
\label{massloss-rate}
\end{equation}
Keeping the coefficient explicitly, one finds
\begin{equation}
\dot{M}
= -2.8\times10^{-7}
\left( \frac{1 M_{\odot}}{M} \right)^2 \left( \frac{\ell}{10\mu
\mathrm{m}} \right)^2 M_{\odot} \mathrm{yr}^{-1}
=:-\cmdot\left( \frac{\ell}{M} \right)^2 , \label{massloss}
\end{equation}
where $\cmdot$ is defined as the coefficient of the mass loss rate for
later use.  This leads to the evaporation time scale as
\begin{equation}
\tau = 5.93 \times 10^5 \left( \frac{14\mrm{\mu m}}{\ell} \right)^2 \left( \frac{M}{M_\odot} \right)^3 \ \mrm{yr}\,.
\label{lifetime}
\end{equation}
From the above consideration, one finds that the four-dimensional BH
on a brane cannot remain static due to the enhanced Hawking radiation.

When we interpret such a phenomenon from the five-dimensional point of
view, one finds that BHs lose their mass \textit{classically}.  This
is the ``classical BH evaporation conjecture'' proposed by
Emparan \textit{et~al.}~\cite{emparan-conj} and
Tanaka~\cite{tanaka-conj}.  One possible explanation for this
phenomenon in the classical picture is as
follows~\cite{tanaka-conj,tanahashi}.  Let us first consider a
brane-localized BH, where the tip of such a BH is expected to form a
blob having an area $\sim \ell^3$ with the dynamical time scale
$\sim G_4 M$ due to the Gregory-Laflamme instability~\cite{gregory}.
Since a universal acceleration $\sim \ell^{-1}$ exists towards the
bulk direction~\cite{gregory-accel}, such a blob is \emph{effectively}
pinched from the BH localized on a brane.  Therefore, one obtains the
relation
\begin{equation}
\label{eq:mass-loss-5D}
\dot{A}^{(5\mrm{D})} \approx -\frac{\ell^3}{G_4 M}\,.
\end{equation}
On the other hand, the surface area of the five-dimension BH is given
by
\begin{equation}
A^{(5\mrm{D})} = A^{(4\mrm{D})} 2 \int_0^\infty e^{-y/\ell} dy = 2 A^{(4\mrm{D})} \ell\,.
\label{area-5D}
\end{equation}
From Eqs.~(\ref{area-4D}),~(\ref{eq:mass-loss-5D})
and~(\ref{area-5D}), one finds an agreement with the mass loss rate
given by Eq.~(\ref{massloss-rate}) obtained from the four-dimensional
CFT side.

\subsection{GW Tests}
\label{sec:large-extra-gw-tests}

In this section, we review possible future constraints with GW
observations.  One example is the work by Inoue and
Tanaka~\cite{inoue}, in which the authors derived the leading
correction to the GW phase of compact binaries due to the correction to the gravitational
potential given by Eq.~(\ref{pot}), which is proportional to
$\ell^2/a^2 = (\ell^2/m^2) (m^2/a^2)$ relative to the Newtonian term.
When the Planck scale is reduced to 
the electroweak scale ($\sim 1$TeV), the early Universe may had a violent
``mesoscopic'' activity~\cite{Hogan:2000is} when the temperature was 1TeV and
large-amplitude fluctuations were produced, forming primordial BHs 
with masses $\sim 10^{-7}M_\odot$.
The authors found a rather weak
upper bound on $\ell$ assuming that third-generation GW
interferometers detect GW signals from binaries with such primordial BHs. 
Primordial BH binaries have an advantage over astrophysical ones as the
relative correction to the waveform phase
becomes larger for a fixed $\ell$ and the velocity (or $m/a$).
Below, we review other proposed constraints in detail.

We note that since static, brane-localized BH solutions have recently been
numerically constructed in~\cite{Figueras:2011gd,Abdolrahimi:2012qi}, 
which contradict with the above prediction
that brane-localized BHs cannot be static due to the enhanced Hawking radiation,
the validity of the Emparan-Fabbri-Kaloper-Tanaka conjecture is now questionable.
However, we proceed and present gravitational waveforms 
from BH binaries in RS-II model
because such waveforms can easily be modified and applied to a 
binary with BHs losing their mass in general,
such as due to phantom energy accretion\footnote{%
Assuming the Schwarzschild spacetime with a time-varying mass for the metric
and a perfect fluid for matter with pressure $p$ and energy density $\rho$, 
one can show from the Einstein equations
that the mass accretion rate is proportional to 
$\rho+p$~\cite{Babichev:2004yx,Babichev:2005py,Babichev:2014lda}.
Therefore, the accretion of the phantom energy with $\rho+p < 0$ decreases the BH mass.
Since it violates the weak energy condition, the theorem by Christodoulou~\cite{Christodoulou:1970wf} 
and Hawking~\cite{hawking-uniqueness0}  does not apply, which 
proves that the horizon area of a BH cannot decrease with any classical processes if 
the weak energy condition is satisfied.
}~\cite{Babichev:2004yx,Babichev:2005py,Babichev:2014lda}
(one may also use GWs including the effect of 
gas~\cite{Barausse:2007dy,kocsis-disk} and dark matter~\cite{Macedo:2013qea} accretion
onto binary constituents).
Such waveforms are also similar to those in varying $G$ theories~\cite{yunespretorius}.

\subsubsection{Monochromatic Signals}
\label{sec:monochr-sign}

LISA may detect GW signals that are almost monochromatic from a
galactic binary composed of a BH and a NS with mass $m_1$ and $m_2$
respectively.
Although GW emission shrinks the orbital separation $r_{12}$ (inspiral),
the BH mass loss due to Hawking radiation increases $r_{12}$ (outspiral)
at a rate of~\cite{mc}
\be
\dot{r}_{12,H} = -\frac{\dot{m}_1}{m} a = 3.2 \times 10^{-9} \left( \frac{r_{12}}{1 \mrm{AU}} \right)
 \left( \frac{m}{7M_\odot} \right)^{-1} \left( \frac{m_1}{5M_\odot} \right)^{-2}
 \left( \frac{\ell}{14 \mu \mrm{m}} \right)^{2} \ \frac{\mrm{AU}}{\mrm{yr}}
 \label{rhawk1}
\ee
due to the conservation of the specific orbital angular momentum
$\sqrt{m r_{12}}$,
where we used the mass loss rate given by Eq.~(\ref{massloss}).
Equating $\dot{r}_\mrm{12,GW}$ and $\dot{r}_{12,H}$
with the former calculated from the GW quadrupole formula,
one finds a critical separation as~\cite{mc}
\begin{equation}
r_\mrm{12,crit} \equiv 1.1 \times 10^{-2} \left( \frac{m_1}{5M_\odot} \right)^{3/4}
\left( \frac{m_2}{2M_\odot} \right)^{1/4} \left( \frac{m}{7M_\odot} \right)^{1/2}
\left( \frac{\ell}{14 \mu \mrm{m}} \right)^{-1/2} \ \mrm{AU}\,.
\end{equation}

If the separation is larger than $r_\mrm{12,crit}$, the mass loss effect
dominates the GW emission and the separation becomes larger.  On the
other hand, if $r_{12}$ is smaller than $r_\mrm{12,crit}$, GW emission
dominates and the separation becomes smaller.  Typically, a galactic
BH binary forms
at a GW frequency slightly outside of the LISA sensitivity band.
Thus, if its signal is detected at e.g.~$f=10^{-4}$Hz, one immediately
finds $r_{12}(f=10^{-4}\mathrm{Hz}) \le r_\mrm{12,crit}$, which then leads
to~\cite{mc}
\begin{equation}
\ell \le 22 \left(\frac{m_1}{5 M_{\odot}} \right)^{3/2} \left(\frac{m}{7 M_{\odot}} \right)^{1/3} \left(\frac{m_2}{2 M_{\odot}} \right)^{1/2} \left(\frac{f}{10^{-4}\mathrm{Hz}} \right)^{4/3} \mu \mathrm{m}\,.
\end{equation}

\subsubsection{Chirping Signals}
\label{sec:chirping-signals}


Although a typical galactic BH binary forms at a GW frequency slightly
lower than the low frequency limit of the LISA sensitivity band, some
of them may form with a frequency higher than this lower frequency
limit~\cite{belczynski}.  Therefore, a systematic
error may exist when constraining the size of the extra dimension with
a monochromatic GW signal.  One obtains a more robust constraint by
measuring the actual inspiral of a BH binary.  Here, we first review
the correction to the GW phase due to the mass loss effect derived
in~\cite{yagi:brane}.  Then, we explain possible constraints on $\ell$
from a parameter estimation study with LISA and DECIGO/BBO.  We assume
that binaries are quasi-circular and neglect the spins of binary
constituents for simplicity.

Let us first consider GWs from a BH/BH binary with component masses of
$m_1$ and $m_2$ (with $m_1\geq m_2$).  The rate at which the orbital
separation changes due to the mass loss effect can be derived from
Eqs.~(\ref{massloss}) and~(\ref{rhawk1}).
Then, the rate at which the GW frequency changes becomes
\begin{equation}
\dot{f} = \frac{\dot{\Omega}}{\pi}
 =   \frac{96}{5}\pi^{8/3} \mathcal{M}^{5/3} f^{11/3} \left( 1 - \frac{5}{48}\cmdot\frac{1-2\eta }{\eta^3} \frac{\ell^2}{m^2} v^{-8} \right)\,.
\label{fdot-brane}
\end{equation}
From this equation, one derives the correction to the GW phase to
leading order in PN approximation, which can be mapped to the
PPE waveform in Eq.~\eqref{eq:simplest-PPE} as~\cite{yagi:brane}
\be
\label{eq:GWphase-brane}
\beta_\ppE = - \frac{25}{851968} \eta_0^{8/5} \cmdot C \frac{\ell^2}{m_0^2} \,, \qquad b_\ppE = -\frac{13}{3}\,,
\ee
where $C$ for a BH/BH binary is given by
\begin{equation}
C^{\mrm{(\BH / \BH)}}  \equiv  \frac{3-26\eta_0 +34\eta_0^2}{\eta_0^4}. \label{c}
\end{equation}
The subscript 0 represents the quantity at the time of coalescence.
The second term in brackets in Eq.~(\ref{eq:GWphase-brane})
corresponds to the ``$-4$PN'' relative correction due to the mass loss effect
valid to $O(\ell^2)$.
For a BH/NS binary, the coefficient $C$ in Eq.~(\ref{c}) changes to
\begin{equation}
C^{\mrm{(\BH / \NS)}}  =  \frac{(3-26 \eta_0 +34 \eta_0^2 )+(-3+20 \eta_0 ) \sqrt{1-4\eta_0}}{2 \eta_0^4}.
\end{equation}
Notice that the GW phase in Eq.~(\ref{eq:GWphase-brane}) is similar to
that for varying $G$ theories derived in~\cite{yunespretorius}.

\begin{figure}[t]
\begin{center}
\includegraphics[width=7.cm,clip=true]{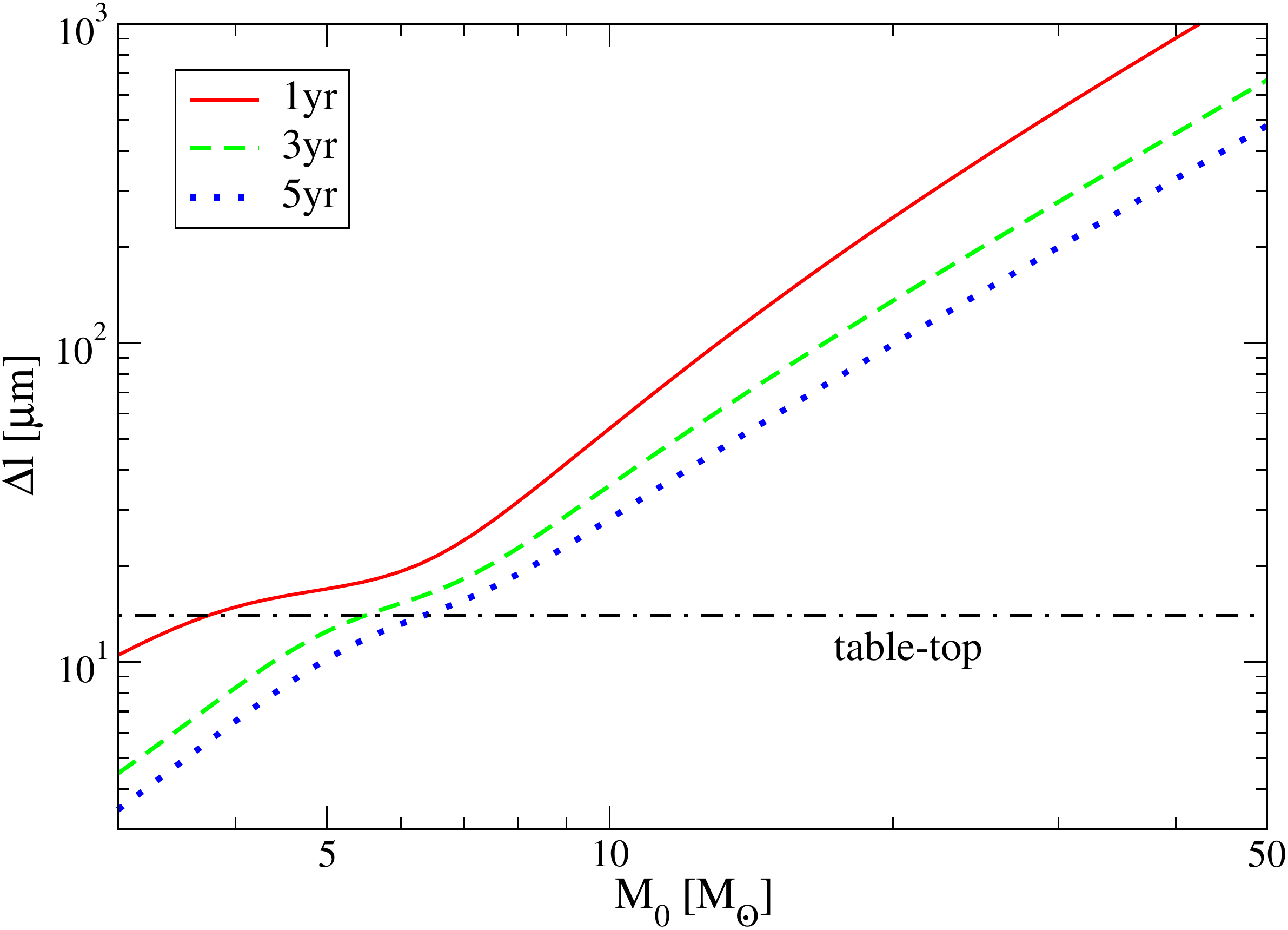}
\includegraphics[width=7.cm,clip=true]{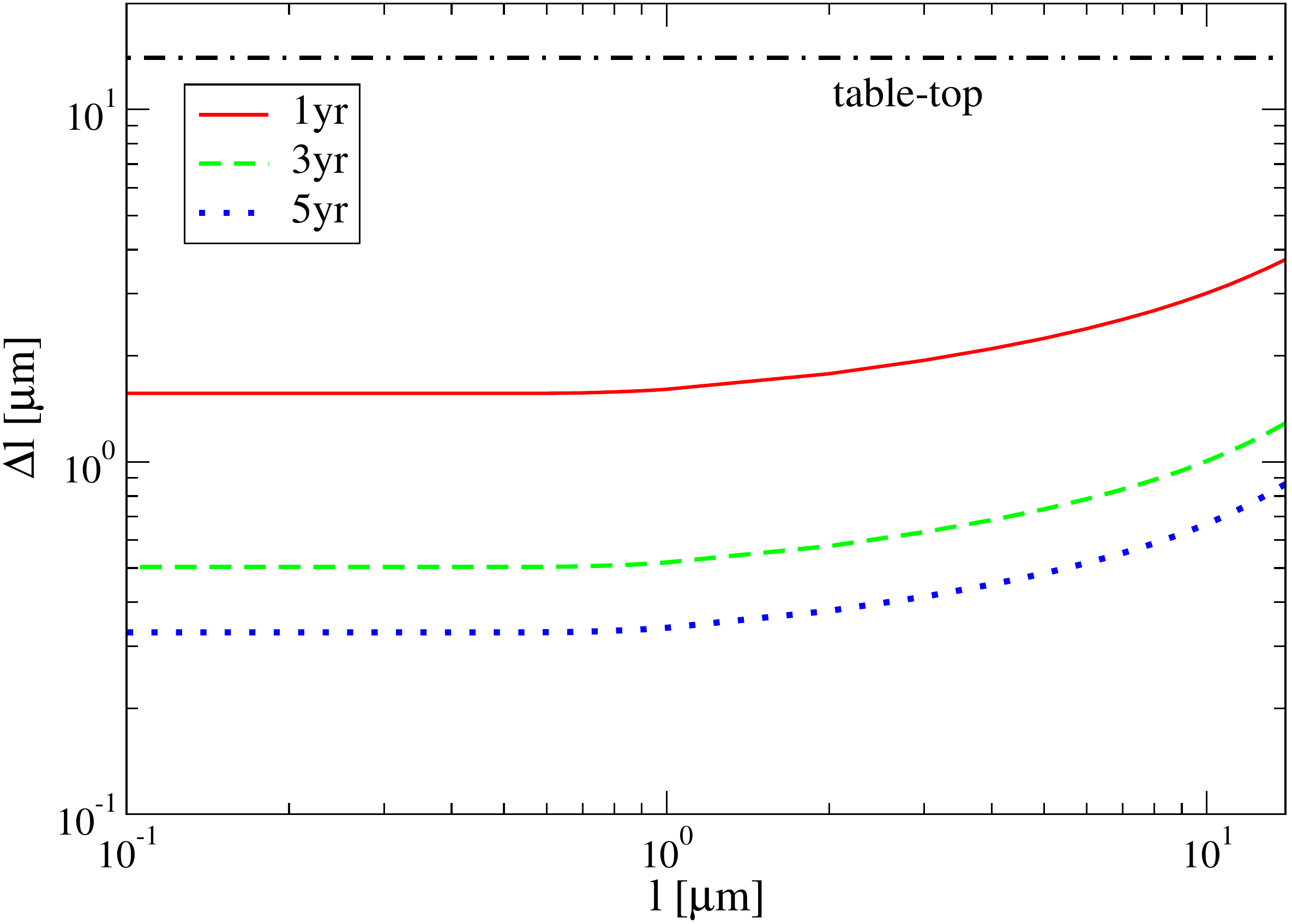}
\caption{\label{bbo-equal}
  (Left) Proposed upper bound on $\ell$ with DECIGO/BBO from an equal
  mass BH/BH binary at $D_L=3$Gpc as a function of the BH mass $M_0$ with various
  observation period. The horizontal dotted-dashed line shows the
  current strongest bound from table-top
  experiments~\cite{adelberger}. Observe that the bounds from GWs can
  be stronger than the current bound for low-mass binaries.  (Right)
  Proposed upper bound on $\ell$ with DECIGO/BBO from a large number
  of NS/BH binaries detected as a function of fiducial $\ell$. Observe that if
  GR is correct, the bounds from GWs can be more than 10 times
  stronger than the current bound as shown by the dotted-dashed line.
  This figure is taken and edited from~\cite{yagi:brane}.
}
\end{center}
\end{figure}

Reference~\cite{yagi:brane} carried out a Fisher analysis and derived
proposed bounds on $\ell$ assuming future space-borne GW
interferometers detect GW chirp signals from BH/BH or BH/NS binaries.
The authors showed that LISA can only place constraints that are 5--10
times weaker than the current bound from table-top experiments.  On
the other hand, DECIGO/BBO can place stronger constraints than the
current bound.  The left panel of Fig.~\ref{bbo-equal} presents one
example, where constraints on $\ell$ with DECIGO/BBO are shown as a
function of the BH mass at the time of coalescence $
M_0 (= m_1 = m_2)$ for an equal mass BH/BH binary with various
observation time. Observe that the constraints become stronger for
low-mass binaries.  This is because the effect of enhanced Hawking
radiation becomes larger for BHs with smaller masses.

DECIGO/BBO can place stronger constraints on $\ell$ by taking the
advantage of the fact that the expected detection rate of BH/NS
binaries is $\sim 10^4$. Such a detection rate depends on $\ell$,
because the detection rate is reduced from that in GR if the enhanced
Hawking radiation exists.  The right panel of Fig.~\ref{bbo-equal}
presents the proposed bound $\ell$ as a function of fiducial $\ell$, taking the
large number of detection events into account.  The bounds are flat
for small $\ell$ because in such a case, the evaporation time exceeds
the age of the universe, and hence, the detection rate is the
same as that in GR. Observe that if GR is correct, DECIGO/BBO can
place constraints that are more than 10 times stronger than the
current bound.

\subsubsection{Event Rate}
\label{sec:event-rate}

McWilliams also derived a proposed constraint on $\ell$ from the
averaged EMRI event rate $\lla \rch\rra_{\mathrm{EMRI}}$.  Such a rate
in GR is estimated in e.g.~\cite{gair}.
In the RS-II braneworld model, this rate is modified due to the BH
mass loss effect and becomes
\begin{equation}
\lla \rch \rra_{\mathrm{H}}  =  r_\HH  \lla \rch \rra_{\mathrm{EMRI}}\,, \qquad r_\HH \equiv \frac{\tau}{T_\mrm{univ}}\,,
\end{equation}
with $\tau$ and $T_\mrm{univ}$ representing the BH lifetime given by
Eq.~(\ref{lifetime}) and the age of the universe respectively. The
above expression is only valid for $r_\HH \leq 1$.  In~\cite{mc},
McWilliams assumed a Poisson probability distribution for the event
rate and derived a constraint on $\ell$ if the GR prediction of
$\lla \rch\rra_{\mathrm{EMRI}}$ is actually observed.

\Table{
\label{table:event}
Proposed constraints on $\ell$ from detection rates with LISA and
DECIGO/BBO.  We also present other parameters used to derive these
bounds. This table is taken from~\cite{yagi:brane}.
}
 \hline\hline
 Detectors, binaries, masses &   $\lla N \rra$ & $r_\mrm{H}$ & $\tau$ & $\ell$ \\
 & (1/yr)  & &(yr) &  ($\mu$m)  \\ \hline\hline
LISA, EMRI, $(5+10^6)\so$ &  3.3 $\times 10^2$ & 0.1 & $10^9$ & 3.9 \\ \hline
DECIGO/BBO, BH/NS, $(1.4+5)\so$ &  9.0-13$\times 10^4$ & 0.1 & 6.9$\times 10^7$ & 15  \\ \hline\hline
\endTable

Reference~\cite{yagi:brane} extended the above analysis by taking the
uncertainties in the prediction of the event rate into account.  The
authors in~\cite{yagi:brane} assumed that due to the enhanced Hawking
radiation, the detection rate $\langle N \rangle_\HH$ of GWs emitted
from BH binaries with future space-borne interferometers is reduced
from the prediction in GR $\langle N \rangle$ by
$\langle N \rangle_\HH = r_\HH \langle N \rangle$.  If the event rate
of $\langle N \rangle$ is detected as predicted in GR with
an order of magnitude uncertainty, one
obtains a lower bound on $r_\HH$ as $r_\HH \geq 0.1$.  Such a bound
sets a bound on $\tau$, which in turn places a bound on $\ell$ through
Eq.~(\ref{lifetime}).  The proposed bounds on $\ell$ with LISA and
DECIGO/BBO are summarized in Table~\ref{table:event}.  Observe that
such proposed constraints are weaker than those from chirp GW signals
mentioned in the previous subsection.

\subsection{EMW Tests}
\label{sec:large-extra-electr-wave-tests}

Current astrophysical BH observations place constraints on $\ell$ from
(i) the estimate of the BH mass and the age, and (ii) the measurement
of the orbital decay rate of BH binaries.  Regarding the former,
Psaltis~\cite{psaltis} derived a constraint of $\ell \le 80\mu$m from
the estimate of the age of the BH in the X-ray binary XTE J1118+480.
Gnedin \textit{et al}.~\cite{gnedin} also placed a constraint using an
estimate of the age of the BH in the extra-galactic globular cluster
RZ2109 as $\ell \le 10\mu$m.
Regarding the latter, Johannsen \textit{et
  al}.~\cite{johannsen1,johannsen2} used the orbital decay rate
measured in the X-ray binaries A0620-00 and XTE J1118+480 and derived
constraints as $\ell \le 161\mu$m and $\ell \le 970\mu$m,
respectively.
Recently, Simonetti \et~\cite{simonetti} derived a proposed, 5
$\sigma$ upper bound on $\ell$ as 0.17$\mu$m, assuming one measures
the orbital decay rate of a BH-pulsar binary in future with the same
accuracy as a 30-year observation of PSR
B1913+16~\cite{weisberg2004,weisbergtaylor}.
%

\section{Open Questions}
\label{sec:open-questions}

We conclude this review by presenting a selected list of open questions in
generic ways of testing GR and in each theory
described above.

\subsection{Generic Ways of Testing GR}

\begin{itemize}
\item Bumpy spacetime that can describe e.g.~slowly rotating BH solutions
in dCS~\cite{Yagi:2012ya} and dQG~\cite{Ayzenberg:2014aka}
that are not of Petrov type D is currently lacking.
%
\item Consistency tests of GR with parameterized PN waveform proposed 
in~\cite{Arun:2006yw,arun-model-indep} have not been extended to spinning BHs.
\item PPE waveforms~\cite{Yunes:2009ke} have not been extended to precessing binaries.
\end{itemize}

\subsection{Scalar-tensor Theories}

\begin{itemize}
\item The stability of hairy BH solutions in scalar-tensor theories is currently unexplored.
\item Previous studies deriving proposed constraints on scalar-tensor theories with GW observations
of compact binaries focus on the inspiral phase only. Since merger simulations of compact binaries
in scalar-tensor theories have recently been performed~\cite{Barausse:2012da,
Palenzuela:2013hsa,Shibata:2013pra,Taniguchi:2014fqa,Healy:2011ef,Berti:2013gfa,Ponce:2014hha}, future studies may 
derive new proposed bound on these theories from GW observations by including the merger and 
ringdown phases.  
\item Gravitational radiation from inspiraling compact binaries in $f(R)$ theories, 
which are equivalent to BD theory with a certain potential, is calculated in~\cite{Naf:2011za,DeLaurentis:2011tp}.
However, the sensitivities (or scalar charges) of compact objects in such theories are lacking, which are 
crucial for deriving proposed constraints on the theories from future GW observations.
\end{itemize}

\subsection{Massive Gravity Theories}

\begin{itemize}
\item Literature that derives proposed constraints on massive gravity theories with GW observations focuses 
mostly on how the propagation of GWs are modified. However, GW generation is also modified, which affects the 
gravitational waveform. Although it may be likely that the correction to the GW propagation dominates that to the 
GW generation, it would be important to explicitly calculate the latter effect and compare which effect dominates. 
\item EMW observables for BHs in massive gravity theories remain to be calculated.
\item One needs to clarify whether a linear stability analysis of BH solutions are valid
in these theories, where the nonlinear Vainshtein screening mechanism plays an important
role.
\end{itemize}

\subsection{Quadratic Gravity}

\begin{itemize}
\item BH solutions in dCS gravity have been constructed only 
within the slow rotation approximation. Such solutions with arbitrary rotation
is currently missing.
\item Stability analysis of BH solutions in quadratic gravity is incomplete. 
Missing pieces include polar gravitational perturbations of non-rotating BHs in EdGB gravity
and any perturbations of slowly-rotating BHs to second order in spin in dQG. 
\item Proposed constraints on dCS gravity with GW observations derived in~\cite{Yagi:2012vf}
focused on spin-aligned systems and used a Fisher analysis. Important extension of such a study
includes taking spin precessions into account and carrying out a Bayesian analysis.
\item EMW observables of BHs in EdGB gravity, in particular the X-ray continuum spectrum, 
Fe line emissions and BH shadow,
remain to be studied.
\item EMW observables of slowly-rotating BHs to second order in spin in dCS gravity is unexplored.
\item Whether geodesic motions of a test particle around a slowly-rotating BH 
to quadratic order in spin in dQG and dCS become chaotic or not
due to the absence of the Carter-like constant has not been studied yet.
\end{itemize}

\subsection{Other Theories}

Important non-GR theories not covered in the main text of this review include
Einstein-\AE{}ther theory~\cite{Jacobson:2000xp,Jacobson:2008aj}, which is the most general
Lorentz-violating (yet diffeomorphism invariant) theory of gravity with a time-like unit vector and
at most second derivatives in the action. This theory includes khronometric gravity~\cite{Blas:2009qj,Blas:2010hb} 
as a certain limit, which coincides with the low-energy limit of Ho\v{r}ava-Lifshitz gravity~\cite{Horava:2009uw}.
Non-rotating BH solutions in Einstein-\AE{}ther theory (which are also solutions to khronometric gravity)
are found in~\cite{Eling:2006ec,Barausse:2011pu}, while slowly-rotating BH solutions in
khronometric gravity are constructed in~\cite{Barausse:2012ny,Wang:2012nv}.
Barausse and Sotiriou~\cite{Barausse:2013nwa} attempted to construct slowly-rotating BH solutions in Einstein-\AE{}ther theory
but found that such solutions do not seem to admit \emph{universal horizons} for scalar, vector and tensor propagating modes.
Selected open questions regarding BHs in Einstein-\AE{}ther theory and khronometric gravity are as follows:

\begin{itemize}
\item Stability analysis of BH solutions in these theories are currently lacking.
\item The BH sensitivities in these theories are unknown.
\item Once the BH sensitivities are calculated, one can derive current and proposed constraints on these theories
with BH-LMXBs and BH-pulsar observations. One can also calculate proposed constraints from BH binaries
with future GW observations.
\item EMW observables of BHs in these theories remain unexplored.
\end{itemize}

\subsection{Common Open Problems}

\begin{itemize}
\item Theories like massive gravity and quadratic gravity in the non-decoupling limit
admits various modes with different propagation speed. In such cases, multiple horizons 
should be present for these modes, like in Einstein-\AE ther theory.
To the best of our knowledge, this subject has not been fully studied yet.
\item More detailed studies of systematic errors in testing GR with GW and EMW observations 
of BHs are necessary. 
\end{itemize}

\section*{Acknowledgments}

We greatly thank Paolo Pani and Helvi Witek for inviting us to write
this article.
We also thank Enrico Barausse for carefully reading the manuscript
and giving us a lot of useful comments.
We further thank Alikram Aliev, Carlos Herdeiro and Nemanja Kaloper
for their important feedback on the manuscript.
K.Y~acknowledges support from NSF CAREER Award PHY-1250636,
NSF grant PHY-1305682
and JSPS Postdoctoral Fellowships for Research Abroad. 
L.C.S.~acknowledges that support for this work was provided by the
NASA through Einstein
Postdoctoral Fellowship Award No.~PF2-130101 issued by the Chandra
X-ray Observatory Center, which is operated by the Smithsonian
Astrophysical Observatory for and on behalf of the National
Aeronautics Space Administration under Contract No.~NAS8-03060, and
further acknowledges support from NSF Grant No.~PHY-1068541.

\section*{References}

\bibliographystyle{iopart-num}
\bibliography{ref}

\end{document}